\documentclass[letterpaper,twocolumn,10pt]{article}
\usepackage{usenix-2020-09}

\usepackage{cprotect}
\usepackage{amsmath}
\usepackage{amssymb}
\usepackage{lipsum}
\usepackage{tcolorbox}
\usepackage{xspace}
\usepackage{enumitem}
\usepackage{graphicx}
\usepackage{multirow}
\usepackage{listings}
\usepackage[noabbrev,capitalize]{cleveref}
\usepackage{cleveref}[noabbrev,capitalize]
\usepackage{filecontents}
\usepackage{tikz}
\usetikzlibrary{arrows.meta, positioning, calc}
\usepackage{tikzpagenodes}
\usepackage{graphicx,subcaption}
\usepackage{booktabs}
\usepackage{xurl}
\usepackage{appendix}
\usepackage[absolute,overlay]{textpos}
\usepackage[available,functional,reproduced]{usenixbadges}
\begin{document}
\include{pkgs}
\include{cmds}
%-------------------------------------------------------------------------------

\newcommand{\attack}{\text{GPUHammer}\xspace}
\newcommand{\TODO}[1]{{\footnotesize\color{red}[TODO: #1]}}
\newcommand*\circled[1]{\tikz[baseline=(char.base)]{
            \node[shape=circle,draw,inner sep=0.75pt] (char) {#1};}}
\newcommand{\cf}{\textit{c.f.},}

%\newcommand{\TODO}[1]{}
%don't want date printed
%\date{}
\title{\Large \bf \attack{}: Rowhammer Attacks on GPU Memories are Practical
\thanks{\textsuperscript{*}The full paper is "GPUHammer: Rowhammer Attacks on GPU Memories are Practical", Chris S. Lin, Joyce Qu, and Gururaj Saileshwar. In the Proceedings of the USENIX Security Symposium (SEC), August 2025.}}

% %for single author (just remove % characters)
\author{
 % Anonymous Authors
{\rm Chris S. Lin$^{\dagger}$}\\
University of Toronto\\
shaopenglin@cs.toronto.edu
\and
{\rm Joyce Qu$^{\dagger}$}\\
University of Toronto\\
joyce.qu@mail.utoronto.ca
\and
{\rm Gururaj Saileshwar}\\
University of Toronto\\
gururaj@cs.toronto.edu
} % end author
\maketitle

\newcommand{\REVISIONTEXTCOLOR}{black}
\newcommand{\REVISIONFIGURECOLOR}{white}
\newcommand{\REVISIONBOXCOLOR}{white}
\newcommand{\REVISIONFILLCOLOR}{white}
\newcommand{\REVISION}[1]{{\color{\REVISIONTEXTCOLOR}#1}}

\newcommand{\OTHERCHANGETEXTCOLOR}{black}
\newcommand{\OTHERCHANGEFIGURECOLOR}{white}
\newcommand{\OTHERCHANGEBOXCOLOR}{white}
\newcommand{\OTHERCHANGEFILLCOLOR}{white}
\newcommand{\OTHERCHANGE}[1]{{\color{\OTHERCHANGETEXTCOLOR}#1}}

\newcommand{\NEWSUGGESTIONSTEXTCOLOR}{black}
\newcommand{\NEWSUGGESTIONSBOXCOLOR}{white}
\newcommand{\NEWSUGGESTIONSFILLCOLOR}{white}
\newcommand{\NEWSUGGESTIONS}[1]{{\color{\NEWSUGGESTIONSTEXTCOLOR}#1}}

\newcommand{\TRC}{\text{tRC}\xspace}
\newcommand{\REF}{\text{REF}\xspace}
\newcommand{\TRH}{$\text{T}_{\text{RH}}$\xspace}
\newcommand{\ACT}{\text{ACT}\xspace}
\newcommand{\TREFI}{\text{tREFI}\xspace}
\newcommand{\TREFW}{\text{tREFW}\xspace}
\newcommand{\TRFC}{\text{tRFC}\xspace}

\definecolor{codegreen}{rgb}{0,0.6,0}
\definecolor{codegray}{rgb}{0.5,0.5,0.5}
\definecolor{codepurple}{rgb}{0.58,0,0.82}
\definecolor{backcolour}{RGB}{245,245,245}
\definecolor{keywordcolour}{RGB}{173, 16, 16}

\lstdefinestyle{mystyle}{
    language=bash,
    backgroundcolor=\color{backcolour},   
    commentstyle=\color{codegreen},
    keywordstyle=\color{keywordcolour},
    numberstyle=\tiny\color{codegray},
    stringstyle=\color{codepurple},
    basicstyle=\ttfamily\bfseries\footnotesize,
    breakatwhitespace=false,         
    breaklines=true,                 
    captionpos=b,                    
    keepspaces=true,                 
    numbers=left,                    
    numbersep=5pt,                  
    showspaces=false,                
    showstringspaces=false,
    showtabs=false,                  
    tabsize=2,
    frame=single,
    morekeywords={size_t},
    rulecolor=\color{black},
}

\lstset{style=mystyle}

\tikzset{
  revision box/.style args={#1/#2}{
    draw=#1,
    text=#1,
    font=\bfseries\color{#1},
    rounded corners,
    minimum height=0.5cm,
    text width=1.3cm,
    align=center,
    thick,
    fill=#2!20,
    inner sep=2pt
  }
}

% Define answerbox: position relative to current location
\NewDocumentCommand{\revisionbox}{O{} O{} O{} m}{

\begin{tikzpicture}[overlay, remember picture, baseline]
\node[revision box=\REVISIONBOXCOLOR/\REVISIONFILLCOLOR] at ($(0, 0) + #1$){#4};
\draw[\REVISIONFILLCOLOR!50, line width=1.5pt]
      ($ (0,0) + #2 + (0.6cm, 0) $) -- 
      ($ (0,0) + #2 + (10.07cm, 0) $);
\draw[\REVISIONFILLCOLOR!50, line width=1.5pt]
      ($ (0,0) + #3 + (10.05cm, 0) $) -- 
      ($ (0,0) + #3 + (10.05cm, 0.33cm) $);
\end{tikzpicture}

}

\NewDocumentCommand{\otherchangebox}{O{} O{} O{} m}{

\begin{tikzpicture}[overlay, remember picture, baseline]
\node[revision box=\OTHERCHANGEBOXCOLOR/\OTHERCHANGEFILLCOLOR] at ($(0, 0) + #1$){#4};
\draw[\OTHERCHANGEFILLCOLOR!50, line width=1.5pt]
      ($ (0,0) + #2 + (0.6cm, 0) $) -- 
      ($ (0,0) + #2 + (10.07cm, 0) $);
\draw[\OTHERCHANGEFILLCOLOR!50, line width=1.5pt]
      ($ (0,0) + #3 + (10.05cm, 0) $) -- 
      ($ (0,0) + #3 + (10.05cm, 0.33cm) $);
\end{tikzpicture}

}

\NewDocumentCommand{\newsuggestionsbox}{O{} O{} O{} m}{

\begin{tikzpicture}[overlay, remember picture, baseline]
\node[revision box=\NEWSUGGESTIONSBOXCOLOR/\NEWSUGGESTIONSFILLCOLOR] at ($(0, 0) + #1$){#4};
\end{tikzpicture}

}

\newcommand{\leftx}{-1.3cm}
\newcommand{\lefty}{0.9cm}
\newcommand{\rightx}{9cm}
\newcommand{\righty}{0.9cm}
\newcommand{\rightrevisionbox}[1]{\revisionbox[(\rightx, \righty)][(\rightx-10cm, \righty-0.6cm)]
[(\rightx-10cm, \righty-0.6cm)]{#1}}
\newcommand{\leftrevisionbox}[1]{\revisionbox[(\leftx, \lefty)][(\leftx - 0.6cm, \lefty - 0.6cm)]
[(\leftx-10.03cm, \lefty-0.6cm)]{#1}}
\newcommand{\rightotherchangebox}[1]{\otherchangebox[(\rightx, \righty)][(\rightx-10cm, \righty-0.6cm)]
[(\rightx-10cm, \righty-0.6cm)]{#1}}
\newcommand{\leftotherchangebox}[1]{\otherchangebox[(\leftx, \lefty)][(\leftx - 0.6cm, \lefty - 0.6cm)]
[(\leftx-10.03cm, \lefty-0.6cm)]{#1}}
\newcommand{\rightnewsuggestionsbox}[1]{\newsuggestionsbox[(\rightx, \righty+1.1cm)][(\rightx-10cm, \righty-0.6cm)]
[(\rightx-10cm, \righty-0.6cm)]{#1}}
\newcommand{\leftnewsuggestionsbox}[1]{\newsuggestionsbox[(\leftx, \lefty+1.1cm)][(\leftx - 0.6cm, \lefty - 0.6cm)]
[(\leftx-10.03cm, \lefty-0.6cm)]{#1}}

\newcommand{\circlednum}[1]{%
  \tikz[baseline=(char.base)]{
    \node[shape=circle, draw=black, fill=black, text=white, minimum size=1em,
      inner sep=0pt,] (char) {\textbf{#1}};
  }%
}

% Define a new counter for observations
\newcounter{observationcounter}
\renewcommand{\theobservationcounter}{\arabic{observationcounter}} % Use plain numbering (1, 2, 3,...)

% Define a custom environment for observations without automatic title
\newtcolorbox{observation}[1][]{
    colback=black!10, colframe=black!85!black, coltitle=black,
    left=1mm, right=1mm, top=1mm, bottom=1mm, boxrule=0.5mm,
    #1
}

% Define a new counter for results
\newcounter{resultcounter}
\renewcommand{\theresultcounter}{\arabic{resultcounter}} % Use plain numbering (1, 2, 3,...)

% Define a custom environment for observations without automatic title
\newtcolorbox{result}[1][]{
    colback=yellow!10, colframe=black!85!black, coltitle=black,
    left=1mm, right=1mm, top=1mm, bottom=1mm, boxrule=0.5mm,
    #1
}

\newtcolorbox{revLabel}[1][]{
    colback=yellow!10, colframe=black!85!black, coltitle=black,
    left=1mm, right=-10mm, top=1mm, bottom=1mm, boxrule=0.5mm,
    #1
}

\begin{abstract}
Rowhammer is a read disturbance vulnerability in modern DRAM that causes bit-flips, compromising security and reliability. While extensively studied on Intel and AMD CPUs with DDR and LPDDR memories, its impact on GPUs using GDDR memories, critical for emerging machine learning applications, remains unexplored. Rowhammer attacks on GPUs face unique challenges: (1) proprietary mapping of physical memory to GDDR banks and rows, (2) high memory latency and faster refresh rates that hinder effective hammering, and (3) proprietary mitigations in GDDR memories, difficult to reverse-engineer without FPGA-based test platforms.

\begingroup
\renewcommand\thefootnote{{\footnotesize†}}
\footnotetext{Equal contribution.}
\endgroup
%This work introduces \attack, the first Rowhammer attack on NVIDIA GPUs with GDDR6 DRAM. \attack tackles these challenges with novel techniques to reverse-engineer GDDR row mappings and GPU-specific memory access optimizations to amplify hammering intensity and bypass mitigations. Our attack triggers up to 8 bit-flips on 4 DRAM banks of an NVIDIA A6000 GPU with GDDR6 memory, and shows how an attacker co-located on the GPU can use these to tamper with ML models, causing significant accuracy drops (by up to 80\%).

%attackers to tamper with ML models and reduce their accuracy by up to 80%.

%Rowhammer is a read disturbance vulnerability in modern DRAM, that results in bit-flips which undermine security and reliability. While Rowhammer attacks have been extensively studied on Intel and AMD CPUs with DDR4 and LPDDR4 memories, the vulnerability of GPUs using GDDR memories, which run emerging machine learning applications, remains an open question. Launching Rowhammer attacks on GPUs presents unique challenges: (1) unknown proprietary mapping of physical memory to GDDR banks and rows, (2) high memory latency and faster refresh on GPUs that make hammering DRAM with high intensity to induce Rowhammer challenging, and (3) unknown proprietary mitigations in GDDR memories that are difficult to reverse-engineer in the absence of any FPGA-based testing platform for GDDR memories.
%, which are available for other memories.

We introduce \attack, the first Rowhammer attack on NVIDIA GPUs with GDDR6 DRAM. \attack proposes novel techniques to reverse-engineer GDDR DRAM row mappings, and employs GPU-specific memory access optimizations to amplify hammering intensity and bypass mitigations. Thus, we demonstrate the first successful Rowhammer attack on a discrete GPU, injecting up to 8 bit-flips across 4 DRAM banks on an NVIDIA A6000 with GDDR6 memory. We also show how an attacker can use these to tamper with ML models, causing significant accuracy drops (up to 80\%).

%Rowhammer, a well-known read disturbance vulnerability in modern DRAMs, continues to undermine hardware security and reliability as DRAM technology scales down. While Rowhammer attacks have been extensively studied on Intel and AMD CPU systems, their feasibility on GPU systems remains an open question. Launching Rowhammer attacks on GPUs presents unique challenges: (1) circumventing proprietary virtual memory mapping without access to physical addresses, (2) achieving sufficient hammering intensity despite high memory latency, and (3) synchronizing with DRAM refreshes to evade proprietary Rowhammer mitigations. In this work, we introduce \attack, the first Rowhammer attack targeting NVIDIA GPU systems. \attack overcomes these challenges by reverse-engineering DRAM row mappings, employing GPU-specific memory access optimizations to amplify hammering intensity, and modeling novel access patterns to bypass Rowhammer mitigations. Our attack demonstrates, for the first time, the ability to trigger bit-flips on NVIDIA GPUs, providing critical insights into the security implications of Rowhammer in GPU architectures.
\end{abstract}
\section{Introduction}
Modern DRAM systems are increasingly vulnerable to Rowhammer attacks, a hardware vulnerability that enables attackers to induce bit-flips in memory cells by rapidly accessing neighboring rows\cite{Rowhammer2014}. Such attacks have been extensively explored on DDR and LPDDR memories in Intel and AMD CPUs~\cite{TRRespass, smash, Blacksmith, HalfDouble, Eccploit, ZenHammer, RevisitRowhammer}, and even on integrated GPUs in ARM CPUs~\cite{GrandPwning}, where a CPU-based LPDDR memory was attacked using the integrated GPUs. Such attacks demonstrate that attackers can tamper sensitive data, and even escalate to kernel-level privileges on CPUs.
In modern systems, discrete GPUs increasingly play a critical role in running high-performance computing, artificial intelligence applications, and graphics processing; however, thus far, there has been limited investigation of the Rowhammer vulnerability on GPUs. Moreover, machine learning models have been shown to be susceptible to bit-flips, allowing bit-flipping attackers to dramatically lower accuracy~\cite{TBD,Deephammer,li2024_1bitflip,DNNFaultInjection,CrushingFlips} and inject backdoors~\cite{tol2023dontknock,chen2021proflip} into these models.  
As GPUs are heavily used for ML model inference, understanding whether discrete GPUs are susceptible to Rowhammer is essential to ensure the security of ML systems. Thus, this paper seeks to explore the vulnerability of NVIDIA GPUs to Rowhammer.

%, culminating in the development of \attack, the first successful Rowhammer attack targeting NVIDIA GPU platforms.

%% TODO: GPUs use GDDR6 and HBM memories, whose vulnerability to Rowhammer attacks is so far unknown. 
%The GPU architectural and memory technology differences between CPUs and GPUs
Unlike CPUs, NVIDIA GPUs use Graphics-DDR (GDDR) memory in client and server GPUs, and High Bandwidth Memory (HBM) only in server GPUs.
Moreover, with GPUs being designed as throughput-oriented processors, they have a distinct architecture, instruction set, and programming model, compared to CPUs. 
These differences make adopting existing Rowhammer attacks to GPUs challenging.

%We first present solutions to established and GPU-exclusive obstacles, enabling the successful execution of Rowhammer attacks on NVIDIA GPUs.
\smallskip
\noindent \textbf{Challenges for GPU Rowhammer.}
Rowhammer attacks require hammering DRAM rows with rapid activations. To do so, an attacker is required to (a) evict addresses from on-chip caches, (b) access conflicting DRAM rows in a memory bank to ensure DRAM row activations, (c) activate the rows at a sufficiently high rate, and (d) develop access patterns that fool the in-DRAM Rowhammer mitigations.
NVIDIA GPUs, since the Ampere generation (sm-80), support the \texttt{discard} instruction, which evicts any address from all on-chip GPU caches, satisfying (a). However, the other requirements (b-d) introduce several challenges (C1-C3) for a successful attack. 

\smallskip
\noindent \textbf{C1.} First, the mappings of virtual to physical addresses, and of physical addresses to DRAM rows and banks in NVIDIA GPUs are proprietary and unknown. This is necessary to know to activate rows in the same bank for Rowhammer.

\smallskip
\noindent \textbf{C2.} Second, as GPUs are optimized for throughput, they have up to 4$\times$~\cite{chipsncheeseGPUs} the memory latency compared to CPUs, limiting hammering intensity. Moreover, GPU memories have faster refresh periods (e.g., 32ms or less) compared to CPU memories (32ms - 64ms), limiting the time available for hammering. Combined, these factors result in insufficient hammering rates to trigger Rowhammer bit-flips on GPUs. 

\smallskip
\noindent \textbf{C3.} Third, GPU memories may have proprietary in-DRAM defenses that need to be defeated by attacks to trigger bit-flips.

In this paper, we develop techniques to overcome these challenges and launch Rowhammer attacks on NVIDIA GPUs.
We focus on NVIDIA A6000 GPU with GDDR6 memory, a popular workstation GPU also available in the cloud, but our analysis is equally applicable to other client or server GPUs.

\smallskip

\noindent \textbf{GDDR Row Addressing.}
Rowhammer attacks on CPUs typically fully reverse-engineer the DRAM bank address function given a physical address, leveraging  DRAMA~\cite{drama}.
However, unlike CPUs, on GPUs, the physical addresses are not exposed even to a privileged user, making reverse engineering the closed-form bank address function challenging. 
However, we discovered that GPU virtual-to-physical memory mapping is typically stable for large memory allocations. Consequently, we directly reverse engineer the mapping of virtual addresses (offsets within an array) to DRAM banks and rows within the GPU DRAM by utilizing the latency increase on DRAM row-buffer conflicts, similar to DRAMA~\cite{drama}. 
Based on this, we identify addresses that map to unique rows, for all rows in a bank, and use these addresses to perform hammering.
%However, to date, no prior work has demonstrated whether this approach is effective for the proprietary memory-addressing functions of NVIDIA GPUs. In our attempts, we were similarly unable to uncover the addressing function without access to underlying physical addresses. Consequently, we adopt a conservative approach in this work, tackling a simplified version of the problem: reverse-engineering addresses to identify unique rows within the same bank. By utilizing the row buffer conflict insight from \textit{DRAMA}, we successfully recover all rows within the same bank as a selected \textit{pivot} address in the allocated memory layout.

\smallskip
\noindent \textbf{Increasing Activation Rates.} Since GPUs have almost a 4$\times$ higher memory latency~\cite{chipsncheeseGPUs} compared to CPUs, the naive hammering of GDDR memory from a single thread, like in CPU-based attacks, cannot achieve sufficiently high activation intensity to flip bits. This is even more important for GPUs, as the refresh period (\TREFW{}) for GDDR6 is just 22ms, as we reverse engineer in \cref{sec:sync}, 30\% less than that of DDR5, limiting the time available for performing activations. 
We address this challenge by increasing the activation rates through parallelized hammering, leveraging the throughput-oriented programming model in GPUs. We develop $k$-warp $m$-thread-per-warp hammering kernels optimized for GPUs, achieving activation rates close to the theoretical maximum of 500,000 activations per \TREFW{}, which is 5$\times$ higher than naive single-thread hammering kernels adapted from CPU-based attacks.

%Given the reverse-engineered rows, we are nevertheless far from having an effective Rowhammer attack due to high memory latency and shortened \TREFI period\cite{JEDEC-GDDR6}. After naively implementing N-sided Rowhammer from prior work\cite{TRRespass}, we identify that we can barely activate a handful of rows within a \TREFI period. The low activation rate can be insufficient to bypass known defenses\cite{UncoverRowhammer, TRRespass} or reach the commonly observed \TRH to induce bitflips\cite{RevisitRowhammer}. We propose an access scheme that distributes aggressors in unique warps of a GPU Streaming Multiprocessor to hide memory latency. The updated access scheme reaches the maximum number of activations possible within a \TREFI period.

\smallskip
\noindent \textbf{Synchronization to Refreshes.} As shown by prior attacks\cite{TRRespass, Blacksmith, smash}, 
defeating in-DRAM mitigations requires not only hammering N-aggressors~\cite{TRRespass,Blacksmith}, but also bypassing the in-DRAM TRR mitigation consistently on each \TREFI{} by synchronizing the attack pattern to \REF{} commands~\cite{smash}. 
Similar to prior work~\cite{smash}, we insert NOPs to each thread to add delays to the hammering kernel, synchronizing it with \REF{} commands. 
However, we observe that synchronization weakens as the number of warps increases. 
To address this, we design GPU kernels that utilize multiple warps and multiple threads per warp, to effectively hammer more aggressors while maintaining synchronization with \REF{} commands.

%Synchronization with the warp-based accessing scheme presents challenges unique due to parallelization: 1) synchronization barrier may re-shuffle warp order which is essential to mitigation bypass\cite{UncoverRowhammer} and 2) synchronization weakens as warp number increases. We resolve the issues by adding NOPs to each warp while reducing the number of warps required for larger aggressor counts with GPU threads.

\smallskip
\noindent \textbf{Demonstrating Rowhammer Exploits.}
Based on the techniques above, we develop our attack \attack{}, with which we launch the first systematic Rowhammer campaign on NVIDIA GPUs, specifically on 
an A6000 with 48GB GDDR6 DRAM.
%, 
%an A100 with 40GB HBM2e DRAM, 
%and a RTX3080 with 16GB GDDR6 DRAM.
We use $n$-sided synchronized hammering patterns $(n=8, 12, 16, 20, 24)$, similar to TRResspass~\cite{TRRespass} and SMASH~\cite{smash}, targeting 4 DRAM banks on the GDDR6 memory. On the A6000, we observe for the first time a total of 8 bit-flips (one per row) in GDDR6 DRAM, including at least one bit-flip in each bank we hammered.

Using these bit-flips, we demonstrate for the first time accuracy degradation attacks on ML models running on a GPU, that affect a wide range of deep learning models. 
We show that bit-flips can occur in the most-significant bit of the exponent in FP16-representation weights, significantly altering the value of the parameter and degrading the accuracy. 
Across five different models, including AlexNet, VGG16, ResNet50, DenseNet161, and InceptionV3, we observe accuracy drops of between 56\% to 80\% due to Rowhammer bit-flips. 
Our observations highlight the pressing need for both system-level mitigations against Rowhammer attacks on GPUs and algorithmic resilience in ML models against bit-flipping attacks, given the real-world threat posed by Rowhammer on GPUs.

\smallskip
\noindent Overall, this paper makes the following contributions:
\begin{enumerate}[label={\arabic*)},itemsep=0pt]
  \item We demonstrate the first Rowhammer attack on discrete GPUs, capable of flipping bits in GDDR6 memories.
  \item We reverse-engineer unknown details about the address to bank and row mappings, to efficiently activate DRAM rows from the same bank in GPU memories.
  \item We propose new techniques for parallelized hammering on GPUs that achieve high activation rates and precise synchronization with refresh operations to bypass in-DRAM Rowhammer defenses in GDDR6 memory.
  \item Using \attack{}, we demonstrate the \textit{first} Rowhammer exploit on a discrete GPU (NVIDIA A6000 GPU), that targets ML models and degrades model accuracy by 56\% to 80\% with a single bit-flip.    
  %by leveraging the insights presented in this work, and evaluate it on Ampere architecture NVIDIA GPUs \textcolor{red}{A, B, and C} with DRAM types \textcolor{red}{X, Y, and Z}. Our attack has demonstrated its effectiveness on the listed GPU devices, enabling Rowhammer exploits on NVIDIA GPUs for the first time.
\end{enumerate}

%\noindent \TODO{Concluding statement?}

\noindent \textbf{Responsible Disclosure.}
We responsibly disclosed the Rowhammer vulnerability on A6000 to NVIDIA on 15th January, 2025, and subsequently also to the major cloud service providers (AWS, Azure, GCP) who may be affected.
%, providing the Proof-Of-Concept code and the manuscript draft.
\OTHERCHANGE{NVIDIA has confirmed the problem, and at the time of writing is still investigating a fix.
%suggested ECC as a suitable mitigation.
}
After the expiry of an embargo requested by NVIDIA till 12th August, 2025, our code will be available on Github at \url{https://github.com/sith-lab/gpuhammer}.
%\REVISION{as requested by the affected companies}.
\rightotherchangebox{CC-Q1}\vspace{-0.1in}

\section{Background and Motivation}
%In this section, we present background information on GPU architecture, the GDDR6 memory used in client and workstation GPUs, and the Rowhammer vulnerability.

\subsection{GPU Execution Model and Architecture}
GPUs are throughput-oriented processors designed for large-scale data processing through parallel execution. Their architecture, composed of numerous smaller cores, allows for the concurrent execution of thousands of threads, making them well-suited for operations like matrix multiplications, image processing, and machine learning.

\smallskip
\noindent \textbf{Execution Model.}
%NVIDIA GPUs' execution model is centered on parallel execution, where computations are broken down into threads, the smallest unit of execution. These threads are organized into warps, with each warp consisting of 32 threads that execute the same instruction in lockstep. 
%Multiple warps are grouped into thread-blocks, with each thread-block assigned to a single Streaming Multiprocessor (SM) on the GPU and each warp in the block executed in a time-multiplexed manner on the same SM. Different thread blocks are spatially mapped to distinct SMs (e.g., A6000 has 80 SMs), and run concurrently. The grid represents the entire computational workload, consisting of multiple blocks with each block containing multiple warps running in a time-multiplexed manner. When a GPU kernel (a function executed on the GPU) is launched, the grid of blocks is distributed across the available SMs for execution.
NVIDIA GPUs are designed for parallel execution, where computations are divided into \textit{threads}, the smallest units of execution. Threads are grouped into \textit{warps}, each consisting of 32 threads that execute the same instruction in lockstep. Multiple warps are further organized into thread \textit{blocks}, with each block assigned to a single Streaming Multiprocessor (SM) on the GPU. Different thread blocks are spatially mapped across the available SMs (e.g., the A6000 GPU has 80 SMs) and run concurrently. Within an SM, the warps of a thread block execute in a time-multiplexed manner. 
%The entire computational workload is represented as a grid of blocks x warps x threads, each containing multiple warps. 
When a GPU kernel (a function to be executed on the GPU) is launched, the computational workload, represented as a grid of blocks $\times$ warps $\times$ threads, is distributed across the available SMs for execution.

\smallskip
\noindent \textbf{Memory Hierarchy.}
Like CPUs, the GPU memory hierarchy includes L1 and L2 caches on-chip, and an off-chip DRAM, which could be a GDDR or HBM-based memory. Typically, GDDR memories are used in client (RTX3080) or server (e.g., A6000) GPUs, whereas HBM is used exclusively in server-class GPUs (e.g., A100 or H100).
The L1 cache is private to each Streaming Multiprocessor (SM) and the L2 cache is shared across all SMs on the GPU.
As GPU systems are optimized for memory bandwidth and have massive die sizes, their memory system latencies can be an order of magnitude higher than CPUs. A recent study~\cite{chipsncheeseGPUs} reports the L1 latency for an RTX 3090 close to 20ns, L2 latency close to 100ns, and memory latency close to 300ns, with the memory latency on GPUs being almost $4\times$ the latency on CPUs\cite{chipsncheeseGPUs}.
%their cache and memory latencies can often be significantly higher than CPU latencies, with memory accesses  

\subsection{GPU DRAM Architecture} \label{subsec:gddr6_arch}
\begin{figure}[ht]
\centering
\fcolorbox{\REVISIONFIGURECOLOR}{white}{ % Red border with a white background
\includegraphics[width=3.3in,height=\paperheight,keepaspectratio]{"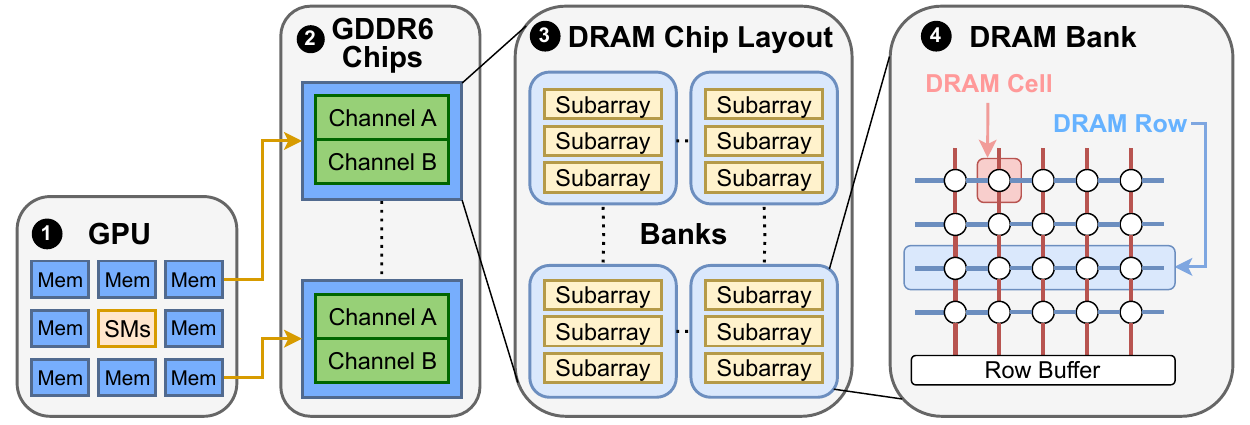"}
}
\caption{\REVISION{GDDR6 DRAM Organization}}
\label{fig:dram_org}
\end{figure}
%\TODO{I only know that GDDR6 is similar to DDRs in its memory layout, but I have yet to find infographics on its distinct components and how it is exactly laid out.}
%\TODO{Uncomment Memory Org and Relevant Operations}
%\noindent \textbf{Memory Organization}

%\noindent \textbf{Relevant Operations}
%\TODO{DRAM Architecture: Rows, Banks, Channels}

\noindent \textbf{DRAM Architecture.} 
% Unlike CPUs that have a 64-bit bus, GPUs have a wider memory bus to achieve higher memory bandwidths: e.g., the A6000 with GDDR6 DRAM has a 384-bit bus and the A100 with HBM2e has a 1024-bit bus.
\REVISION{Modern GPUs typically use GDDR or HBM-based memories.} 
While the GDDR DRAM has a planar structure, the HBM has a 3D structure with DRAM dies layered one above the other. 
\REVISION{However, within the DRAM chip, the architecture is similar across form factors.}

\REVISION{\cref{fig:dram_org} shows the organization of GDDR6 DRAM in A6000 GPU, where multiple DRAM chips surround the GPU core containing the SMs. 
%In a GPU, memory access begins at the Streaming Multiprocessors (SMs), which issue load and store requests to memory. These requests are sent to memory controllers (MCs) on the GPU cores, which manage communication with the DRAM chips via a memory bus. 
The memory controllers (MCs) on the GPU core route memory requests from the SMs to the appropriate DRAM chip.
Each MC connects to a DRAM chip via a memory bus, operating as an independent memory channel.
Each DRAM chip is organized into 16 banks, and each bank consists of arrays of DRAM cells. Cells are grouped into DRAM rows that 
are 1KB to 2KB in size. On a memory access, an entire DRAM row is activated and read into a row buffer, before the requested portion is sent over the data bus. 

The A6000 GDDR6 has a 384-bit memory bus with 12 DRAM chips, each providing 32 bits. A chip has two channels, which may be used as a single 32-bit channel or two 16-bit memory channels, resulting in 12 to 24 channels. Each channel has 16 banks, and each bank has 16K to 32K rows.}
\rightrevisionbox{R1}
% However, within the DRAM die, the architecture is similar across all form factors, with the cells hierarchically divided into rows, banks, and channels.  Each row is a group of DRAM cells (1KB to 2KB in size) that are activated simultaneously during a single access.
% A group of rows (16K to 32K rows) form a bank, with each device consisting of 8 to 16 banks per channel that can be activated in parallel.
% Each DRAM channel has a separate memory bus, which is shared by the banks in a channel.
% Each bank has a row-buffer into which the data of the entire row is read out on an access.
%\TODO{Memory Controller Commands, ACT, PRE, and Refresh.}

\smallskip

\noindent \textbf{DRAM Operations.}
To access data in DRAM, the memory controller (MC) \OTHERCHANGE{issues} a sequence of commands. 
%First, it ensures the targeted bank is precharged by issuing a PRECHARGE command, which deactivates the row buffer. 
First, an ACTIVATE command (ACT) is sent to activate or open a row and read the charge into the row-buffer. 
Then, the MC can perform READ or WRITE operations by specifying the column within the row-buffer that is to be read or written to.
%Lastly, before accessing another row from the same bank, the MC has to issue a PRECHARGE command (PRE) to close the specific row. 
Finally, before a new row can be activated in the same bank, the MC must send a PRECHARGE command (PRE) to close and deactivate the row, before accessing another row in the bank.
Between two ACTs to a single bank, the MC is required to wait for at least \TRC{} (typically about 45ns).
To maintain data integrity, the memory controller periodically issues REFRESH commands (REF) on average at least once every 1.9 $\mu$s (\TREFI) \cite{JEDEC-GDDR6}, which replenishes the charge in a subset of rows at a time. 
Typically, the entire DRAM is refreshed within 8K or 16K REF commands, ensuring that each row is refreshed at least once every 16ms or 32ms\cite{JEDEC-GDDR6}.
%This bounds the amount of time available for a successful Rowhammer attack. 
%The retention time is standardized as 32 ms for GDDR6 DRAM\cite{JEDEC-GDDR6}, which defines the minimum duration that cells can reliably hold data without refresh.

\subsection{Rowhammer}
Rowhammer is a read-disturbance phenomenon in which rapid activations of an aggressor row in DRAM can cause charge leakage in a physically adjacent victim row, ultimately inducing bit-flips~\cite{Rowhammer2014}. 
%The minimum number of accesses required to trigger bit flips, known as \TRH{}, has been shown to be decreasing over time, driven by the continuous scaling down of DRAM chip sizes \cite{Rowhammer2014, RevisitRowhammer}.
The minimum number of ACTs to a row required to trigger bit-flips, known as \TRH{}, has been shown to be decreasing over time in CPU-based DRAM. Prior works~\cite{Rowhammer2014,RevisitRowhammer} using DRAM testing infrastructure show that the \TRH{} has dropped from 139K in DDR3~\cite{Rowhammer2014} to 10K in DDR4 and 4.8K in LPDDR4~\cite{RevisitRowhammer}.
Similar FPGA-based testing~\cite{HBMRowhammer} has shown HBM2 to have a \TRH of around 20K.
No prior work has studied the Rowhammer vulnerability on GPU memories like GDDR6 or HBM2e/3, due to the lack of any open-source testing infrastructure for such memories.

\smallskip
\noindent \textbf{Mitigations in CPU-Based DRAM.}
In DDR4 and DDR5 memories, Target Row Refresh (TRR)~\cite{TRRespass, UncoverRowhammer,ZenHammer} has been widely adopted as an in-DRAM Rowhammer mitigation.
%, while ECC (Error Correction Code) is commonly employed for detecting and correcting bit flips \cite{ECC, CosmicRay}.
TRR operates by tracking frequently activated aggressor rows and proactively refreshing their neighboring victim rows during \REF commands (e.g., once every $n$ \REF), to prevent bit-flips.
%In contrast, ECC can detect and correct single-bit errors and identify two-bit errors, though it is not guaranteed to handle errors involving three or more bits. Notably, TRR implementations remain proprietary, as the mechanism has not been standardized.

Recently, JEDEC, the consortium of DRAM vendors, has standardized more advanced mitigations such as RFM (Refresh Management) \cite{JEDEC-DDR4} and PRAC (Per Row Activation Counter) \cite{jedec_ddr5_prac}, to address Rowhammer vulnerabilities. However, thus far, to our knowledge, no commercial systems have adopted such sophisticated mitigations.

%as these mitigations were not observed in the devices studied for this paper, they are considered beyond the scope of our investigation.

\smallskip
\noindent \textbf{Bypasses.}
%While TRR has proven effective against basic one-sided \cite{Rowhammer2014} and double-sided \cite{doubleSidehammer} Rowhammer attacks, 
Recent attacks on DDR4 and DDR5 have demonstrated that TRR is vulnerable to sophisticated attack patterns.
TRRespass~\cite{TRRespass} showed that $n$-sided attack patterns could evade the TRR mitigation and induce Rowhammer bit-flips in DDR4 memories, by overflowing its tracking mechanism
%evicting certain aggressor rows from the tracker, 
and preventing certain aggressor rows from being tracked and mitigated. 
Blacksmith~\cite{Blacksmith} observed that TRR samples aggressor rows only at certain time instances within \TREFI, and thus creates patterns where certain aggressor rows can evade the tracker.
SMASH~\cite{smash} further shows that aligning the hammering patterns with 
\TREFI intervals, when mitigations are issued, can increase the chance of a Rowhammer bit-flip, as TRR continues to be fooled in a predictable manner.
Zenhammer~\cite{ZenHammer} shows that similar attacks continue to be applicable to some DDR5 DRAM.

%Furthermore, the state-of-the-art non-uniform Rowhammer pattern \cite{Blacksmith} has revealed that TRR may only sample aggressor rows during specific intervals within \TREFI. This behavior can be exploited by carefully tuning the frequency, phase, and amplitude of the hammering pattern.

%On the other hand, ECC has been circumvented by triggering precisely three-bit flips using carefully crafted data patterns \cite{Eccploit}, thereby bypassing its error detection and correction mechanisms.

\subsection{Motivation and Goal}
The growing demand for machine learning models has caused a surge in GPU deployment in the cloud,
%\cite{AIReport2024}
with NVIDIA GPUs commanding a market share of almost 90\%\cite{marketShare2}.
While machine learning model inference is vulnerable to a wide variety of threats, such as adversarial examples~\cite{advexamples}, model inversion attacks~\cite{modelinv}, and jailbreaking attacks~\cite{jailbreakingLLMs}, more recently, they have also been shown to be vulnerable to tampering via bit-flipping attacks like Rowhammer, that can cause accuracy degradation~\cite{TBD,Deephammer,li2024_1bitflip,DNNFaultInjection,CrushingFlips} or backdoor injection into these models~\cite{tol2023dontknock,chen2021proflip}. 
As ML inference is predominantly performed on GPUs, it is important to examine whether GPUs are susceptible to Rowhammer to fully evaluate the threat landscape for machine-learning systems. 
%However, thus far, due to the lack of any FPGA-based testing infrastructure~\cite{DRAMBender} for GPU memories, this vulnerability has remained unexplored. 
Thus, this paper studies the potential for Rowhammer attacks on NVIDIA GPUs.

\REVISION{

\subsection{Threat Model for GPU Rowhammer}\label{subsec:rowhammer_threat_model}

\textbf{Setting.} We assume an attacker aiming to tamper with a victim’s data in GPU memory via Rowhammer. This requires co-location of multi-tenant data within the same DRAM bank. Such a setting is possible in the case of time-sliced usage of GPUs in the cloud using NVIDIA GPU Operator~\cite{TimeSlicing}. Tools such as Google Kubernetes Engine (GKE)~\cite{GKE} can also enable memory isolation across time-slices allowing kernels of different users to time-share on a GPU. Multi-tenant data co-location is also feasible in case GPUs are spatially shared in a fine-grained manner, as in emerging serverless settings~\cite{MIGnificient}.
%for different users, enabling kernels of different users to run in time-slices and multi-tenant co-location of data.
%in a GPU in the cloud, where a GPU is used in a time-sliced manner.  
%Rowhammer is widely used to tamper with~\cite{smash, FFS} and leak~\cite{RamBleed} sensitive data by through bit-flips, often requiring co-location of multi-tenant data within the same DRAM bank. On GPUs, 
%During these intervals, data from multi-tenants resides in memory simultaneously, enabling Rowhammer-based exploits.
%Similarly, fine-grained spatial partitioning of the GPUs 
%We elaborate on these while describing exploits in \cref{sec:exploit}.
\NEWSUGGESTIONS{We assume ECC is disabled on GPUs.\footnote{\NEWSUGGESTIONS{Cloud service providers like AWS and GCP allow guest VMs to disable ECC on GPUs. Enabling ECC on GDDR-based GPUs can introduce slowdown~\cite{ECCDisable1,ECCDisable2}, \cf{} \cref{sec:mitigations}, and reduce memory capacity~\cite{IMTSullivan}.}} We leave Rowhammer attacks on GPUs with ECC enabled for future work.}

\smallskip
\noindent \textbf{Attacker Capabilities.} We assume the attacker can execute CUDA kernels natively on the GPU with user-level privileges.
The attacker can execute kernels that generate high-intensity memory access patterns for inducing Rowhammer.
%, and also advanced attack patterns, such as activating multiple dummy rows [17] or synchronizing attack patterns with refresh commands [12].
Additionally, GPU memories might deploy in-DRAM mitigations, like TRR\cite{JEDEC-GDDR6, JEDEC-HBM3, HBMRowhammer} that act in the shadow of a refresh command. 
These kernels may attempt to surpass these mitigations, by activating multiple dummy rows~\cite{TRRespass} and synchronizing their attack patterns with respect to refresh commands~\cite{smash}.

The attacker can identify vulnerable rows in the GPU memory by targeting memory allocated by itself. Then, while co-located with the victim, the attacker may force sensitive victim data to be mapped to vulnerable locations through memory massaging. We provide more details in \cref{sec:exploit}.
%We assume the attacker can first profile the GPU memory offline, identifying the vulnerable rows by performing a Rowhammer attack on its own memory. 
%Subsequently, in the online phase while co-located with the victim, it can perform memory massaging to map sensitive victim data to vulnerable locations.
%We describe these capabilities in more detail while considering exploits in \cref{sec:exploit}.
%To launch a successful Rowhammer attack, we first need to have sufficient DRAM row activation intensity. Rowhammer is a time-constraint problem, as modern DDR DRAMs may require at 5-10K row activations\cite{RevisitRowhammer} before \TREFW to trigger bit-flips. We explore how to achieve this under reduced timings and higher memory latency in \cref{sec:act}.

%Additionally, modern GPU memories \cite{JEDEC-GDDR6, JEDEC-HBM3, HBMRowhammer} may integrate mitigations like TRR and RFM. Evading these defenses demands advanced techniques, notably the $n$-sided attack pattern, where multiple aggressor rows are used to fool TRR samplers. Precise refresh synchronization can predictably fool TRR, further amplifying the success rate. We explore those techniques for GPUs in \cref{sec:act,sec:sync}.
%\rightrevisionbox{R2}\vspace{-0.1in}
%\rightnewsuggestionsbox{NS}

}

\section{Building Blocks of \attack{} Attack} \label{sec:buildingblock}
In this section, we introduce the minimum primitives required to perform a Rowhammer attack on NVIDIA GPU systems, as well as the challenges that come with them. 

%\TODO{Add a figure with different blocks: (a) Reverse Virtual Address to Row and Bank ID, (b) Derive CUDA Kernel to Hammer DRAM with High Intensity, (c) Hammer with Patterns (synchronized, n-sided) to Fool in-DRAM Mitigations. Explain this.}
\subsection{Primitives for Activating DRAM Rows} \label{subsec:overview_unique}
Rowhammer attacks require rapid activations of a DRAM row. To obtain a DRAM row activation on successive loads to the same address, after each load, the address needs to be (a) evicted from the on-chip cache, and also (b) evicted from the row-buffer in the DRAM. 
%As mentioned in \cref{subsec:gddr6_arch}, in order for memory accesses to trigger activations, it requires two conditions: uncached and not already present in the row buffer.

\begin{figure}[ht]
\centering
\includegraphics[width=3.1in,height=2in,keepaspectratio]
{"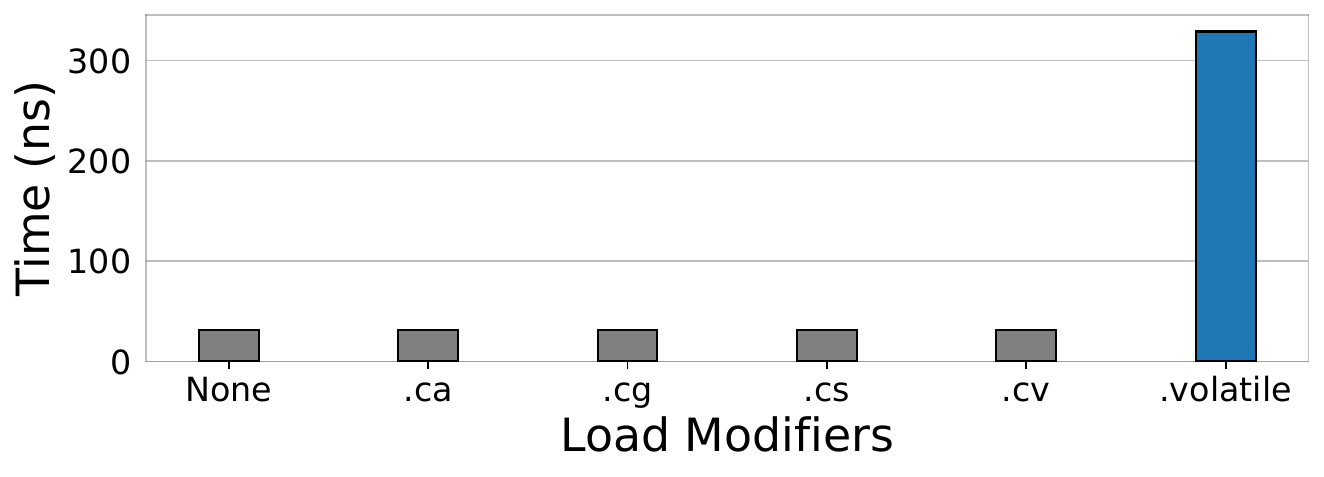"}
\cprotect\caption{\label{fig:ld_access_latency} Latency of \texttt{discard} and load with different modifiers. On Ampere GPUs, only \texttt{.volatile} brings data into L2 and not into L1 cache, thus incurring memory access latency.}
\end{figure}

\smallskip
\noindent \textbf{Cache Eviction.} 
%We discovered that accessing an address with volatile load instructions only brings the data into the shared L2 cache.
Since the \texttt{sm\_80} architecture, i.e., Ampere generation GPUs, NVIDIA PTX introduced the \texttt{discard}\cite{nvidia_ptx} instruction to its ISA. \texttt{discard} clears an address from
the L2 cache, making it an effective primitive for hammering. 
However, we still need to ensure the address is not cached in the L1 cache, to ensure a discard followed by a load is always fetched from memory. 
PTX provides multiple modifiers for load instructions to modify their caching behavior. So we measure the latency for performing a discard, followed by a load with these modifiers, to test which one enables consistent memory accesses.
\cref{fig:ld_access_latency} shows the result.

%\TODO{Add graph with latencies of accessing two (loads + discard), with different load variants below.}

As shown in \cref{fig:ld_access_latency}, of all the \texttt{ld} modifiers in PTX, including cache hints such as \texttt{.ca} (cache at all levels), \texttt{.cg} (cache in L2, not L1), \texttt{.cs} (evict-first policy), \texttt{.cv} (don't cache), and \texttt{.volatile} (no reordering of loads), on the A6000 GPU, all the cache hints are ignored, and only the \texttt{.volatile} modifier ensures the load does not install the address into the L1 cache. Only the combination of \texttt{discard} and \texttt{ld.global.volatile} results in a latency close to memory access (around 300ns~\cite{chipsncheeseGPUs}), and other ld modifiers result in cache access latency (around 40ns~\cite{chipsncheeseGPUs}).
Thus, memory accesses for an address are ensured with the following code: 
%suitable replacement for the \textit{CLFLUSH} instruction for our hammering. An example of an uncached access is as follows:

\vspace{0.1in}
\begin{lstlisting}[caption={Memory access using \texttt{discard} and \texttt{ld.volatile}},label={lst:discard}, numbersep=6pt,xleftmargin=8pt]
discard.global.L2 [%0], 128;
ld.global.volatile %0, [%1];
\end{lstlisting}

\noindent \textbf{Row Buffer Eviction.} To ensure row activations, after each memory access, the accessed row must be evicted from the row buffer. This can happen in two ways: (1) when a row is closed due to a period of inactivity, or (2) when another row in the same bank is accessed, causing a row buffer conflict that closes the current row~\cite{drama}. As the first method can be slow, we rely on the second method to achieve activations. 
If one accesses addresses mapped to different rows in the same bank (\texttt{row1} and \texttt{row2}), a row-buffer conflict can evict \texttt{row1} from the row-buffer as follows:
%To determine the specific addresses that map to different rows within a bank, reverse engineering the virtual-to-physical memory mapping is necessary. The retrieval routine for this process is detailed in \cref{sec:reverse}. As an illustration of using a row buffer conflict to close a row, consider the following example:

\vspace{0.1in}
\begin{lstlisting}[caption={Row-buffer eviction via row-buffer conflict.},label={lst:row_buffer_conflict},numbersep=6pt,xleftmargin=8pt]
ld.global.volatile %0, [%row1];
ld.global.volatile %0, [%row2]; // row1 evicted
\end{lstlisting}

\subsection{Naive Rowhammer CUDA Kernel}
\label{sec:naivehammering}
%Building on top of \cref{subsec:overview_unique}, we can now build a naive N-sided rowhammer kernel:
Using the cache and row-buffer eviction primitives described above, we can construct a CUDA kernel to perform a $n$-sided Rowhammer attack similar to TRRespass~\cite{TRRespass}. We perform a \texttt{discard} and a \texttt{ld} to $n$ successive addresses that map to different rows in a bank, in a loop of $it$ iterations, to ensure $it/n$ activations to each of the $n$ rows. Our CUDA kernel to perform $n$-sided hammers on the GPU memory is as follows:

%\vspace{0.1in}
\begin{lstlisting}[caption={Naive single-threaded Rowhammer kernel},label={lst:simpleRH},numbersep=6pt,xleftmargin=8pt]
__global__ void hammer(size_t **addrs, size_t N, size_t it)
{
  size_t dummy, i = 0;
  for (; it--;)
  {
    asm ("discard.global.L2 [%0], 128;" 
    ::"l"(addrs[i]));
    
    asm ("ld.u64.global.volatile %0, [%1];" 
    : "=l"(dummy) : "l"(addrs[i]));
    
    i = (i + 1) % N;
  }
}
\end{lstlisting}

%This kernel can sequentially hammer $n$ aggressors, and result in $n/it$ activations per row. 
Thus, GPUs provide all the major building blocks required for a Rowhammer attack on GPU memories.
%However, to launch a successful Rowhammer attack on GPU memories, there are a few additional challenges 
%introduced by GPUs 
%that need to be overcome, as we discuss next.
%However, in practice, to enable a successful Rowhammer attack, we observe that there are several challenges we need to overcome, as we discuss next.
%However, due to GPU systems' high memory latency (200-400ns) as seen in \cref{sec:reverse} and reduced tREFI(1.9$\mu$s)\cite{JEDEC-GDDR6}, this scheme cannot achieve the desired activations. We analyze in detail the shortcomings of potential access shcemes and propose a solution to achieve the maximum activation rate in \cref{sec:act}.

\subsection{Challenges}
To launch a successful Rowhammer attack, there are a few additional challenges introduced by the architectural constraints of GPUs, that we need to overcome. 

\smallskip
\noindent \textbf{Challenge-1: Reversing GPU Address to Bank Mappings}. To engineer row buffer conflicts, we first need to reverse engineer the mapping of addresses to DRAM banks and rows. Unlike CPUs, whose functions have been reverse-engineered in prior works~\cite{drama,ZenHammer,wang2020dramdig}, no prior work has reverse-engineered the proprietary mapping functions used in GPUs. Moreover, unlike CPUs, where the physical addresses are exposed to privileged software, NVIDIA GPUs do not expose the virtual to physical mappings even to privileged software, further complicating this task. In \cref{sec:reverse}, we show how to recover the bank and row address mappings, overcoming this challenge.

\smallskip
\noindent \textbf{Challenge-2: Achieving High Activation Rates on GPUs.} 
As DRAM cells are refreshed periodically, the Rowhammer attack is a time-bound attack. For a successful attack, the number of activations must cross the \TRH{} for a given row, before the refresh period (e.g., 32 ms) completes.
However, achieving such rates on GPUs is challenging for two reasons. 

First, GPU
%s are optimized for throughput rather than latency, leading to 
memory latencies (around 300ns) are around 4$\times$ \REVISION{to} 5$\times$ higher than CPUs (around 60ns)~\cite{chipsncheeseGPUs} -- due to this, our naive CUDA hammering kernel in \cref{sec:naivehammering} is only able to achieve 100K activations in 32ms, and less than 7K activations per row in a 16-sided pattern, which is lower than \TRH{} for most CPU-based DRAM.
%, almost 5x lower than similar CPU-based hammering.
%This inherently reduces the frequency of activations significantly on GPUs compared to CPUs. 
Second, the refresh intervals in GPU memory (e.g., 32ms or lower for GDDR6) are shorter compared to CPU DRAM (32–64 ms), further limiting the time available for hammering. Together, these factors contribute to hammering rates on GPUs that are insufficient to trigger Rowhammer-induced bit-flips.
In \cref{sec:act}, we show how we can overcome such limitations and hammer with significantly high intensity to induce bit-flips.
%.by parallelizing the hammering. 

\smallskip
\noindent \textbf{Challenge-3: Synchronizing Hammering to REFs.} 
Recent CPU-based attacks have demonstrated that Rowhammer aggressor patterns should not only be $n$-sided~\cite{TRRespass}, but also be synchronized to memory REF commands~\cite{smash} to fool the in-DRAM mitigations in an effective manner. 

Like CPU-based DDR memories, GPU memories like GDDR6 are also likely to be equipped with proprietary in-DRAM Rowhammer mitigations, that may be fooled by synchronized hammering patterns. 
However, little is known about the \REF{} command frequency on NVIDIA GPUs, or the capability of memory accesses in CUDA kernels to be synchronized to REF commands of GPU memory controllers.
In \cref{{sec:sync}}, we reverse-engineer details about GPU memory refresh, and show how one can achieve synchronized many-sided aggressor patterns on GPUs. 

\smallskip
\OTHERCHANGE{In subsequent sections (\cref{sec:reverse,sec:sync,sec:act}), we present our attack primitives that address these challenges. 
%We describe them in the context of GDDR6 memory on the NVIDIA RTX A6000 GPU, but these primitives are applicable to other modern NVIDIA GPUs regardless of whether they successfully induce bit flips or not (we have tested these primitives on an RTX-3080 GDDR6 and an A100 HBM2e memory).
}
\section{Reversing GPU DRAM Row Addressing}
\label{sec:reverse}
Understanding the mapping of memory addresses to DRAM banks and rows in GPU memory is crucial to generating row-buffer conflicts and activations. For this, we adopt the approach from DRAMA~\cite{drama}, tailoring it to NVIDIA GPUs. However, one key challenge is that unlike CPUs, where the virtual to physical memory translations are accessible to privileged software, allowing the reverse engineering of the closed-form function mapping the physical addresses to bank and row IDs, NVIDIA GPUs make the physical addresses inaccessible making the reversing of the exact functions challenging.

\begin{figure}[ht]
\centering
\includegraphics[width=3.3in,height=\paperheight,keepaspectratio]
{"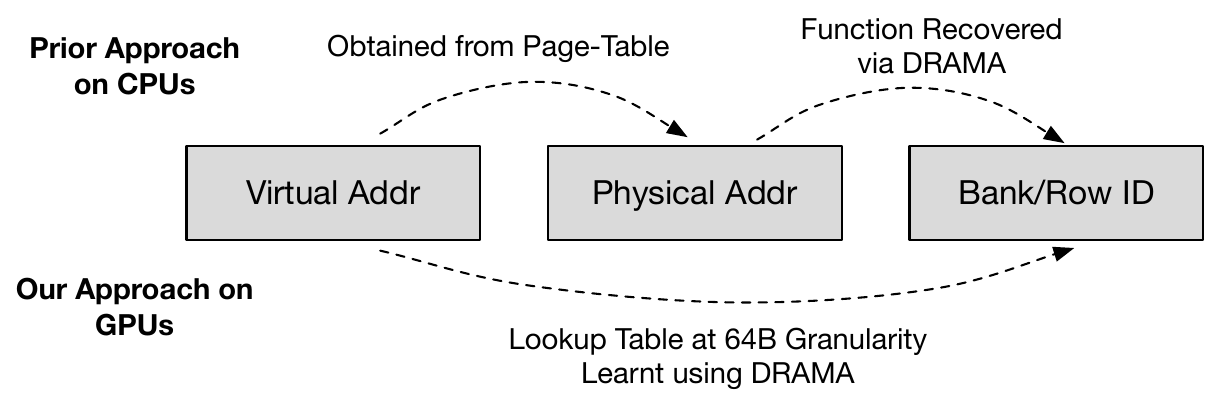"}
\caption{Prior approaches on CPUs reverse engineer the exact function of physical addresses to DRAM bank/row IDs. Without access to physical addresses on GPUs, we directly learn the mapping of virtual addresses to row and bank IDs.}
\label{fig:addr_mapping}
\end{figure}
%\TODO{Need a figure showing conventional approach - directly observe physical address, recover physical to Bank-IDs via DRAMA. Our approach, construct a lookup table to Bank/Row-ID for each address in a large allocation/}

To overcome this challenge, our approach instead directly learns a lookup table of virtual addresses to bank and row IDs, as shown in \cref{fig:addr_mapping}, by measuring the increase in latency due to row-buffer conflicts similar to DRAMA~\cite{drama}. This enables us to identify virtual addresses mapping to distinct rows in the same bank without recovering the exact row-addressing function itself. Below, we outline our approach in more detail. 
%, the challenges encountered  and solutions, followed by evaluations.

\begin{comment}
To uncover the mapping of memory addresses to DRAM rows in GPU memory, we adopt the row buffer conflict approach from \textit{DRAMA}\cite{drama}. Unlike conventional systems, where physical memory addresses can be leveraged for such analyses, we lack access to these addresses on NVIDIA GPUs. Furthermore, it remains unclear whether NVIDIA employs a deterministic row addressing function in their GPUs.

Given these constraints, we take a conservative approach: our goal is to identify unique addresses corresponding to distinct rows, even if the precise row order remains unresolved. This methodology allows us to investigate the underlying DRAM structure without relying on assumptions about the row-addressing function's behavior. We briefly define and describe \textit{DRAMA}'s terminologies and our main steps.
\end{comment}

\subsection{Approach}
To construct a lookup table mapping addresses at a 64B granularity to DRAM rows and banks, we use the following steps.

\smallskip

\noindent \circled{1} \textbf{Allocate a large array.} We allocate an array just smaller than the GPU DRAM, using \texttt{cudaMalloc}. The virtual addresses of this array let us access the entire GPU DRAM.

\begin{figure}[ht]
\centering
\includegraphics[width=3in,height=2in,keepaspectratio]
{"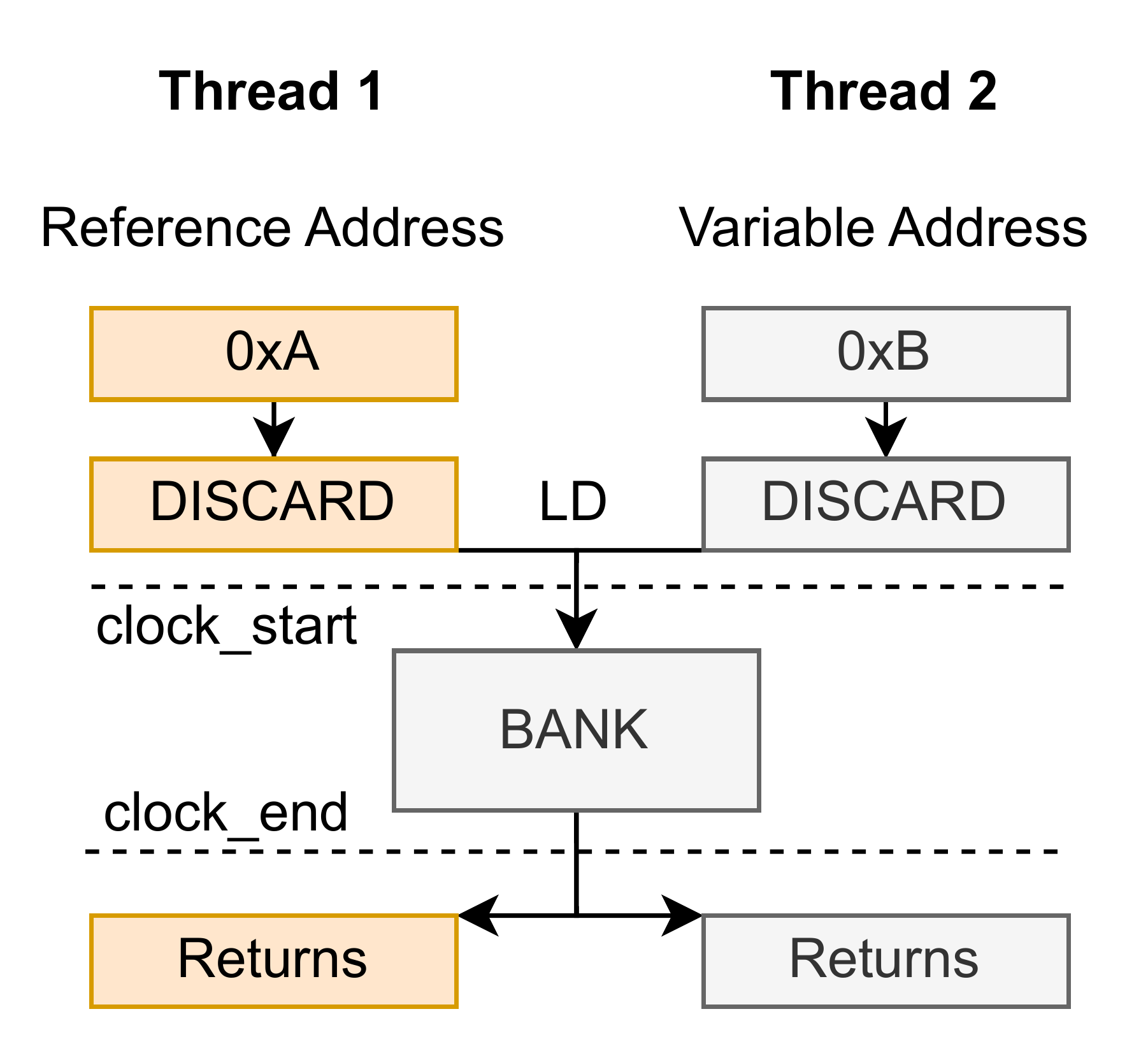"}
\cprotect\caption{\label{fig:conflict_primitive} Conflict testing primitive. We measure the time to access a pair of addresses: the first address is a reference address, and the second is varied by iterating through the entire memory. As instructions in both threads execute in lockstep, the \verb|clock_end| records the time when the longer of the two memory access completes.
%\TODO{change fixed address (pivot) to reference address.}
%With lockstep, memory load is simultaneous, and consequently, \verb|clock_end| records the slowest of the two loads.
}
\end{figure}

\smallskip
\noindent \circled{2} \textbf{Check for conflicts with a reference address.} We fix a reference address from the array, and identify all the addresses that have row-buffer conflicts with it. We do this by accessing pairs of memory addresses and measuring their latency, as shown in \cref{fig:conflict_primitive}. Here the first address is the fixed reference address, and the second is varied by iterating through the memory in 64B increments. 
When both addresses map to the same bank, the row-buffer-conflict causes a higher latency, compared to when both addresses map to different banks.

\smallskip
\noindent \circled{3} \textbf{Generate a Conflict-Set.}
We collect all addresses in memory that have a high latency when accessed with the reference address, i.e., have a row-buffer conflict, and map to the same bank as the reference address. We call this set a \textit{Conflict-Set}.

\smallskip
\noindent \circled{4} \textbf{Generate a Row-Set.}
From the conflict-set, we filter out duplicates, i.e., addresses that map to the same row. 
We do this by iterating through the conflict-set, and only retaining an address if it conflicts with all the addresses already present in the set.
The set of remaining addresses, consists of one address per row, which we call a \textit{Row-Set}.
We map each address in the Row-Set linearly to successive RowIDs.
For each RowID, we also maintain a vector of addresses (\textit{RowSet-Vec}) that do not conflict with the row-identifying address, \textit{i.e.}, thus obtaining all the addresses that map to the same row.

The \textit{Row-Set} obtained using a reference address from one bank, provides a lookup table mapping an address to a RowID for that bank. 
By changing the reference address, and repeating this process, we can get the Row-Set for successive banks.

\begin{figure*}[ht]
\centering
\begin{subfigure}[t]{.45\linewidth}
\centering
\includegraphics[width=3.1in,height=\paperheight,keepaspectratio]
{"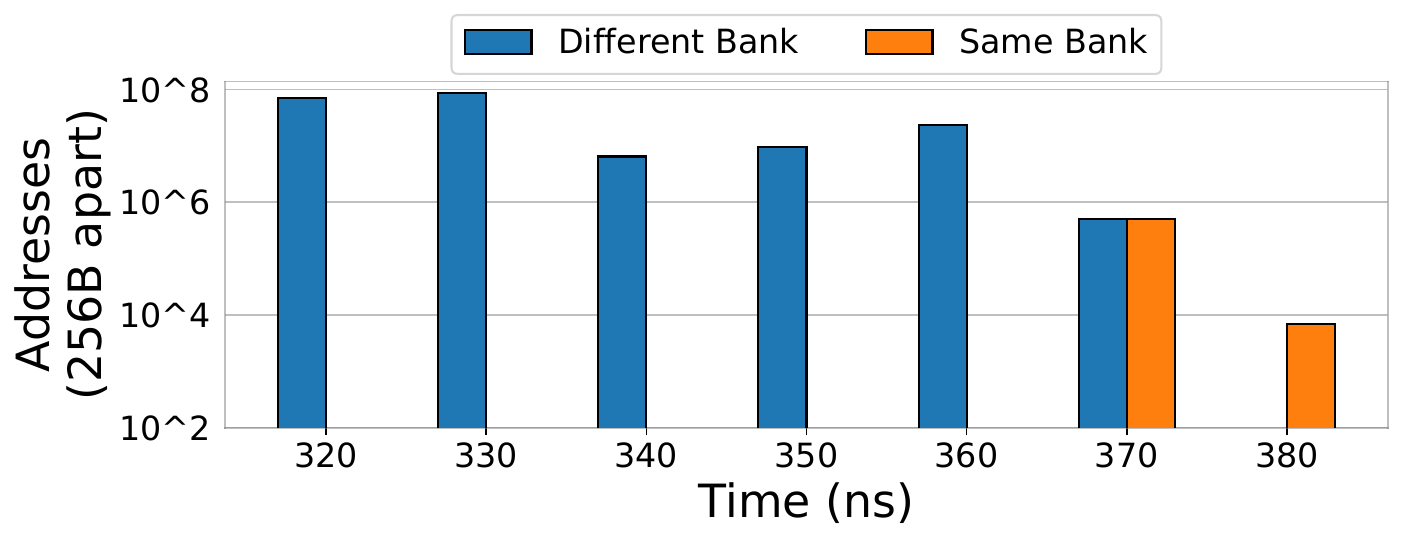"}
\caption{Before Accounting for NUMA effects}
\label{fig:before_numa}
%(a) Before Noise Removal
%\caption{\label{fig:same_diff_timing}Memory access latency measurements for address-pairs, categorized as same-bank and different-bank. Row buffer conflicts incur a delay of $\approx 360-380ns$, but as well as a subset of addresses in different banks.}
\end{subfigure}
\begin{subfigure}[t]{.45\linewidth}
\centering
\includegraphics[width=3.1in,height=\paperheight,keepaspectratio]
{"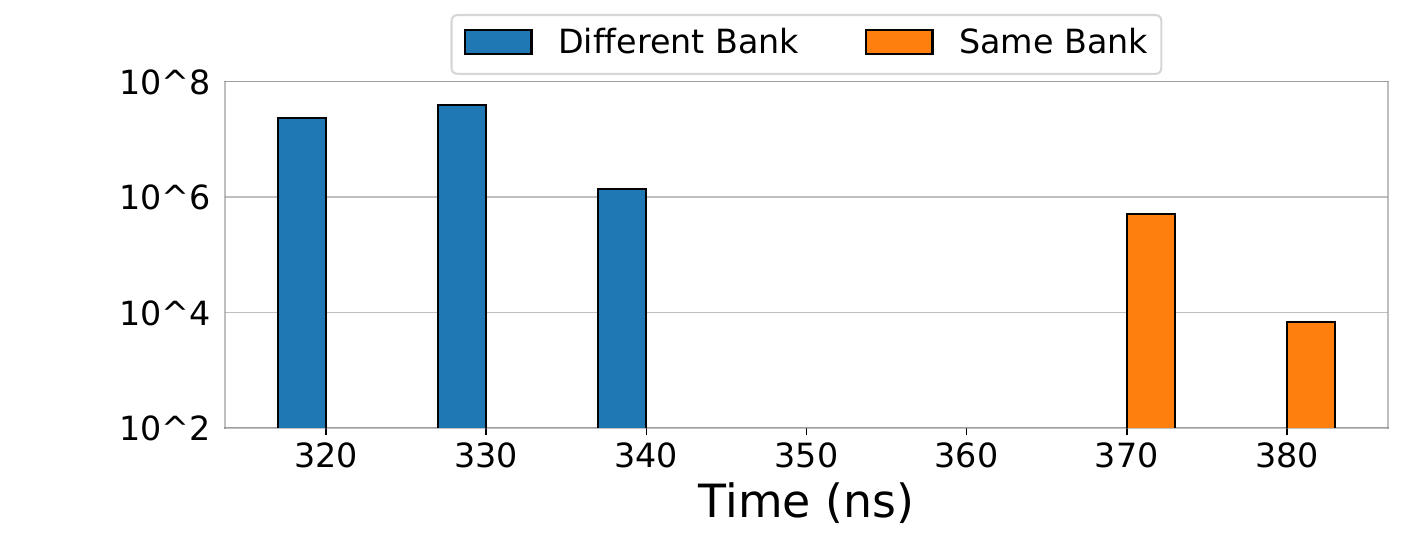"}
\caption{After Eliminating NUMA effects}
\label{fig:after_numa}
\end{subfigure}
\caption{Memory access latency measurements for address-pairs, categorized as same-bank and different-bank. (a) Naively considering pairs of reference addresses with all addresses in the memory, causes variation in the different-bank latencies (320ns - 370ns) due to NUMA effects, that overlap with the same-bank latencies (370-380ns). (b) Only considering addresses that have similar single memory access latency as the reference address, allows us to eliminate NUMA effects. 
}
\label{fig:same_diff_timing}
%Row buffer conflicts incur a delay of $\approx 360-380ns$, but as well as a subset of addresses in different banks.
%}
%\TODO{Generate Fig-(b)}
\end{figure*}

\smallskip
\noindent \textbf{Results of Timing Measurements.} We evaluated our conflict testing primitive on an NVIDIA RTX A6000 GPU equipped with 48GB of GDDR6 DRAM. We allocated a 47GB array and selected the starting address of the array as the \textit{reference} address.
%and executed both the \textit{conflict set} and the \textit{row set} routines on 47GB of the allocated memory.
\cref{fig:before_numa} shows the results of our timing measurements. 
Although pairs of addresses mapping to the same-bank with row-buffer conflicts have higher latency (360 ns - 380 ns), we observe that pairs without row-buffer conflicts\footnote{We identified the absence of row-buffer conflicts and row-activations in these pairs by adding an additional address from this set, and not observing an increase in latency corresponding to \TRC{}, which is expected on an activation.} 
have a wide range of latencies (from 320 ns to 370 ns), and some overlap with conflicting pairs.
%Although we expect the same-bank address-pair with row-buffer conflicts to have significantly higher latency than other address-pairs, that does not appear to be the case. In fact, we observe a continuous distribution of the latencies, 
This makes separating the same-bank address pairs from different-bank pairs using a cutoff threshold difficult.

We discover that this is because of non-uniform memory access latency in GPUs, similar to Non-Uniform Memory Access (NUMA) effects in servers. 
In server-grade GPUs with large memory, DRAM is often deployed in a distributed manner around the GPU SoC, introducing variation in memory access latency.
As our conflict-testing primitive measuring latency of address-pairs measures the greater of the two latencies, it is highly susceptible to such NUMA effects.
We discuss and address this next.

\subsection{Eliminating Noise due to NUMA Effects}\label{sec:chip_noise}
\begin{figure}[hbt]
\centering
\includegraphics[width=3.1in,height=\paperheight,keepaspectratio]
{"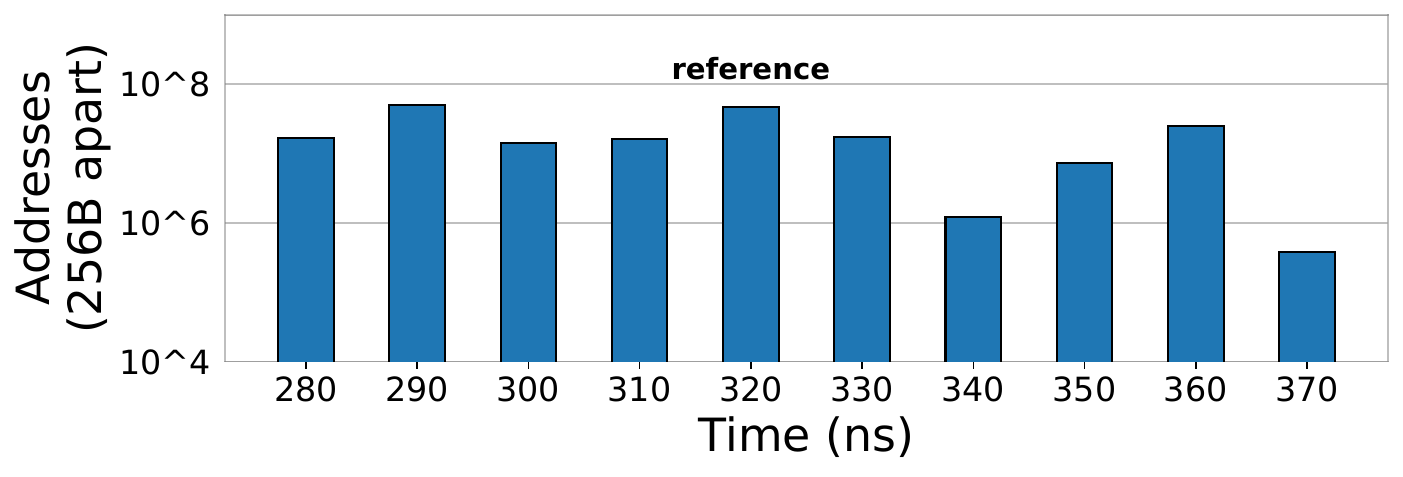"}
\caption{\label{fig:access_delay} NUMA effects in A6000 GPU. Memory access latency for single memory access for each address in allocated memory (47GB). There is significant variation in the latencies, due to NUMA effects.
%is significant enough to cause false nega/positives. 
Our \textit{reference} address
% the start of the memory space, resides in a chip that 
has a memory access latency of 320ns.}
\end{figure}

%timing and behavior of memory accesses, adding noise to our existing timing routines.
%This makes it challenging to determine a single threshold to separate the row-buffer-conflicts with row-buffer-hits, since row-buffer-hits to a remote memory chip can be higher latency than a row-buffer-conflict. 
%As shown in \cref{fig:same_diff_timing}, this noise complicates the determination of an accurate cutoff threshold. 
%To illustrate the impact of this noise, we conducted a simple experiment: fixing a single SM and performing uncached memory accesses across the memory space. 
We characterize the NUMA effects in our A6000 GPU, by performing single uncached memory accesses to different addresses in the GPU DRAM and measuring their latencies.
\cref{fig:access_delay} shows the histogram of measured memory access latency.
There is a significant variation in the latency from 280ns to 370ns, with the higher end of this range overlapping with the latency of row-buffer conflicts on our original bank (360-380ns).
However, we observe that our reference address has a latency of 320ns, and therefore all the addresses belonging to the same bank as the reference address, must also have a similar memory access latency in \cref{fig:access_delay}.

Thus, we eliminate the noise due to NUMA effects in our conflict testing process, by only considering the addresses that have a similar latency as our \textit{reference} address (within  $\pm 10$ ns of the \textit{reference}'s latency). This leverages a key observation that addresses belonging to the same bank and therefore the same DRAM chip should exhibit similar access delays. 
%By applying this heuristic to group addresses based on similar timings, we reduce the impact of noise due to the NUMA effects.
\cref{fig:after_numa} shows the results of our conflict-testing process after we only consider the addresses that have a similar latency as the reference address. Now, we see a clear separation between the row-buffer-hit latency and the row-buffer-conflict latency. We set the threshold as 350ns, i.e.,  30ns higher than the reference address latency (320ns), and automatically classify any address pairs with higher latency as row-buffer conflicts and mapping to the same bank.

\subsection{Evaluation of Row Addressing Recovery}
%We evaluated our timing routine using CUDA 12.4 on an NVIDIA RTX A6000 GPU equipped with 48GB of GDDR6 DRAM. The GDDR6 memory on this GPU features 64K rows per bank and a 2KB page size. For our experiment, we selected the starting address of the allocated memory space as the \textit{pivot} and executed both the \textit{conflict set} and the \textit{row set} routines on 47GB of the allocated memory.

%\noindent \textbf{Results}: As illustrated in \cref{fig:same_diff_timing}, the proportion of correctly identified conflict sets relative to the entire memory space is shown. 
%After collating the \textit{conflict set} into the \textit{row set}, we accurately identified $64170 \approx 64K \times (47/48)$ rows for the bank corresponding to the \textit{pivot}. 

After allocating 47GB out of 48GB GDDR6 DRAM on our A6000, we derive the conflict set and row set for 4 different banks using different reference addresses. 
Analysis of the \textit{row set} revealed the following observations: 

\begin{observation}\refstepcounter{observationcounter}\label{obs:256}
\textbf{Observation~\theobservationcounter.} In GPU memories, contiguous chunks of 256 bytes are mapped to a single row and bank, and successive 256-byte chunks map to different banks.
%decided by an unknown, possibly non-linear hash function. 
\end{observation}

%that addresses mapping to a specific row are grouped in contiguous 256-byte chunks. 
Leveraging the observation that row and bank mappings were performed at 256-byte granularity, we significantly reduced the execution time of our row set recovery, by testing at 256B granularity instead of 64B; this reduced the runtime for our row set recovery to 4 hours per bank.

\begin{observation}\refstepcounter{observationcounter}\label{obs:rbsize}
\textbf{Observation~\theobservationcounter.} The row-sets show that an A6000 GDDR6 has a row size of 2KB and 64K rows per bank.
%decided by an unknown, possibly non-linear hash function. 
\end{observation}
We repeat a similar row-set recovery for an RTX3080 with GDDR6 memory and an A100 with HBM2e memory and find that both have 32K rows per bank and a row size of 2KB and 1KB, respectively.

\begin{observation}\refstepcounter{observationcounter}\label{obs:nonlinear}
\textbf{Observation~\theobservationcounter.} The gap between two successive 256B chunks mapping to the same row varies from row-to-row; as does the gap between addresses of two rows. 
%This implies the address to row and bank ID mapping might be a non-linear hash.
%decided by an unknown, possibly non-linear hash function. 
\end{observation}
We hypothesize that the mapping of addresses to DRAM banks uses a \OTHERCHANGE{complex XOR function with inputs from many bits of the address}, although this is hard to definitively conclude without access to physical addresses. We provide a visualization of this behavior in \cref{app:non-linear}.

\begin{observation}\refstepcounter{observationcounter}\label{obs:vp_consistent}
\textbf{Observation~\theobservationcounter.} For large memory allocations, such as 47GB out of 48GB, the Virtual-to-Physical memory mapping is consistent, allowing the row-sets to be reused.
\end{observation}

Repeated executions of the row-set recovery process, across system or program restarts, resulted in different virtual addresses\REVISION{, as ASLR is enabled}, but the mapping of the offsets within the array to the rows remained the same. 
%\REVISION{This observation holds without disabling ASLR, and even well as across program and system restarts. \leftrevisionbox{R2}}
This implies that the physical memory allocated remains the same, and that we can simply reuse the row-sets for subsequent hammering.
%steps and reducing the overall time required.

By accessing the addresses from a single row set in a loop, we are able to activate rows and launch Rowhammer attacks.
%% %---------------------------

% Figure~\ref{fig:vectors}.
\section{Hammering with High Activation Rates} \label{sec:act}

%% PROBLEM

%% DIFFERENT HAMMERING APPROACHES 

%% RESULTS
Successful Rowhammer attacks require a high activation rate that crosses the \TRH{} for a given row within a refresh period. 
%High activation rate is a prerequisite to launching effective Rowhammer attacks, but 
The high memory latency (around 300ns as shown in \cref{fig:access_delay}) and the reduced refresh window (32ms or lower)\cite{JEDEC-GDDR6} in GPUs make it difficult to achieve sufficiently high activation rates. 
%deemed this difficult to achieve. We describe in this section the motivation and design of our warp pipelining approach to hide memory latency and reach the highest possible activation rate.

\subsection{Problem with Single-Thread Hammering} \label{subsec:act_motivation}
A naive CUDA kernel for hammering GPU memory, inspired by CPU-based hammering, uses a single thread to activate aggressor rows in a loop, as shown in \cref{lst:simpleRH}.
However, with GPU memory latency of around 300 ns, such a kernel can only achieve a maximum of around 106,000 activations within 32ms. 
For $n$-sided attacks which require accessing multiple aggressor rows simultaneously to bypass in-DRAM mitigations like TRR~\cite{TRRespass}, this rate is insufficient. For example, a 20-sided attack activating 20 rows at a time results in only 5K activations per row, whereas the \TRH{} for modern memories is at least 10K to 20K~\cite{RevisitRowhammer,HBMRowhammer}.

\begin{figure}[ht]
\centering
\includegraphics[width=2.8in,height=\paperheight,keepaspectratio]
{"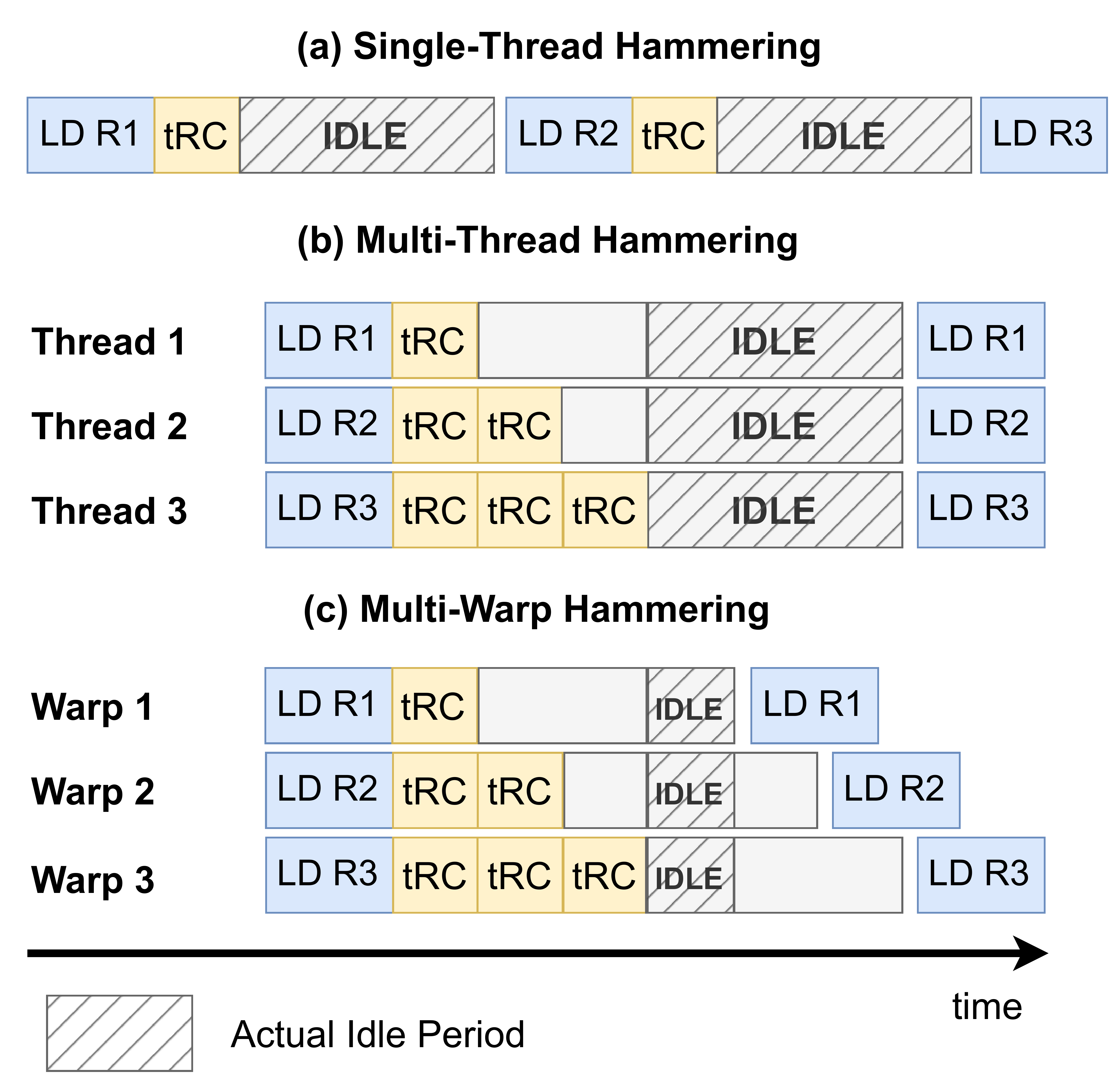"}
\caption{\label{fig:hammer_technique} Memory controller utilization with (a) single-thread, (b) multi-thread, and (c) multi-warp hammering. Multi-warp hammering minimizes idle time, maximizing activation rates.}
\end{figure}

Notably, the \TRC{}, \textit{i.e.}, time between two activations for the GPU memories GDDR6~\cite{JEDEC-GDDR6} is only around 45ns, implying that it is theoretically possible to achieve over 700,000 activations in 32ms, \textit{i.e.}, $7 \times$ the activation rates of our naive single-thread hammering.
This discrepancy indicates significant idle time at the memory controller, caused by the high round-trip latency from SMs to the memory controller due to the GPU’s large chip size, as shown in \cref{fig:hammer_technique}(a). 
We seek to reduce this idle time at the memory controller by parallelizing the hammering process.

\begin{comment}
Most modern DRAM systems implement proprietary versions of Target Row Refresh (TRR) mechanisms \cite{TRRespass}, which often require larger sets of aggressors to bypass. As a result, the total activation load is distributed across the aggressors, necessitating a higher number of activations to reach the Rowhammer threshold (\TRH), which is crucial for inducing bit flips in DDR4 memory \cite{RevisitRowhammer}. To maximize activation in a DRAM bank, it is essential to keep the memory occupied with a sufficient number of bank accesses. To illustrate the evolution of our approach, we first examine traditional hammering techniques and their limitations in the context of high memory latency.

\noindent \textbf{Single Threaded} 
The simplest hammering implementation uses a basic for-loop to repeatedly activate aggressor rows, as depicted in \cref{fig:hammer_technique}. This approach suffers from significant inefficiencies due to idle memory periods, which result from high turnaround times. The inability to hide this latency limits the activation rate, achieving only approximately 64K total row activations, or roughly 12\% of the theoretical maximum.
\end{comment}

\subsection{Parallel Hammering for More Activations}
\label{subsec:multi-warp}
To reduce the idle time at the memory controller and increase DRAM activation rates, we explore two techniques for parallelized hammering: multi-thread and multi-warp hammering. 

\smallskip
\noindent \textbf{Multi-Thread Hammering.} 
\begin{comment}
To address these limitations, we explore a multi-threaded design in which each aggressor is assigned to a dedicated thread, allowing GPU threads to process memory accesses in lockstep. 
As illustrated in \cref{fig:act_rate}, this design effectively hides the idle window of earlier accesses by utilizing \TRC, leading to a significant improvement in activation rates (approximately 370K activations with 32 threads). The performance scales well with the number of aggressors, making this approach far more effective than single-threaded execution.
However, the lockstep execution model introduces its own inefficiencies. When some threads complete their memory requests earlier than others, they must wait for the slowest request to finish, creating idle periods in the hammering process. These gaps reduce overall efficiency. Ideally, memory requests should arrive with minimal delay between them to maintain continuous bank activity, with interruptions only during \REF. This underscores the need for a finer-grained access pattern to further optimize activation rates.
\end{comment}
In this approach, each aggressor is assigned a dedicated thread in a single warp. 
As shown in \cref{fig:hammer_technique}(b), the \texttt{load} instructions to each hammered row are executed in parallel in lockstep.
Thus, the ACT to each row is issued back-to-back every \TRC{}, reducing the idle time at the memory controller and increasing activation rates. 
However, this still incurs some idle time due to the lock-step execution of the loads, as the SM must wait for all loads in the current iteration to return before issuing the next round of loads.

\smallskip
\noindent \textbf{Multi-Warp Hammering.} 
%To address the limitation of multi-thread hammering, 
To address the above limitation, we adopt multi-warp hammering, where each aggressor is handled by a thread in a different warp. 
When one warp is stalled due to memory access, the SM schedules the next warp in a round-robin manner with minimal overhead.
Unlike threads within a warp, threads in different warps do not execute in lock-step; thus, the SM can issue a memory request as soon as a request from a previous iteration returns, without waiting for all the memory requests of the current iteration to complete. 
This further reduces idle time at the memory controller and increases the activation rate during hammering. 

We implement $n$-sided multi-warp hammering, \textit{i.e.}, one aggressor per warp hammering with $n$ aggressors, using a CUDA kernel with $32 \times n$ threads (as each warp has 32 threads). Each aggressor is assigned to a thread $0, 32, 64, \dots$, and so on.

\begin{comment}
The primary limitation of multi-threaded hammering is its inability to allow completed memory operations to proceed without unnecessary delays. To address this, we leverage the concept of warps to effectively mitigate memory access latency. In NVIDIA GPUs, a warp consists of a group of 32 threads that execute instructions in lockstep. These warps are scheduled by the Streaming Multiprocessor (SM) in a manner similar to CPU thread scheduling. When a warp encounters a stall, modern GPUs can seamlessly switch to another ready warp, ensuring continuous execution. This warp-switching process incurs negligible overhead, enabling us to efficiently mask memory latency in our hammering workload. 

\noindent \textbf{Design} On the RTX A6000 and many other GPUs, a warp consists of 32 threads, with thread blocks typically containing more than 1024 threads. We assume that consecutive groups of 32 threads are optimized to form warps, such that Warp 1 includes threads [0–31], Warp 2 includes threads [32–63], and so on. With 1024 threads allocated, we can utilize 32 warps, assigning one aggressor per warp. Specifically, we designate threads 0, 32, 64, and so forth to serve as aggressors, while the remaining threads are allowed to exit. Once a sufficient number of warps are allocated to saturate the DRAM bank, the memory accesses generated by each warp will be evenly spaced, with their return times staggered by \TRC after the initial warm-up phase.

\end{comment}

\begin{figure}[ht]
\centering
\includegraphics[width=3.1in,height=\paperheight,keepaspectratio]
{"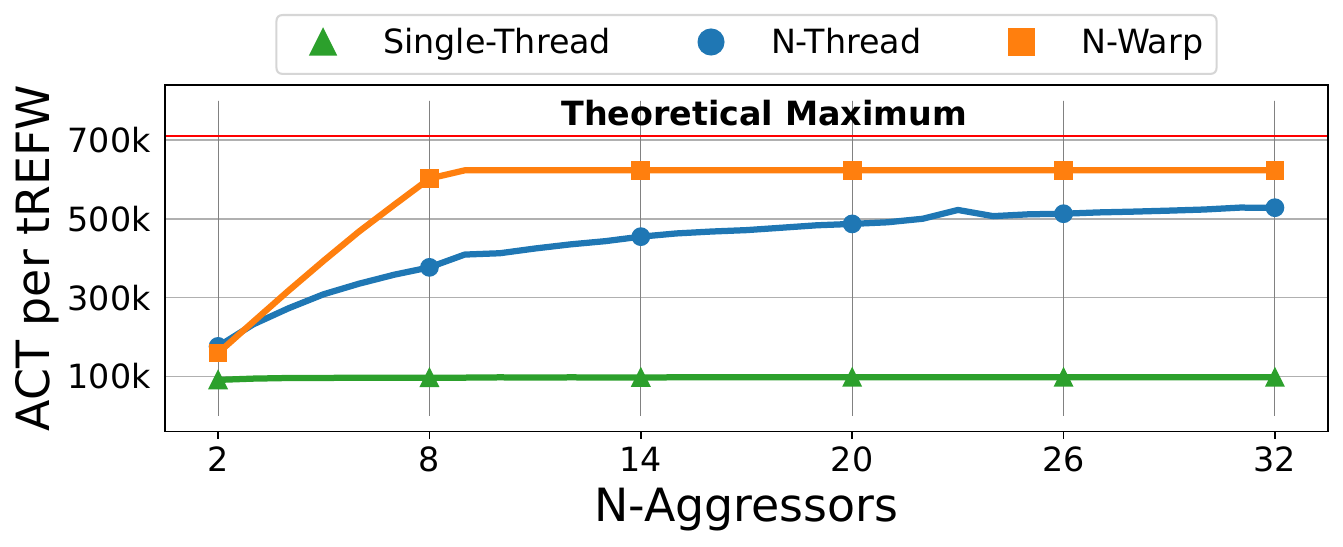"}
\caption{\label{fig:act_rate} Activation rates (ACTs per \TREFW) with three different hammering techniques. Multi-warp hammering achieves the best activation rate of 620K per \TREFW (32ms), reaching close to the theoretical maximum (around 700K).
}
%We assumed \TRC = 45ns, \TREFI = 1.407$\mu$s (explained in \cref{subsec:sync_eval}) and calculated \textbf{Theoretical Maximum} with equation: $\TREFW / \TRC$}
\end{figure}

%\noindent 
\subsection{Evaluation of Parallelized Hammering}
%\textbf{Results} 
We evaluate the different hammering techniques, including single-thread, multi-thread, and multi-warp hammering, on an NVIDIA A6000 GPU with 48GB GDDR6 DRAM, and compare the number of activations in a refresh period (ACTs per tREFW) as the number of aggressor rows increases. 
%We assume an ACT time (\TRC{}) of 45ns. 
We assume a refresh period (tREFW) of 32ms and a time per ACT (\TRC{}) of 45ns.
Thus, the theoretical maximum ACTs per tREFW is around 700K.

As shown in \cref{fig:act_rate}, the single-thread hammering only reaches an ACT per tREFW of approximately 90K, or roughly 12\% of the theoretical maximum.
The multi-thread hammering (N-Thread) significantly increases the activation rate, which grows as the number of aggressor rows increases. However, even with 32 aggressors, it reaches a maximum of less than 520K ACTs per \TREFW. In contrast, multi-warp hammering (N-Warp) reaches around 620K ACTs per \TREFW with just 8 aggressors, reaching close the theoretical maximum.
This is due to the reduced idle time at the memory controller compared to the other approaches.

Multi-warp hammering, with as few as 8 warps, achieves an activation rate almost 7$\times$ higher than single-thread hammering, and sustains close to 20K activations per \TREFW per row even with 32 aggressors.
Therefore, we adopt multi-warp hammering with 8 warps by default to ensure the maximum ACT intensity for our attack.
%still allowing around $10K$ activations at 32 aggressors.

%The sufficiency of 8 or more warps is tied to the conflict timing detailed in \cref{sec:reverse}. With fewer than 8 aggressors, the maximum memory latency gap from a single aggressor (360ns), cannot be fully hidden. For instance, assuming a \TRC of 45 ns and 3 aggressors, we can keep the memory busy for at most $3 \times 45 = 135ns$. However, there is still at least $360 - 135 = 225ns$ of gap before another request can be issued from one of the aggressors. At 8 aggressors, the accumulated latency of $45 \times 8 = 360ns$ is sufficient to bridge this gap, allowing us to achieve the maximum activation rate.

\begin{comment}
we achieve almost the theoretical maximum after having 8 warps. The gap between the theoretical maximum and our design is explained by the \TRFC required per \TREFI. We achieve around $450K$ activations, still allowing around $10K$ activations at 32 aggressors.

The sufficiency of 8 or more warps is tied to the conflict timing detailed in \cref{sec:reverse}. With fewer than 8 aggressors, the maximum memory latency gap from a single aggressor (360ns), cannot be fully hidden. For instance, assuming a \TRC of 45 ns and 3 aggressors, we can keep the memory busy for at most $3 \times 45 = 135ns$. However, there is still at least $360 - 135 = 225ns$ of gap before another request can be issued from one of the aggressors. At 8 aggressors, the accumulated latency of $45 \times 8 = 360ns$ is sufficient to bridge this gap, allowing us to achieve the maximum activation rate.
\end{comment}

\section{Synchronization with Refresh Commands}\label{sec:sync}

%%% Importance of Synchronization. Key requirements - (a) Bubble at the memory controller, and (b) Act of synchronization should not affect the order of the aggressors in the pattern.

%%% Prior work synchronizes in a single-threaded setting. To emulate this naively, one needs to add __syncthreads, to synchronize all the threads and then add the delay. But this causes the ordering to be affected. Therefore, we experiment adding a delay (a non-memory operation like ADD) directly in each of the threads, without explicit thread synchronization primitives.

%% Leads to loss of synchronization or very weak synchronization, beyond 8 warps. 
%% Need more delays that overlap. 
Synchronizing hammering patterns with DRAM refresh commands (REFs) is crucial for bypassing in-DRAM mitigations like TRR~\cite{smash,ZenHammer}. \REVISION{TRR mitigates Rowhammer by tracking frequently accessed rows and issuing victim refreshes coupled with regular refreshes. Without synchronization, the TRR sampler state is unpredictable, giving each victim row a chance to be refreshed. As a result, our attack can be mitigated even if we overflow the sampler’s capacity with multi-warp hammering in \cref{subsec:multi-warp}. By contrast, a synchronized pattern forces the TRR sampler into a well-defined state before regular refresh. This technique ensures that specific aggressor rows repeatedly escape mitigation in the presence of TRR.}\rightrevisionbox{R2}\vspace{-0.1in}

% As mitigative refreshes in TRR are issued at \TREFI intervals, in the shadow of regular refreshes~\cite{UncoverRowhammer}, a synchronized many-sided attack pattern can fool the TRR sampler deterministically in every \TREFI interval, ensuring a particular aggressor consistently escapes mitigation.
To synchronize with REF commands, there are two key requirements: (1) A bubble must be inserted at the memory controller at the end of a hammering iteration, sparing time for the REF command, and (2) synchronization must not alter the order of aggressors. 
Prior works~\cite{smash,ZenHammer} achieved synchronization in single-thread setting by strategically inserting NOPs after a hammering iteration, timed to align with \TREFI intervals. 
However, in multi-warp hammering on GPUs, we observe that simply using \texttt{\_\_syncthreads} for thread-synchronization at the end of a hammering iteration introduces uncertainty to warp ordering, complicating the synchronization to REF\REVISION{s}. 

\begin{figure}[ht]
\centering
\includegraphics[width=3.1in,height=\paperheight,keepaspectratio]
{"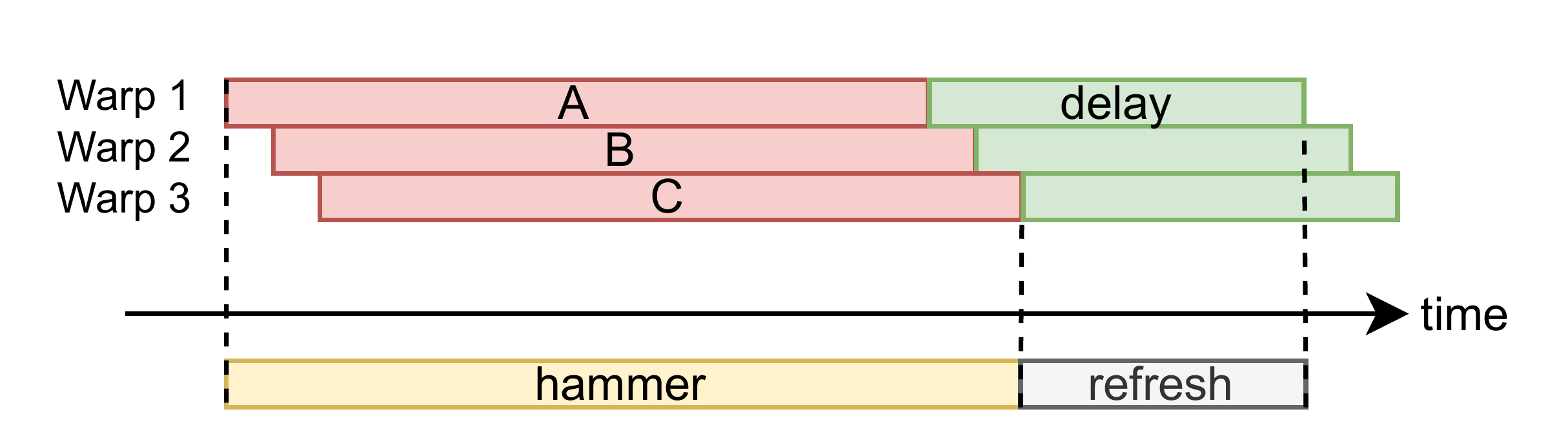"}
\caption{\label{fig:per_warp_nop} Per-Warp delays inserted by \texttt{add} for synchronizing to \REF. When the delays overlap for all the warps, the REF is inserted in alignment with the hammering pattern.}
\end{figure}

\smallskip
\noindent \textbf{Per-Warp Synchronization Design.} 
In our design, we avoid using \texttt{\_\_syncthreads} and instead add delays within each warp after $i$ hammering iterations using the \texttt{add} instruction. When delays overlap across all warps, a bubble is implicitly created at the memory controller, allowing a \REF{} to be inserted in alignment with the hammering pattern.
\Cref{fig:per_warp_nop} illustrates how this overlap achieves synchronization without disrupting warp order.
We insert \texttt{add} instructions every $i = 1, 2, \text{or } 3$ hammering iterations, depending on the $n$-sided pattern, to ensure that a whole number of iterations are completed within a  \TREFI{} while limiting activations to no more than 24 per \TREFI{}.
%, to retain the synchronization.
Thus, each \textit{round} of hammering consists of $n \times i$ activations followed by some \texttt{add} delays. We provide a sample implementation in \cref{app:Nwarp_Kthread}.

\begin{figure*}[ht]
\centering
\begin{subfigure}[t]{.49\linewidth}
\centering
\includegraphics[width=3.5in,height=\paperheight,keepaspectratio]
{"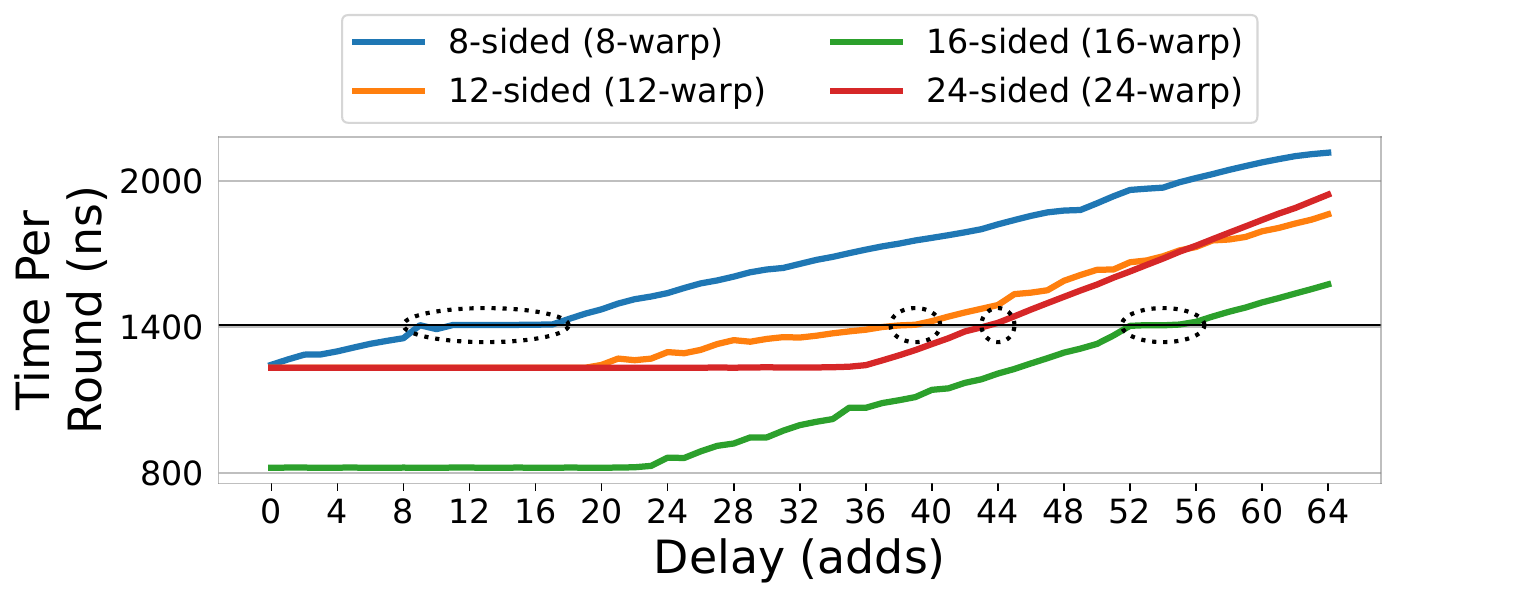"}
\caption{Multi-Warp Hammering (1 thread per warp).}
\label{fig:1threadperwarp}
\end{subfigure}
\begin{subfigure}[t]{.49\linewidth}
\centering
\includegraphics[width=3.5in,height=\paperheight,keepaspectratio]
{"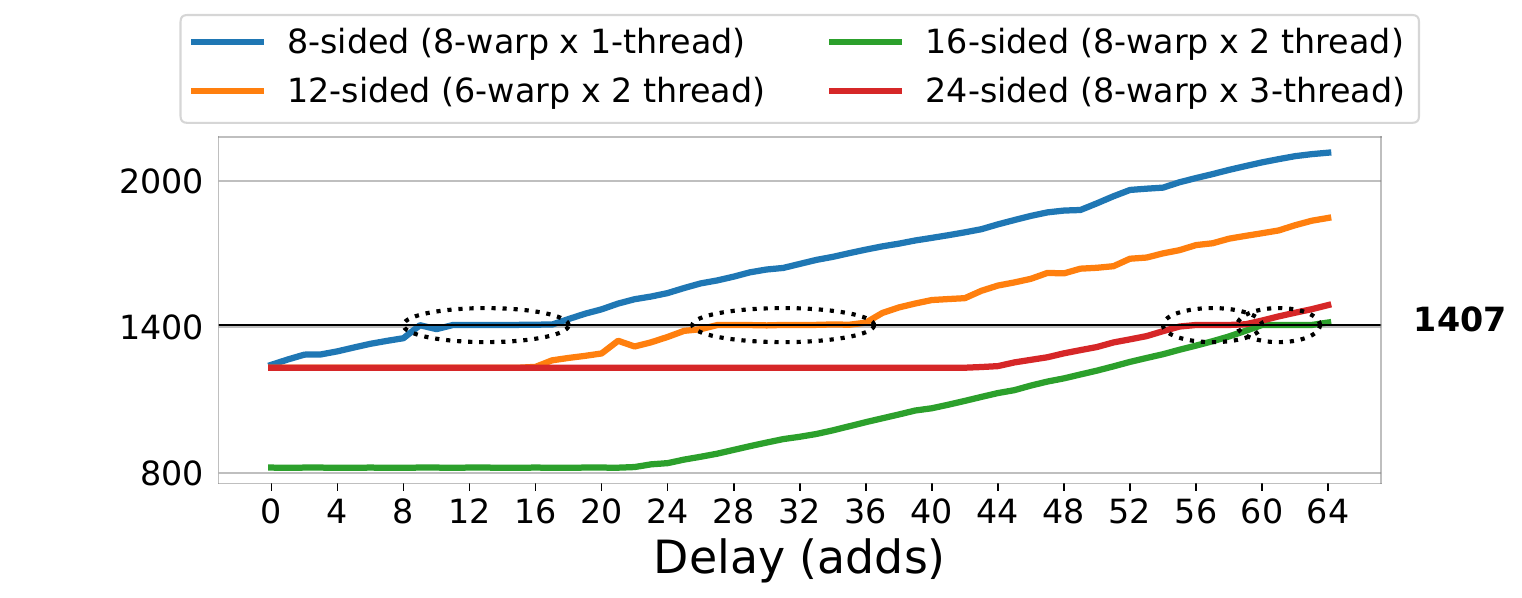"}
\caption{Multi-Warp \& Thread Hammering ($k$-warps, $m$-threads-per-warp)}
\label{fig:manythreadsperwarp}
\end{subfigure}
\caption{Evaluation of \TREFI Synchronization plotting Time per Round (ns), where a round consists of 1, 2, or 3 iterations of n-sided hammering with up to 24 activations. (a) With delays added to naive multi-warp hammering, synchronization works well for 8-sided (8-warp), but is weaker for 12-sided or 24-sided (12/24 warps), likely due to delays not overlapping sufficiently. (b) By using 8 warps or less and multiple threads per warp, we ensure synchronization for all $n$-sided patterns.
}
\label{fig:sync_results}
%\vspace{-0.1in}
\end{figure*}

\smallskip
\noindent \textbf{Evaluation of Synchronization.} 
Our synchronization approach relies on the overlap of delays across warps, which is heavily influenced by the warp configuration.
\cref{fig:sync_results}(a) shows synchronization results for multi-warp hammering patterns (1 thread per warp) using 8-,12-,16-, and 24-sided aggressor patterns.
We observe that the 8-warp configuration achieves strong synchronization with tREFI, as shown by a flat-lined time-per-round at approximately 1407ns. 
However, the 12- to 24-warp configurations have weak or no synchronization, likely because the overlap of delays diminishes, and eventually vanishes, as the number of warps increases.

To ensure synchronization for patterns with a larger number of aggressors ($n$), we adopt configurations with $k$ warps and $m$ threads per warp, such that $n = k \times m$. We further restrict the number of warps $k$ to be 8 or lower, in order to maintain a sufficient overlap of delays for synchronization.
\cref{fig:sync_results}(b) shows the synchronization for $n$-sided patterns, using $k$-warp $m$-thread-per-warp configurations. We achieve sufficient synchronization for all tested $n$-sided patterns.
We list the code for our final kernel with synchronized, many-sided hammering with $k$ warps and $m$ threads per warp, in \cref{lstlisting:synchammering} in \cref{app:Nwarp_Kthread}.

Additionally, based on the synchronization instances, we observe that the GDDR6 memory in the A6000 GPU has a \TREFI of 1407ns and 16K tREFIs per refresh period. Thus, our $n$-sided hammering can hammer up to 24 aggressor activations per \TREFI{} and a total of 393K activations per refresh period.
%Additionally, a key statistic for Rowhammer attacks is the maximum number of aggressor activations that can be issued within a single \TREFI interval. Without accounting for \TRFC, and given $\TRC = 45 \text{ ns}$, up to approximately 31 activations (ACTs) can occur per \TREFI. To refine this estimate, we leverage the activation count from \cref{fig:act_rate} alongside the fact that there are 16K refreshes per \TREFW~\cite{JEDEC-GDDR6}. This calculation yields $\lfloor 450K / 16K \text{ refreshes} \rfloor \approx 27$ ACTs per \TREFI.
\begin{observation}\refstepcounter{observationcounter}
\textbf{Observation~\theobservationcounter.} The refresh interval (tREFI) for A6000 GDDR6 is 1407ns, leading to a refresh period of 23ms. 
\end{observation}

\section{\attack Evaluation}\label{sec:eval}

Using the row-sets from \cref{sec:reverse}, the high-intensity hammering kernels from \cref{sec:act}, and the synchronized many-sided hammering from \cref{sec:sync}, we launch \attack{} attacks. To evaluate our attack, we answer the following questions:
\begin{enumerate}[itemsep=-2pt,topsep=1pt]
    \item In systematic hammering campaigns, which NVIDIA GPUs are vulnerable to bit flips?~(\cf{}~\cref{subsec:eval_campaigns}) 
    \item What aggressors (\cf{}~\cref{subsec:aggr_patterns}) and data patterns (\cf{}~\cref{subsec:datapattern}) are most effective in flipping bits?
    \item What is the directionality of the bit-flips (\cf{}~\cref{subsec:bitflip_direction}), the Rowhammer threshold (\cf{}~\cref{subsec:TRH}), and the in-DRAM TRR sampler size~(\cf{}~\cref{subsec:TRR_size})?
    \item Are the bit-flips realistically exploitable? (\cf{}~\cref{sec:exploit})
    \item What are the possible mitigations for \attack{} and their associated costs?~(\cf{}~\cref{sec:mitigations}) 
\end{enumerate}

\subsection{Systematic Hammering Campaigns}\label{subsec:eval_campaigns}

% - 8, 12, 16, 24 sided hammering. 4 banks on A6000 (GDDR6)
% - RTX 3080 (GDDR6)
% - A 100 (HBM2e)
% - Backup RTX 4090 (GDDR6X)
We conduct systematic hammering campaigns across four DRAM banks on three GPUs with ECC disabled: an RTX A6000 GPU with GDDR6 memory (in a workstation), an A100 with HBM2e memory (in the cloud), and an RTX 3080 with GDDR6 (in a laptop). 
We used $n$-sided patterns, starting with distance-2 between aggressors (hammering rows $R_{i}, R_{i+2}, \dots, R_{i+2n}$) where we did not observe bit-flips, and then with distance-4 between aggressors (hammering $R_{i}, R_{i+4}, \dots, R_{i+4n}$) when we started to see bit-flips.

\cref{table:sweep} shows the results for 8-,  12-, 16-, 20-, and 24-sided hammering with distance-4 aggressors, using victim data of \texttt{0x55$\dots$} or \texttt{0xAA$\dots$} (aggressor data was inverted). For each $n$-sided pattern, we iterated through the entire DRAM bank, incrementing 3 rows at a time. We hammered each pattern for 128ms ($\sim$5 refresh periods), with a total run time of 30 hours per bank per GPU.
The A6000 with GDDR6 memory exhibited a total of 8 unique Rowhammer bit-flips (one per row), with at least one bit-flip in each hammered bank and all of them were single bit-flips. The A100 (HBM2e) and the RTX3080 (GDDR6) GPUs did not exhibit any bit-flips --
\OTHERCHANGE{we discuss potential reasons in \Cref{sec:discussion}.}

\begin{result}\refstepcounter{resultcounter}
\REVISION{\textbf{Takeaway~\theresultcounter.} A6000 GDDR6 DRAM is vulnerable to Rowhammer, with bit-flips observed on every bank hammered; No flips were observed on the A100 or RTX3080.}
\end{result}
%\vspace{-0.075in}\rightrevisionbox{R2}\vspace{-0.25in}

\begin{table}[htb]
\centering
\caption{Number of bit-flips observed across 4 banks on different GPUs. 
A6000 with GDDR6 \REVISION{had} bit-flips on every bank.
%with 24-sided hammering.
}
%\vspace{-0.1in}
\footnotesize
\begin{tabular}{cc|ccccc}
\hline \hline

\multirow{2}{*}{\begin{tabular}[c]{@{}c@{}}GPU\end{tabular}} & \multirow{2}{*}{\begin{tabular}[c]{@{}c@{}}Bank\end{tabular}} & \multicolumn{5}{c}{$n$-Sided Patterns}    \\
                        &    & 8 & 12  & 16 & 20 & 24 \\ \hline \hline
\multirow{4}{*}{RTX A6000}  & A  & 0  & 0  & 0  & 0           & \textbf{1}    \\
                            & B  & 0  & 0  & 0  & \textbf{1}  & \textbf{3}    \\
                            & C  & 0  & 0  & 0  & 0           & \textbf{1}    \\
                            & D  & 0  & 0  & 0  & 0           & \textbf{3}    \\
\hline
\hline
\multirow{1}{*}{A100}  & E~/~F~/~G~/~H  & 0  & 0  & 0  & 0  & 0   
\\
                        %& F  & 0  & 0  & 0  & 0    \\
                        %& G  & 0  & 0  & 0  & 0    \\
                        %& H  & 0  & 0  & 0  & 0    \\
                        \hline \hline
\multirow{1}{*}{RTX 3080}  & I~/~J~/~K~/~L  & 0  & 0  & 0  & 0  & 0    \\
%                        & J  & 0  & 0  & 0  & 0    \\
%                        & K  & 0  & 0  & 0  & 0    \\
 %                       & L  & 0  & 0  & 0  & 0    \\
                        \hline \hline
                        
\end{tabular}
%\vspace{-0.2in}
\label{table:sweep}
\end{table}

\subsection{Critical Aggressors Characterization}\label{subsec:aggr_patterns}
For the A6000 GPU where we observed bit-flips, we characterize the critical aggressor rows whose presence is sufficient to trigger a bit-flip in the victim.  
%This provides some hints on the nature of the vulnerability, and the neighborhood
For this, we hammered the 8 vulnerable locations we identified in the A6000 (banks A, B, C, and D), starting with distance-4 sequential aggressor patterns around the victim, and replacing rows from the pattern with random rows, to identify the critical aggressor rows whose presence is sufficient to trigger a bit-flip.
\cref{table:bitflips} shows the critical aggressor rows for each victim.
%We further characterized the bitflips by repeatedly launching sequential hammering near rows with observed bitflips. This allowed us to identify the critical aggressor row(s) for each victim, whose presence in a sequential pattern is sufficient to trigger a bitflip.

\begin{table}[htb]
\centering
\caption{Critical Aggressor Rows. Assuming the vulnerable victim row is $R_i$, a tick under $R_{j}$ indicates that hammering $R_{j}$ in the aggressor pattern is sufficient to trigger bit-flip at $R_i$.}

\footnotesize
\begin{tabular}{c|cccccc}
    \multirow{2}{*}{\begin{tabular}[c]{@{}c@{}}Bit-flip\\ No.\end{tabular}} & \multicolumn{6}{c}{Critical Aggressor Row} \\
     & $R_{i-3}$ & $R_{i-2}$ & $R_{i-1}$ & $R_{i+1}$ & $R_{i+2}$ & $R_{i+3}$ \\ \hline \hline
    $A_1$ &  & \checkmark &  &  &  &  \\
    $B_1$ &  & \checkmark & \checkmark &  &  &  \\
    $B_2$ &  &  &  &  & \checkmark & \checkmark \\
    $B_3$ &  &  &  &  & \checkmark &  \\
    $C_1$ &  & \checkmark & \checkmark &  &  &  \\
    $D_1$ &  & \checkmark & \checkmark &  &  &  \\
    $D_2$ &  &  &  &  & \checkmark &  \\
    $D_3$ &  &  &  & \checkmark & \checkmark & 
\end{tabular}
\label{table:bitflips}
\end{table}

Our first observation is that all our bit-flips are due to \textit{single-sided} hammering: our critical aggressor rows are either $+1$, $+2$, $+3$ neighbors of the victim row $R_i$, or $-1$, $-2$ neighbors of $R_i$, but never on both sides.
Second, some of our bits ($B_2$) flip with aggressors at $R_{i+2}$ and $R_{i+3}$, but not with $R_{i+1}$. This indicates that physical rows may not be linearly laid out within the DRAM, and there could be some remapping internally.

Finally, we observe that hammering the $R_{i + 2}$ or $R_{i - 2}$ rows produces the most reliable bit-flips, whereas $R_{i \pm 1}$ or $R_{i + 3}$ produces bit-flips less reliably. Thus, we hypothesize that $R_{i \pm 2}$ may be neighbors of $R_i$, and $R_{i \pm 1}$ or $R_{i + 3}$ may be in the neighborhood, but not directly adjacent to $R_i$.

\begin{result}\refstepcounter{resultcounter}
\REVISION{\textbf{Takeaway~\theresultcounter.} On the A6000, bit-flips are triggered by single-sided hammering on critical aggressor rows from only one side of the victim, with $R_{i \pm 2}$ being the most effective, suggesting non-contiguity in DRAM row layout.}
\end{result}
%\leftrevisionbox{R2}\vspace{-0.2in}

\subsection{Rowhammer Threshold Characterization}\label{subsec:TRH}

\OTHERCHANGE{
The Rowhammer threshold ($T_{RH}$) is the minimum number of activations to an aggressor row required to induce bit-flips in adjacent rows. We characterize the $T_{RH}$ for the bit-flips on the A6000 using 24-sided aggressor patterns.
%with some gaps in hammering.
We do so 
%we 
%by 
%.identifying the lowest hammering intensity that still triggers each flip. Specifically, 
by hammering the aggressor pattern for $x$ \TREFI{}s out of every $x + y$ \TREFI{}s, issuing dummy accesses for the remaining $y$ \TREFI{}s.
We gradually lower the hammering intensity until it stops triggering bit flips.
%We use 24-sided sequential patterns, where each aggressor is activated once per \TREFI{}. 
Thus, each aggressor receives $x/(x + y)$ times the activations it originally received in a refresh window (tREFW), \textit{i.e.}, $16\text{K} \times x/(x + y)$. 
%We vary $x$ and $y$ from 1 to 100, applying binary search to narrow the search space.
%, and repeat each configuration 1200 times. 
We vary $x$ and $y$ from 0 and 100, to find the lowest value of $x/(x+y)$ for which the bit-flip is triggered, to estimate the $T_{RH}$ for a bit flip.}

\begin{figure}[ht]
\centering
\includegraphics[width=3.4in,height=\paperheight,keepaspectratio]
{"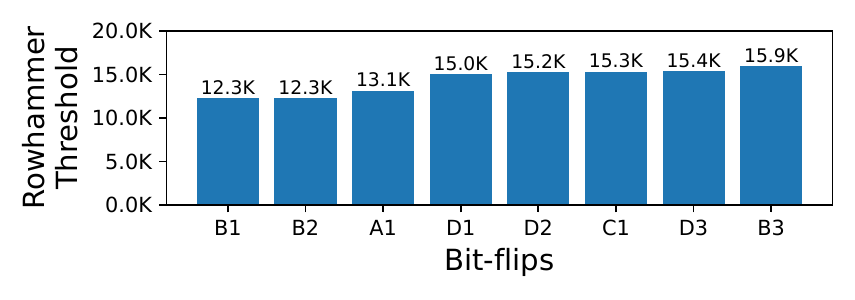"}
\vspace{-0.2in}
\caption{\label{fig:trh}Rowhammer threshold ($T_{RH}$) for each of the 8 observed bit-flips on the A6000. The minimum $T_{RH}$ observed for the A6000 GPU across all the bit flips is 12.3K activations.}
\vspace{-0.1in}
%The dotted line at 16.4K marks the number of accesses to the critical aggressor per refresh window when no dummy activations are inserted. Lower bars indicate that fewer activations are sufficient to induce bit-flips under reduced hammering intensity.}

\end{figure}

\Cref{fig:trh} shows the $T_{RH}$ for the different bit flips. Across all bit flips, the minimum $T_{RH}$ is around 12.3K, close to the previously reported values for DDR4~\cite{RevisitRowhammer}. We observe that 3 bit-flips (in banks A and B) have $T_{RH}$ of 12–13K, while 4 others (in banks C and D) have it around 15–16K. This suggests that $T_{RH}$ may have some correlation with the bank.
%or sub-array in which the flip occurs, as flips from the same bank exhibit similar thresholds. 
One outlier is B3, whose $T_{RH}$ of 15.9K is higher than that of B1 and B2 (12.3K). However, B3’s bit-flip direction is opposite to that of B1 and B2, as shown in \Cref{subsec:bitflip_direction}. Therefore, it might be in a different sub-array within the bank. 
%The bit-flip direction analysis will be presented .

\begin{result}\refstepcounter{resultcounter}
\REVISION{\textbf{Takeaway~\theresultcounter.} The A6000 GPU has an observed $T_{RH}$ of 12.3K activations. An attacker requires at least 12.3K activations to a single DRAM row to observe a bit-flip.}
\end{result}

\subsection{TRR Sampler Size Characterization}\label{subsec:TRR_size}
\OTHERCHANGE{Prior works show that DRAM implements Target Row Refresh (TRR), an in-DRAM Rowhammer mitigation that tracks frequently activated rows by a fixed-size sampler and accordingly issues mitigative refreshes~\cite{TRRespass, Blacksmith, UncoverRowhammer}. 
Following prior works like TRRespass and SMASH~\cite{TRRespass, smash}, we design our aggressor patterns to test for the memory’s Rowhammer defenses. We observe, in GDDR6, bit-flips only occur when using patterns with a large number of distinct aggressor rows issued in synchronization with REF commands, suggesting the presence of a TRR-like mitigation that can only track a limited number of aggressors.}

% Our aggressor patterns, designed similarly to prior works like TRRespass and SMASH~\cite{TRRespass, smash}, can only induce bit-flips in GDDR6 memory when hammering a large number of aggressors synchronized to REF commands.}
% Thus, we can conclude that GDDR6 also has an in-DRAM mitigation, similar to the TRR sampler.
%overflowing this sampler, hammering more aggressor rows than its capacity. Therefore, the minimum number of aggressor rows required to trigger a bitflip reveals the sampler's capacity.

\begin{figure}[ht]
\centering
\includegraphics[width=3.4in,height=\paperheight,keepaspectratio]
{"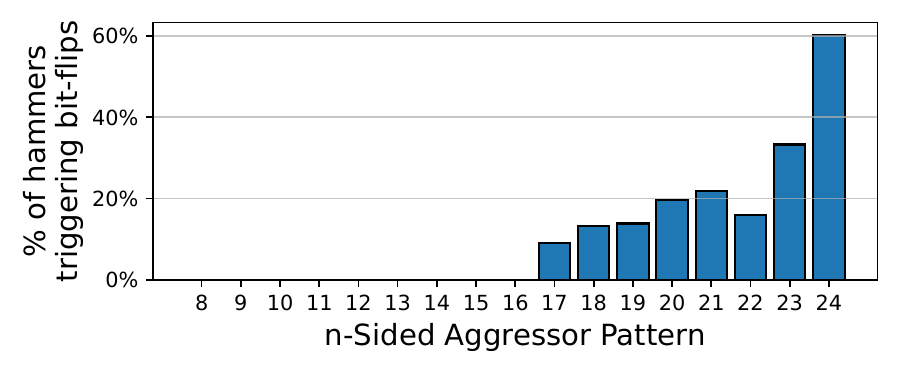"}
\caption{\label{fig:trr_sampler}Fraction of hammers that trigger bit-flips with different $n$-sided aggressor patterns. No bit-flips are observed when the number of aggressors is less than 17.}
\vspace{-0.1in}
\end{figure}

We studied 8- to 24-sided attack patterns on a victim row with a known bit-flip, $A_1$. In each attack pattern, we included a critical aggressor row and filled the remaining positions with random dummy rows from the same bank.
%, minimizing the impact of locality on the sampler. 
For each $n$-sided pattern, we also varied the position of the critical aggressor row within the pattern randomly and hammered each configuration 50 times. \Cref{fig:trr_sampler} shows the fraction of hammers that triggered a bit-flip with a given $n$-sided pattern. %triggered across all positions of the critical aggressor row. 

\REVISION{
We observe an absence of bitflips for patterns with 16 or fewer aggressor rows ($n \leq 16$);
whereas patterns with 17 or more aggressor rows (($n \geq 17$) consistently induce bit-flips.
%, while those with 16 or fewer aggressors fail to do so across all tested configurations. 
This behavior mirrors that in prior work on TRR-protected DDR4 modules~\cite{TRRespass}, where insufficient aggressors fail to trigger bitflips. The presence of bitflips only after $n = 16$ suggests that the TRR sampler in A6000’s GDDR6 can track at most 16 rows per bank and issue mitigative refreshes.
}
\begin{result}\refstepcounter{resultcounter}
\REVISION{\textbf{Takeaway~\theresultcounter.} A TRR-like fixed-size sampler is present in A6000’s GDDR6 memory, tracking 16 rows per bank.}
\end{result}
%\vspace{-0.05in}\rightrevisionbox{R2}\vspace{-0.25in}

\subsection{Bit-Flip Direction Characterization}\label{subsec:bitflip_direction}
\Cref{table:flip_data} shows the bit-flip directions for each of the bit-flips we observed on A6000. A majority of the bit-flips are $0 \rightarrow 1$ flips (in anti cells) compared to flips from $1 \rightarrow 0$ (true cells), which are fewer in number.
Interestingly, we observe Bank $B$ has bit-flips in both directions, indicating that the GDDR6 DRAM bank architecture probably consists of multiple sub-arrays with different types of cells, similar to that observed for DDR4 DRAM~\cite{DRAMScope}.
%We present the bit values and their corresponding bit locations for each bitflip in \Cref{table:data_pattern}.
%In terms of the relative location of these bit flips within the word,  
Additionally, we notice that a majority of these flips are in the 0th, 6th, or 7th bit in a given byte, which may impact their exploitability, as we discuss in \cref{sec:exploit}.
%covers 7 out of 8 flips. 

%the flip locations, since each address stores 8 bits, we numbered them from 1 to 8, starting from the least significant bit to the most significant bit.

\begin{table}[htbp]
\centering
\caption{Direction of Bit-Flips and Bit Location in a Byte.}
\footnotesize{}
\begin{tabular}{c|cc|cc|c}
\multirow{2}{*}{\begin{tabular}[c]{@{}c@{}}Bit-flip\\ No.\end{tabular}} 
& \multirow{2}{*}{\begin{tabular}[c]{@{}c@{}}Original \\ Data\end{tabular}} 
& \multirow{2}{*}{\begin{tabular}[c]{@{}c@{}} Flipped \\ Data\end{tabular}} 
& \multicolumn{2}{c|}{Direction} 
& \multirow{2}{*}{\begin{tabular}[c]{@{}c@{}}Bit\\ Location\end{tabular}} \\
 &  & & $0 \rightarrow 1$ & $1 \rightarrow 0$ &  \\ \hline \hline
    $A_1$ & \texttt{0x55} & \texttt{0x45} &  & \checkmark & 4 \\
    $B_1$ & \texttt{0xAA} & \texttt{0xAB} & \checkmark &  & 0 \\
    $B_2$ & \texttt{0xAA} & \texttt{0xEA} & \checkmark &  & 6 \\
    $B_3$ & \texttt{0xAA} & \texttt{0x2A} &  & \checkmark & 7 \\
    $C_1$ & \texttt{0xAA} & \texttt{0xAB} & \checkmark &  & 0 \\
    $D_1$ & \texttt{0xAA} & \texttt{0xEA} & \checkmark &  & 6 \\
    $D_2$ & \texttt{0xAA} & \texttt{0xAB} & \checkmark &  & 0 \\
    $D_3$ & \texttt{0xAA} & \texttt{0xEA} & \checkmark &  & 6 
\end{tabular}
\label{table:flip_data}
\end{table}
\vspace{-0.1in}

%We observed both $0 \rightarrow 1$ and $1 \rightarrow 0$ flips. We also observed bitflips in both directions on bank $B$, \TODO{Cite}.

\subsection{Data Pattern Characterization}\label{subsec:datapattern}
%We analyzed the impact of data values in victim and aggressor rows on the frequency of bit-flips, using one representative $0 \rightarrow 1$ () and $1 \rightarrow 0$ bit flip. Our findings reveal that the aggressor bits diagonally adjacent to the victim bit influence the occurrence of $0 \rightarrow 1$ bit-flips but have no effect on $1 \rightarrow 0$ flips. Additionally, victim data patterns play a more significant role in triggering bit-flips than aggressor patterns. Detailed experiments and results are provided in Appendix \Cref{subsec:data_pat_appdx}.
We analyzed the impact of data values in victim and aggressor rows on the frequency of bit-flips, using a representative $0 \rightarrow 1$ flip ($D_1$) and a $1 \rightarrow 0$ flip ($A_1$). We observe that the aggressor bits directly above or below the flippy victim bit can influence the bit-flip frequency for both directions. But, the aggressor bit diagonal to the flippy victim bit can only influence the $0 \rightarrow 1$ bit-flip frequency. 
Moreover, victim data patterns around the flippy bit influence the bit flip frequency to a greater extent than aggressor data.
These insights can be used to launch Rambleed~\cite{RamBleed} attacks on GPU memories to leak sensitive bit values based on data-dependence in bit-flip frequency.
We provide more details about these results in \Cref{subsec:data_pat_appdx}.
%occurrence of $0 \rightarrow 1$ bit-flips but have no effect on $1 \rightarrow 0$ flips. Additionally, victim data patterns play a more significant role in triggering bit-flips than aggressor patterns. Detailed experiments and results are provided in Appendix \Cref{subsec:data_pat_appdx}.
\section{End to End Exploit with GPU Rowhammer}\label{sec:exploit}
%To the best of our knowledge, no prior work has demonstrated practical exploits that leverage bit flips in GPU memory. 
%Here, we demonstrate a proof-of-concept exploit using the Rowhammer bit-flips we achieve on A6000 GPU. 
%introduce a proof-of-concept exploit adapted from an existing CPU-side Rowhammer exploit and provide a comprehensive evaluation of its effectiveness.

\subsection{Attack Overview} \label{subsec:vector}
In this section, we demonstrate a proof-of-concept exploit using the Rowhammer bit-flips and use them to tamper with Deep Neural Network (DNN) weights on an A6000 GPU. 
We achieve an \textit{accuracy degradation} attack during DNN inference like Terminal Brain Damage (TBD)~\cite{TBD}, which simulated such attacks.
TBD showed that a single bit-flip ($0 \rightarrow1$) surgically inserted in the most significant bit (MSB) of the exponent of a DNN weight, can cause significant damage to model accuracy, reducing it by up to 99\%, and almost 50\% of the weights are susceptible to such attacks. 
In our exploit, we show for the first time that such an attack can be executed using our Rowhammer-induced bit-flips on GPUs, and the resultant tampering of the DNN weights resident in the GPU memory can impact the DNN accuracy significantly.
%Observations from \cref{sec:eval} indicate that GPU memory can be exploited to target DNNs weights stored on the GPU. We extend this insight to demonstrate TBD by targeting weights stored in GPU memory.

%Terminal Brain Damage (TBD) \cite{TBD}, which showed that a single bit flip ($0 \rightarrow1$) can cause significant damage to model accuracy, reducing it by up to 99\%. Observations from \cref{sec:eval} indicate that GPU memory can be exploited to target DNNs weights stored on the GPU. We extend this insight to demonstrate TBD by targeting weights stored in GPU memory.

%The original study highlighted that flipping the most significant exponent bit is critical for causing widespread damage. Specifically, this corresponds to the 31\textsuperscript{st} bit for FP32 and the 15\textsuperscript{th} bit for FP16 representations \cite{IEEE_FP16}. By leveraging these insights, we explore the feasibility of executing TBD on GPU-resident data, emphasizing its potential to compromise DNN integrity.

\subsection{Attack Setup} \label{subsec:exploit_model}

\noindent \textbf{Threat Model.}
\REVISION{We assume a multi-tenant setting where the attacker and victim are co-located on the same GPU and execute CUDA kernels in a time-multiplexed manner. Such co-location is feasible on time-sliced GPUs, as enabled by NVIDIA’s GPU Operators and schedulers like RunAI\cite{TimeSlicing}, which default to 250ms slices ($\sim$10 \TREFI{}s). This interval is enough for multiple hammering iterations and inducing bit-flips. Memory allocations can persist across slices, allowing attackers to retain access. 
We assume the GPU memory allocations are managed via the RAPIDS Memory Manager (RMM)~\cite{RAPIDS}, a widely used allocator in ML applications endorsed by NVIDIA for its speed and efficiency\cite{RAPIDS_endors}.
\NEWSUGGESTIONS{We assume ECC is disabled, as discussed in \cref{subsec:rowhammer_threat_model}.}
% We assume the ECC is disabled, as many cloud  providers disable ECC for GDDR based GPUs  for performance reasons. 
Using our \attack{} primitives, an attacker can profile DRAM for bit-flips, identifying both their location and direction.
}
\smallskip
\noindent\textbf{Memory Massaging.}
\REVISION{To map victim data to vulnerable bits in GPU DRAM, the attacker must perform memory massaging. 
%For this, we assume the GPU memory is managed via the RAPIDS Memory Manager (RMM)~\cite{RAPIDS}, a widely used allocator in ML applications endorsed by NVIDIA\cite{RAPIDS_endors}. 
We achieve this by exploiting the fact that RAPIDS immediately reuses any freed memory, making it easier for an attacker to control the victim's memory allocation.

%We transfer memory allocated via RAPIDS across processes using CUDA IPC, allowing the attacker to massage victim allocations. 
Specifically, the attacker first allocates large contiguous memory regions, then frees the chunk that contains a known flippy bit. When the victim runs and requests memory, this chunk is allocated to the victim process (e.g., PyTorch), mapping the victim’s data adjacent to attacker-controlled aggressor rows. As RAPIDS can allocate in 256-byte increments, the attacker has fine-grained control over this placement.

\NEWSUGGESTIONS{Unlike RAPIDS, which reuses memory immediately, \texttt{cudaMalloc} reuses memory after a longer period; exploration of similar attacks with \texttt{cudaMalloc} is left for future work.}
%We leave the detailed analysis of allocator-specific reuse patterns for future work.}
%While we demonstrate memory massaging with  RAPIDS, \texttt{cudaMalloc} may also be exploitable with sufficient reverse engineering of its reuse patterns.
%We leave the analysis of allocator-specific reuse behavior for future work.
%While we focus on RAPIDS for its allocator behavior, we
%note that cudaMalloc may also be exploitable. With sufficient
%reverse engineering of its reuse patterns, attackers could
%similarly perform memory massaging. We leave the analysis
%of allocator-specific reuse behavior to future work.
}

\smallskip
\noindent\textbf{Attack Targets and Metrics.} As the victim, we choose an ML inference process in PyTorch using pre-trained ImageNet models: AlexNet~\cite{AlexNet}, VGG16~\cite{VGG16}, ResNet50~\cite{ResNet50}, DenseNet161~\cite{DenseNet}, and InceptionV3~\cite{InceptionV3}, whose weights we tamper with bit-flips. We use ILSVRC2012 dataset~\cite{ImageNet} as validation set to calculate the accuracy before ($\text{Acc}_\text{pristine}$) and after ($\text{Acc}_\text{corrupted}$) the bit-flip, and the Relative Accuracy Drop (RAD) as, $RAD = (\text{Acc}_\text{pristine} - \text{Acc}_\text{corrupted}) / \text{Acc}_\text{pristine})$. 

%We define indiscriminate damage on a model accuracy similar to prior work, as when the attack incurs RAD $> 0.1$. 
%Note that we used the Sampled Validation (SV) set technique \cite{TBD} to speed up the attack verification process.

%\noindent\textbf{Metrics.} We similarly select the latest pre-trained ImageNet models from Pytorch: AlexNet\cite{AlexNet}, VGG16\cite{VGG16}, ResNet50\cite{ResNet50}, DenseNet161\cite{DenseNet}, and InceptionV3\cite{InceptionV3}. We the ILSVRC-2012\cite{ImageNet} validation set to calculate the Relative Accuracy Drop $(RAD = (\text{Acc}_\text{pristine} - \text{Acc}_\text{corrupted}) / \text{Acc}_\text{pristine})$. We define indiscriminate damage on a model accuracy similar to prior work, as when the attack incurs RAD $> 0.1$. Note that we used the Sampled Validation (SV) set technique \cite{TBD} to speed up the attack verification process.
\subsection{Attack Evaluation}
We launch the bit-flipping attack on PyTorch models using FP16 weights.
We make 50 attempts for the attack, which
takes around 30 minutes per model, moving the victim memory by a randomly chosen distance in each attempt so that the bit flip maps to a new model weight.
We report the highest accuracy degradation (highest RAD) across all 50 attempts and repeat it for 4 different $0 \rightarrow 1$ bit flips on the A6000 GPU.

\cref{table:exploit} shows that bit flip $D_1$ and $D_3$ both cause considerable damage, reducing top-1 accuracy from 80\% to less than 0.5\% (RAD of 0.99). 
% Whereas, $D_3$ achieves a RAD of 0.89, with top-1 accuracy reducing to 2.7\% or less for four out of five models (RAD of 0.99) and a top-1 accuracy of 37\% for AlexNet (RAD of 0.34). 
The attack is highly effective with $D_1$ and $D_3$ as PyTorch stores FP16 weights as 2-byte aligned, and both map to the Most Significant Bit (MSB) of the FP16 exponent (the 4\textsuperscript{th} bit of exponent, $E^4$).\footnote{In NVIDIA GPUs, with little-endian architecture, the 6\textsuperscript{th} bit of the 2\textsuperscript{nd} physical byte, where $D_1$ and $D_3$ are located, maps to the Most Significant Bit (MSB) of the FP16 exponent, i.e.,  4th bit of the exponent, $E^4$.} A $0 \rightarrow 1$ flip here exponentially increases the weight's value, causing much damage. 
On the other hand, the other two bit-flips, $B_1$ and $B_2$, map to bits in the Mantissa ($M^0$ and $M^6$), and only cause a small increase in the weight\OTHERCHANGE{s},  with negligible accuracy impact.

\begin{figure}[h]
\centering
%\fcolorbox{\OTHERCHANGEFIGURECOLOR}{white}{ % Red border with a 
\includegraphics[width=3in,height=\paperheight,keepaspectratio]
{"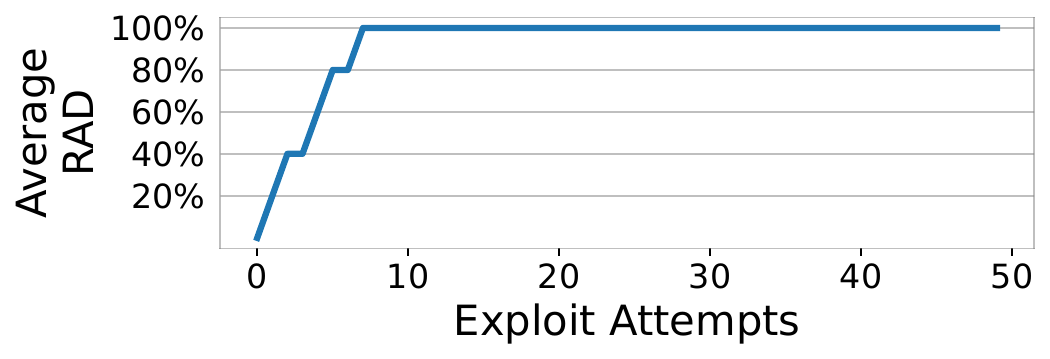"}
%}
\caption{\label{fig:attack_vs_attempt} \OTHERCHANGE{Average Relative Accuracy Drop (RAD) across all models as exploit attempts vary for bitflip $D_1$. Average RAD is $20\%$ after 2 attempts, and $>99\%$ after only 8 attempts.}}
%\vspace{-0.1in}
\end{figure}

\OTHERCHANGE{For the bit flips that map to the MSB of the exponent (e.g., $D_1$ or $D_3$), not all 50 attempts are necessary. \cref{fig:attack_vs_attempt} compares the average RAD across all models with $D_1$, as the number of exploit attempts increases. We see substantial degradation (RAD of 20\%) after only 2 attempts, and RAD $>$99\% after 8 attempts, highlighting the efficacy of our attack.}
%\leftotherchangebox{CC-Q2}
%\vspace{-0.2in} 

% Attack_VS_Attempt.pdf

We attempted a similar attack on FP32 models but found that no bit-flips mapped to the MSB of the FP32 exponent (6\textsuperscript{th} bit of the 4\textsuperscript{th} byte), limiting accuracy damage. However, with more time and profiling of banks, it is likely that we will find exploitable bit-flips even for FP32 weights.

\begin{table}[h]
\centering
\caption{Accuracy degradation attack on ImageNet models using FP16 weights, with Rowhammer bit-flips on NVIDIA A6000 GPU. We report the top-1~/~top-5 accuracy without (Base Acc) and with (Degraded Acc) the bit-flip, and the Relative Accuracy Drop (RAD) for top-1 accuracy.}
\label{table:exploit}

%\vspace{-0.1in}

\resizebox{3.3in}{!}{
\begin{tabular}{l|lll|r}
\hline
\textbf{Flip} & \textbf{Network} & \textbf{Base Acc. (\%)} & \textbf{Degraded Acc. (\%)} & \textbf{RAD} \\ 
\hline\hline
 & \textbf{AlexNet} & 56.66 / 79.78 & 0.10 / 0.64 & 0.99 \\
 &\textbf{VGG16} & 72.22 / 90.90 & 0.08 / 0.52 & 0.99\\
$D_1$ (E$^4$)  & \textbf{ResNet50} & 80.26 / 94.94 & 0.08 / 0.60 &  0.99 \\
 & \textbf{DenseNet161} & 77.20 / 93.58 & 0.08 / 0.52 & 0.99\\
&\textbf{InceptionV3} & 69.92 / 88.74 & 0.04 / 0.58 & 0.99\\
\cmidrule{2-5}
&\textbf{Average} & 71.26 / 89.59 & 0.08 / 0.58 & 0.99\\

\hline\hline
 & \textbf{AlexNet} & 56.66 / 79.78 & 0.10 / 0.66 & 0.99 \\
 &\textbf{VGG16} & 72.22 / 90.90 & 0.12 / 0.42 & 0.99 \\
$D_3$ (E$^4$)  & \textbf{ResNet50} & 80.26 / 94.94 & 0.02 / 0.34 & 0.99 \\
 & \textbf{DenseNet161} & 77.20 / 93.58 & 0.08 / 0.56 & 0.99 \\
&\textbf{InceptionV3} & 69.92 / 88.74 & 0.10 / 0.50 & 0.99 \\
\cmidrule{2-5}
&\textbf{Average} & 71.26 / 89.59 & 0.09 / 0.50 & 0.99\\

\hline\hline
 & \textbf{AlexNet} & 56.66 / 79.78 & 56.66 / 79.78 & 0.00 \\
 &\textbf{VGG16} & 72.22 / 90.90 & 72.22 / 90.90 & 0.00 \\
$B_1$ (M$^0$)  & \textbf{ResNet50} & 80.26 / 94.94 & 80.24 / 94.94 & < 0.01 \\
 & \textbf{DenseNet161} & 77.20 / 93.58 & 77.20 / 93.58 & 0.00 \\
&\textbf{InceptionV3} & 69.92 / 88.74 & 69.88 / 88.74 & < 0.01 \\
\cmidrule{2-5}
&\textbf{Average} & 71.26 / 89.59 & 71.24 / 89.59 & < 0.01\\

\hline\hline
 & \textbf{AlexNet} & 56.66 / 79.78 & 56.64 / 79.80 & < 0.01\\
 &\textbf{VGG16} & 72.22 / 90.90 & 72.20 / 90.90 & < 0.01 \\
$B_2$ (M$^6$)  & \textbf{ResNet50} & 80.26 / 94.94 & 80.24 / 94.94 & < 0.01 \\
 & \textbf{DenseNet161} & 77.20 / 93.58 & 77.20 / 93.58 & 0.00\\
&\textbf{InceptionV3} & 69.92 / 88.74 & 69.84 / 88.76 & < 0.01 \\
\cmidrule{2-5}
&\textbf{Average} & 71.26 / 89.59 & 71.24 / 89.60 & < 0.01\\

\hline\hline

\end{tabular}
}
\smallskip
\newline
E = Exponent, M = Mantissa, (E/M)$^{\textbf{x}}$ = x$^{\textbf{th}}$ bit in E/M
\end{table}

%\vspace{-0.2in} 
\begin{result}\refstepcounter{resultcounter}
\REVISION{\textbf{Takeaway~\theresultcounter.} A single Rowhammer bit-flip in the MSB of the exponent of a model weight can significantly degrade accuracy from 80\% down to 0.02\% (RAD $>$ 0.99).}
\end{result}
\section{Discussion}\label{sec:discussion}
\noindent\textbf{Additional Targets and Attack Vectors.} 
While we target ML models in our Rowhammer attack, another potential target can be GPU page tables, which are also stored in the device memory~\cite{GPUPT1}, 
similar to how CPU Rowhammer exploits have targeted CPU PTEs~\cite{TRRespass, Blacksmith, HalfDouble, Eccploit}. Another potential attack vector can be the GPU-based renderer in browsers~\cite{webGPU} that may co-locate the pixel data of different websites.
%in GPU memory. 
Future works can also attempt to steal sensitive pixels via Rambleed~\cite{RamBleed}, leveraging the data-dependent flips we observe in \cref{subsec:datapattern}.

%\attack exploits vulnerable, application-specific data to compromise system security. In CPU systems, Page Table Entries (PTEs) have been identified as a frequent target for exploitation\cite{TRRespass, Blacksmith, HalfDouble, Eccploit}. By inducing bit-flips in PTEs, attackers can manipulate virtual-to-physical address translations, gaining unauthorized access to arbitrary physical pages and potentially escalating their privileges to kernel level. Similarly, GPU PTEs are stored in physical memory\cite{GPUvm}, making them equally susceptible to \attack and a potential vector for exploitation. Future work will require more sophisticated memory massaging techniques to co-locate aggressors with PTE.\TODO{I looked around and it doesn't seem user kernel code is stored in the global memory.}

%\smallskip
%\noindent\textbf{Triggering Bit-flips on A100 and RTX3080.} 
%Our row set recovery and hammering synchronization methods, described in \cref{sec:reverse} and \cref{sec:sync}, function as intended on Ampere architecture GPUs such as the A100 and the laptop-grade RTX 3080. 
%While we did not observe any bit-flips under our current hammering pattern. The A100 employs HBM2e memory, while the RTX 3080 Laptop uses GDDR6 memory. However, we identified that the GDDR6 chip in the RTX 3080 Laptop differs from that in the RTX A6000, as it features half the rows per bank. These chip differences may reflect variations in Rowhammer defense implementations, higher sampler size, or a higher threshold (\TRH) required to induce bit-flips.

\smallskip
\noindent\REVISION{\textbf{Absence of Bit-flips on A100 and RTX3080.}
%The Rowhammer threshold for a DRAM device can depend on the technology and process variation, and as well as runtime (e.g., temperature). 
The Rowhammer threshold for DRAM devices varies with technology, process variations, and runtime conditions (e.g., temperature).
%, leading to chip-to-chip and form-factor-dependent vulnerability differences. 
Consequently, the vulnerability levels can have chip-to-chip variations as well as be form factor dependent. This makes pinpointing the root cause of the absence of bit flips difficult.
%making the identification of the exact root cause for the absence of bit flips difficult.
%The RTX3080 being a laptop card, uses different GDDR6 chips compared to the workstation-grade A6000, that possibly has a higher Rowhammer threshold making it less vulnerable to Rowhammer.
The RTX3080, a laptop GPU, uses GDDR6 chips distinct from the workstation-grade A6000, which possibly have a higher Rowhammer threshold, making it less vulnerable. 
The A100, equipped with HBM2e memory, features a fundamentally different DRAM architecture compared to GDDR.
%In case of A100 with HBM2e memory, the DRAM architecture is quite different from GDDR; 
Thus it might not only have a higher Rowhammer threshold, but also a different in-DRAM mitigation mechanism immune to the attack patterns we tested. 
}
Future works can systematically study the vulnerability of these GPUs using more sophisticated attack patterns~\cite{Blacksmith,Rowpress,HalfDouble}.
\REVISION{Additionally, on-die ECC in HBM2e can also mask observable bit flips, as discussed next.}
%Finally, on-die ECC on HBM2e can also mask bit flips from being observed as we explain next.

%Future work may explore more sophisticated attack strategies, such as Blacksmith \cite{Blacksmith}, to bypass existing defenses or overcome elevated \TRH levels. However, the memory coalescing behavior in GPU memory could complicate the direct adoption of these methods, highlighting the need for more intelligent and tailored approaches to achieve effective Rowhammer exploitation on GPUs. Additionally, we encourage research on uncovering Rowhammer vulnerability and defenses in GDDR and HBM chips, as done so by experimentally by Olgun et al.\cite{HBMRowhammer}.

\smallskip
\REVISION{
\noindent\textbf{On-Die ECC in HBM2e and Future GPU Memories.}
GPUs like the A100 (HBM2e), H100/B100 (HBM3/e), and RTX5090 (GDDR7) integrate on-die ECC~\cite{HBM2eECC, GDDR7ECC, HBM3ECC} that can correct single bit-flips and detect double bit-flips by default.
%The A100 GPU (HBM2e) and newer GPUs like H100 and B100 (HBM3/HBM3e) amd RTX5090 (GDDR7) integrate on-die ECC by default that cannot be disabled~\cite{HBM2eECC, GDDR7ECC, HBM3ECC}.
%We evaluated GPUHammer on an NVIDIA A100 with HBM2e memory, which includes optional on-die ECC. While we cannot confirm whether ECC was enabled, its presence could explain the absence of bit-flips. We leave rigorous analysis of Rowhammer with on-die ECC for future work.
% On the other hand, on-die ECC may miscorrect when a Rowhammer-induced disturbance coincides with a natural retention error, potentially escaping correction. 
%Emerging NVIDIA GPUs such as the RTX 5090 (GDDR7) and server-class GPUs like the H/B100 (HBM3/e) integrate on-die ECC by default that cannot be disabled.~\cite{GDDR7ECC, HBM3ECC}. 
%This can correct single-bit errors and 
This can conceal any Rowhammer-based single bit-flips that may occur. 
However, if more than two bit-flips occur within the same code word, the on-die ECC can also mis-correct the data. 
%However, such on-die ECC can also mis-correct the data if more than two bit flips were to occur in the same code word. 
%While on-die ECC increases the complexity of Rowhammer attacks on these GPUs, it does not make them impossible.
Thus, on-die ECC can make Rowhammer attacks more difficult on emerging GPUs, but not impossible.
\rightrevisionbox{R2}\vspace{-0.2in}
%These defenses make newer memories more resilient to Rowhammer attacks, but also complicate the detection and study of underlying vulnerabilities.
}

\smallskip
\REVISION{
\noindent\textbf{Impact on NVIDIA MIG and Confidential Computing.}
In cloud-based GPUs, NVIDIA's Multi-instance-GPU (MIG)\cite{NVIDIA_MIG} allows shared usage of a GPU by spatially partitioning GPU memory along channels for MIG instances assigned to different users. 
Meanwhile, NVIDIA Confidential Computing (CC) assigns an entire GPU or multiple GPUs to a single confidential VM.
Both configurations prevent multi-tenant data co-location in the same DRAM bank required for our exploits, \NEWSUGGESTIONS{thwarting our Rowhammer-based multi-tenant data tampering exploits in such environments.
%In both scenarios, there is no multi-tenant data co-location in the same DRAM-bank, which is required for our Rowhammer-based data tampering attacks, thus preventing our exploits.
% similarly preventing multi-tenant data co-location in 
% However, privilege-escalation attacks, based on tampering with page tables, are possible if GPU page tables are stored in the same DRAM bank as attacker data. This requires reverse engineering~\cite{tunnels} of GPU memory layouts and page tables which can be explored by future works.}
We leave the exploration of Rowhammer exploits on such environments for future work.}
}
%On the other hand, 
%\REVISION{Multi-instance-GPUs (MIG)\cite{NVIDIA_MIG} partitions GPU memory by channels, preventing multi-tenant data sharing the same DRAM-bank, which is required for Rowhammer. However, privilege-escalation attacks (e.g., tampering with page tables) may be possible with further reverse-engineering such as TunneLs for Bootlegging\cite{tunnels}, which we leave for future work.}
\rightrevisionbox{R3}
%\rightnewsuggestionsbox{NS}
%\vspace{-0.1in}
\begin{comment}
\smallskip
\noindent\textbf{Vulnerability of GPU Confidential Compute.} NVIDIA's Confidential Computing (CC)~\cite{NVIDIA_CC} assigns the GPU to a single tenant and scrubs all GPU state and memory on tenant exit. In this configuration, the GPU does not share memory across tenants, effectively eliminating multi-tenancy at the hardware level. This isolation prevents the co-location of adversarial and victim data within the same DRAM bank, required for GPU Rowhammer exploits.
\end{comment}
%Recent NVIDIA architectures, including Hopper \cite{nvidia_hopper_tuning} and Blackwell \cite{nvidia_blackwell_architecture}, have adopted HBM3 and HBM3e memory, respectively. Notably, these DRAM technologies incorporate On-Die ECC as part of their Reliability, Availability, and Serviceability (RAS) features \TODO{cite spec?}. ECCSploit \cite{Eccploit} has demonstrated that precise multi-bit-flips can effectively bypass ECC detection and correction mechanisms, enabling deterministic exploitation. Consequently, achieving multiple bit-flips within a single DRAM row is critical for successful Rowhammer attacks on modern GPU systems. 
\section{Mitigations}\label{sec:mitigations}

\begin{figure}[ht]
\centering
\includegraphics[width=2.4in,height=\paperheight,keepaspectratio]
{"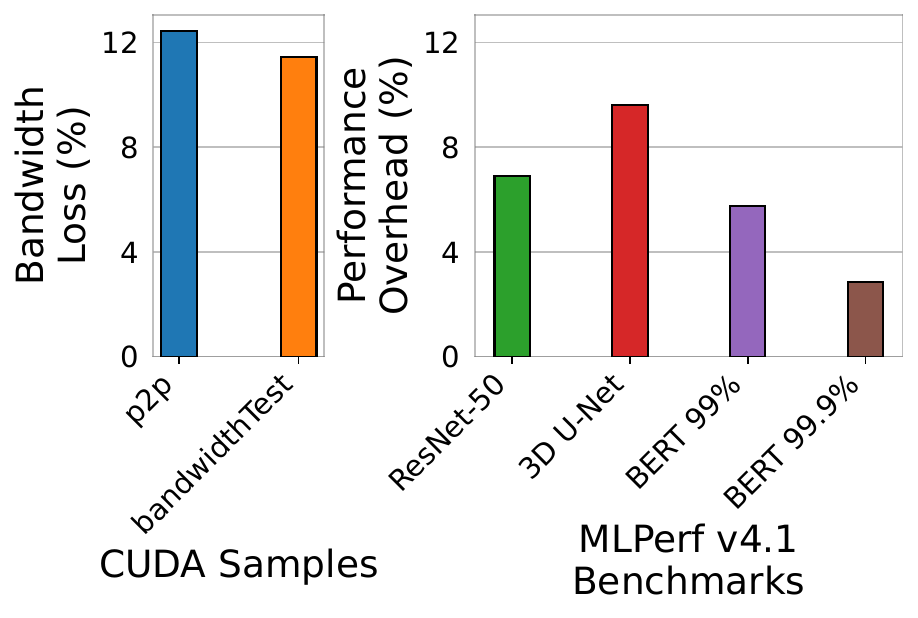"}
\caption{\label{fig:ecc_perf} Overheads of enabling ECC in A6000 GPU for MLPerf Inference and CUDA samples benchmarks.}
%on NVIDIA MLPerf v4.1 Inference implementations in Edge \& Offline setting. We are limited to the listed benchmarks due to system limitations and technical issues in the existing MLPerf codebase.}
\end{figure}

\noindent {\bf Enabling  ECC.} 
GDDR-based GPUs like A6000 can optionally enable memory-controller-based ECC, which can detect and correct single bit-flips and mitigate our attack.
However, systems disable ECC by default as it is stored out-of-band, in a separate region of memory, and enabling it causes a memory overhead of 6.5\%~\cite{IMTSullivan} and slowdown. 
%By default GDDR6 based GPUs have memory-controller-based ECC disabled, as the ECC is stored out-of-band in a separate region of memory, and enabling it introduces memory bandwidth and performance overheads. 
%Thus, production systems are likely to use ECC disabled by default.
We evaluate the slowdown of ECC on our A6000 GPU, using CUDA samples~\cite{nvidia_cuda_samples} and MLPerf v4.1 Inference benchmarks~\cite{mlperf}. \cref{fig:ecc_perf} shows that enabling ECC results in memory bandwidth loss of up to 12\% and slowdown of 3\%-10\% in ML applications.
Thus, ECC can effectively mitigate our attacks at reasonable costs.

%using  \textit{p2pBandwidthLatencyTest}, \textit{BlackScholes}, and \textit{bandwidthTest} from CUDA Samples \cite{nvidia_cuda_samples}, all identifying a $\sim$12\% of bandwidth reduction. On NVIDIA MLPerf v4.1 Inference implementation benchmarks\cite{mlperf}, we show similar results in \cref{fig:ecc_perf}, observing a 3-10\% overhead.

\smallskip
\noindent {\bf Randomizing Virtual to Physical Mappings.} 
In the absence of access to physical addresses on the GPU, our attack relies on directly reversing the virtual address to DRAM row mappings. This leverages the fact that the virtual-to-physical mappings on NVIDIA GPUs are the same across different runs for large memory allocations spanning the entire GPU RAM. If the GPU driver randomized the virtual to physical mapping on each run, it would require the attacker to repeatedly profile the DRAM row mappings each time. 
While this would significantly increase the cost of the attack, this may have some impact on performance due to TLB fragmentation.

%Observation \ref{obs:vp_consistent} demonstrates that the virtual memory of allocated memory is consistently mapped to the same physical addresses across different program executions. This predictable mapping presents an advantage to attackers, as it allows them to easily determine key information, such as the row set and bit-flip locations, after a single execution.

%To counter, employing a randomized mapping scheme can significantly increase the complexity of the attack. By invalidating the row set across different runs, the attacker must reconstruct the address mappings in a single attempt. Furthermore, memory massaging becomes more challenging, as both the victim and attacker’s rows may occupy the same memory space, necessitating additional retries and verification steps before exploitation can succeed.

\noindent {\bf Making Memory Massaging Unpredictable.} 
Our exploit requires precise memory massaging to map victim memory to the vulnerable bits in DRAM. RAPIDS memory manager, designed for memory efficiency, immediately reallocates memory released by one allocation, enabling such precise memory massaging. 
Using allocators that quarantine deallocated memory~\cite{ASAN} or randomize the reallocation makes memory massaging harder; however such allocators may increase the memory usage, making them less desirable.

\begin{comment}
In \cref{subsec:exploit_model}, the attacker can successfully target data residing in vulnerable rows only when equipped with precise memory massaging capabilities. 
This vulnerability can be mitigated through intelligent techniques that disrupt the locality between memory allocations. For example, placing consecutive allocations at random or distant locations in physical memory can effectively thwart naive memory massaging strategies discussed in \cref{subsec:exploit_model}.
\end{comment}

\smallskip
\noindent {\bf Adopting Modern Rowhammer Mitigations.} 
Beyond TRR, advanced Rowhammer mitigations, such as Refresh Management (RFM), have been proposed for HBM3~\cite{JEDEC-HBM3} and DDR5~\cite{JEDEC-DDR5}, and Per Row Activation Counting (PRAC) \cite{jedec_ddr5_prac,QPRAC,MOAT,Chronus} has been proposed in the DDR5 specification. 
Adopting such mitigations in emerging GDDRx memories can address the risk of Rowhammer in future GPUs.

\begin{comment}
Researchers have highlighted the shortcomings of naive Target Row Refresh (TRR) implementations in several studies \cite{TRRespass, Blacksmith, ZenHammer, smash}. Concurrently, novel defense mechanisms have been proposed to address these vulnerabilities more effectively. Some approaches focus on performing targeted refreshes to aggressor rows \cite{PRIDE, Hydra, TWiCe}, while others aim to quarantine highly activated rows \cite{SRS, Blockhammer, AQUA}. Additionally, to mitigate the risks posed by these attacks and the diminishing \TRH thresholds due to DRAM scaling, memory vendors have introduced standardized mitigation techniques, including Refresh Management (RFM) \cite{JEDEC-DDR4} and Per Row Activation Counting (PRAC) \cite{jedec_ddr5_prac}. Adopting these advanced mitigation techniques can significantly enhance the resilience of GPU memory against Rowhammer attacks.
\end{comment}

\section{Related Work}
%\noindent \textbf{\textit{A. Rowhammer Exploits:}} Previous Rowhammer exploits targeted CPU-based DDR3, DDR4, and DDR5 memories. Drammer~\cite{Drammer} enabled privilege escalation on mobile platforms, while Flip Feng Shui (FFS)\cite{FFS} used memory de-duplication to compromise cryptographic systems. ThrowHammer\cite{ThrowHammer} demonstrated network-based attacks, and Rowhammer.js~\cite{RowhammerJS} operated within a Firefox sandbox. Advanced variants like ECCploit~\cite{Eccploit}, TRRespass~\cite{TRRespass}, Blacksmith~\cite{Blacksmith}, SMASH~\cite{smash}, and Zenhammer~\cite{ZenHammer} bypassed defenses like ECC and TRR, exposing vulnerabilities in AMD Zen CPUs and DDR5. Grand Pwning Unit~\cite{GrandPwning} leveraged integrated GPUs to accelerate attacks on shared DRAM. In contrast, we present the first Rowhammer exploits on GDDR6 memory used in discrete GPUs for workstations and cloud servers.

\noindent \textbf{\textit{A. Rowhammer Exploits:}} Prior Rowhammer exploits targeted CPU-based DDRx and LPDDRx memories. 
In DDR3, Flip Feng Shui (FFS)~\cite{FFS} used memory de-duplication in virtualized settings to compromise cryptographic code, while Drammer~\cite{Drammer} enabled privilege escalation on mobile platforms.
%by flipping bits in cryptographic keys.
ThrowHammer~\cite{ThrowHammer} demonstrated Rowhammer exploits over the network, while Rowhammer.js~\cite{RowhammerJS} demonstrated these from JavaScript.
Attacks like 
%New attack variants with more complex attack patterns have also been developed, such as 
ECCploit~\cite{Eccploit} defeated ECC,
%in ECC-DIMMs, 
while 
TRRespass~\cite{TRRespass}, Blacksmith~\cite{Blacksmith} and SMASH~\cite{smash} defeated in-DRAM defenses like TRR in DDR4.
Zenhammer~\cite{ZenHammer} recently showed modern AMD Zen CPUs and DDR5 memories are also vulnerable.
Grand Pwning Unit~\cite{GrandPwning} used integrated GPUs on mobile platforms that share the CPU DRAM to perform Rowhammer attacks on LPDDR3 CPU DRAM.
In contrast to all these attacks, we enable the first Rowhammer attacks on GDDR6 GPU memory in discrete NVIDIA GPUs, widely used in workstations and servers.

\smallskip
\noindent \textbf{\textit{B. Rowhammer Mitigations:}}
Prior software-based Rowhammer mitigations have focused on CPUs. GuardION~\cite{van2018guardion} uses guard rows between security domains, while CATT~\cite{brasser2017can}, RIP-RH~\cite{bock2019rip}, ZebRAM~\cite{konoth2018zebram}, and Siloz~\cite{Siloz} isolate security domains in separate DRAM regions. These methods could potentially defend against GPU-based attacks with NVIDIA driver support. Hardware-based mitigations, including Rowhammer trackers~\cite{park2020graphene, mithril, PRIDE, Hydra, ProTRR}, aggressor relocation~\cite{AQUA, Shadow, RRS, SRS}, delayed activations~\cite{Blockhammer}, and refresh-generating activations~\cite{REGA_SP23}, designed for CPU DRAM, might also be adapted to GPUs, though their overheads may vary depending on the GPU latency and bandwidth constraints.

\smallskip
\noindent \textbf{\textit{C. Bit-Flipping Attacks on DNNs:}} 
Many exploits have shown that bit-flips in CPUs can tamper with DNN weights, causing accuracy degradation~\cite{TBD,Deephammer,li2024_1bitflip,DNNFaultInjection,CrushingFlips} and enabling backdoor injection~\cite{tol2023dontknock,chen2021proflip}. Our GPU-based exploit injects a single bit flip in the MSB of the exponent in DNN weights, similar to TBD~\cite{TBD}. However, more sophisticated exploits with more than one bit-flip can also be adapted to GPUs.
%Many prior exploits have demonstrated that bit-flips in CPUs can tamper with DNNs weights, causing accuracy degradation~\cite{TBD,Deephammer,li2024_1bitflip,DNNFaultInjection,CrushingFlips,kundu2024bitbybit} and injection of backdoors~\cite{tol2023dontknock,chen2021proflip}.
%Similar to our exploit tampering DNN weights using bit-flips on GPUs, numerous attacks in CPUs have shown that bit-flips in DNNs can result in dramatic accuracy degradation in DNN inference, enable injection of backdoors in DNNs, and also impact accuracy at training time. 
%While we demonstrate a GPU-based exploit where bit-flips target the MSB of the exponent of the DNN weights, like TBD~\cite{TBD}, more sophisticated exploits can also be ported to GPUs. 
Recent works propose ML model-based defenses against bit-flips, including usage of early-exits~\cite{Aegis}, honeypot neurons~\cite{NeuroPots}, and checksums~\cite{RADAR}. 
Such defenses can also be used on GPUs to mitigate bit-flipping attacks on ML models. 

\section{Conclusion}
%\lipsum[100]
This paper introduces \attack{}, the first Rowhammer attack on NVIDIA GPUs with GDDR6 DRAM. Overcoming challenges like high memory latency and faster refresh, \attack{} achieves bit-flips by reverse-engineering DRAM bank mappings and leveraging GPU parallelism. We show how these bit-flips dramatically degrade ML model accuracy by up to 80\% on GPUs, highlighting the vulnerability of GPUs to Rowhammer attacks and the need for mitigations.

\section{Acknowledgments}
We thank the anonymous reviewers and our shepherd for their
helpful suggestions.
We also acknowledge NVIDIA and the CSPs for their prompt responses following our report.
This research was supported by Natural Sciences and Engineering Research Council of Canada (NSERC) under funding reference number RGPIN-2023-04796, and an NSERC-CSE Research Communities Grant under funding reference number ALLRP-588144-23.
Any research, opinions, or positions expressed in this work are solely those of the authors and do not represent the official views of the Communications Security Establishment Canada or the Government of Canada.

\section{Ethics Considerations} 
We responsibly disclosed the Rowhammer vulnerability in the A6000 GPU to NVIDIA on 15th January, 2025, sharing proof-of-concept code and a draft of our paper. 
\OTHERCHANGE{
%NVIDIA has confirmed our findings and indicated that users can enable ECC as a suitable mitigation as needed, in line with our suggestions.
Subsequently, we also disclosed our findings to major cloud providers (AWS, Azure, GCP) who may be affected, so that 
%, who are currently investigating the issue.
%To the best of our ability, we have ensured that affected vendors are informed so that the vulnerability is addressed responsibly before public dissemination. 
the vulnerability is addressed before public dissemination.  
Our experiments were conducted locally or in isolated environments, and targeted open-source models, mitigating risks to production systems.
}

\section{Open Science}
\OTHERCHANGE{
%We are committed to advancing open science and reproducibility. Upon acceptance of this paper, we will release the following artifacts: (1) code to reverse-engineer GPU memory address-to-row mappings and derive the row-set and conflict-set for a given DRAM bank, (2) code for hammering DRAM banks to identify Rowhammer-induced bit-flips, and (3) code for our exploit targeting accuracy degradation in ImageNet models. These artifacts will be made public on GitHub after consultation with NVIDIA and the lifting of the embargo.
%Upon acceptance of the paper, 
%we will open source all associated code artifacts, after the expiry of any requested embargo period,
%and participate in the Artifact Evaluation process.
%With this, we aim to promote 
%transparency and 
%further research on mitigating Rowhammer vulnerabilities in GPUs.
Upon expiry of the embargo requested by NVIDIA on 12th August 2025, our code will be publicly available on \url{https://github.com/sith-lab/gpuhammer}. Our artifact is archived at \url{https://doi.org/10.5281/zenodo.15612689} and includes the code to reproduce the key results on the A6000 GPU including the Rowhammer campaigns (Table-1), bit flip characterization (Figure 11, Figure 12, Table 3), and the exploit on ML applications (Figure 13, Table 4). 
}
\bibliographystyle{plain}
\bibliography{refs}

@inproceedings {drama,
author = {Peter Pessl and Daniel Gruss and Cl{\'e}mentine Maurice and Michael Schwarz and Stefan Mangard},
title = {{DRAMA}: Exploiting {DRAM} Addressing for {Cross-CPU} Attacks},
booktitle = {25th USENIX Security Symposium (USENIX Security 16)},
year = {2016},
isbn = {978-1-931971-32-4},
address = {Austin, TX},
pages = {565--581},
url = {https://www.usenix.org/conference/usenixsecurity16/technical-sessions/presentation/pessl},
publisher = {USENIX Association},
month = aug
}

@inproceedings {smash,
author = {Finn de Ridder and Pietro Frigo and Emanuele Vannacci and Herbert Bos and Cristiano Giuffrida and Kaveh Razavi},
title = {{SMASH}: Synchronized Many-sided Rowhammer Attacks from {JavaScript}},
booktitle = {30th USENIX Security Symposium (USENIX Security 21)},
year = {2021},
isbn = {978-1-939133-24-3},
pages = {1001--1018},
url = {https://www.usenix.org/conference/usenixsecurity21/presentation/ridder},
publisher = {USENIX Association},
month = aug
}

@inproceedings {TBD,
author = {Sanghyun Hong and Pietro Frigo and Yigitcan Kaya and Cristiano Giuffrida and Tudor Dumitras},
title = {Terminal Brain Damage: Exposing the Graceless Degradation in Deep Neural Networks Under Hardware Fault Attacks},
booktitle = {28th USENIX Security Symposium (USENIX Security 19)},
year = {2019},
isbn = {978-1-939133-06-9},
address = {Santa Clara, CA},
pages = {497--514},
url = {https://www.usenix.org/conference/usenixsecurity19/presentation/hong},
publisher = {USENIX Association},
month = aug
}

@article{JEDEC-GDDR6,
  title={{Graphics Double Data Rate (GDDR6) SGRAM Standard (JESD250D)}},
  author={JEDEC},
  year={2023},
}

@INPROCEEDINGS{TRRespass,
  author={Frigo, Pietro and Vannacc, Emanuele and Hassan, Hasan and der Veen, Victor van and Mutlu, Onur and Giuffrida, Cristiano and Bos, Herbert and Razavi, Kaveh},
  booktitle={2020 IEEE Symposium on Security and Privacy (SP)}, 
  title={TRRespass: Exploiting the Many Sides of Target Row Refresh}, 
  year={2020},
  volume={},
  number={},
  pages={747-762},
  doi={10.1109/SP40000.2020.00090}}

@INPROCEEDINGS{RevisitRowhammer,
  author={Kim, Jeremie S. and Patel, Minesh and Yağlıkçı, A. Giray and Hassan, Hasan and Azizi, Roknoddin and Orosa, Lois and Mutlu, Onur},
  booktitle={2020 ACM/IEEE 47th Annual International Symposium on Computer Architecture (ISCA)}, 
  title={Revisiting RowHammer: An Experimental Analysis of Modern DRAM Devices and Mitigation Techniques}, 
  year={2020},
  volume={},
  number={},
  pages={638-651},
  doi={10.1109/ISCA45697.2020.00059}}

@INPROCEEDINGS{Blacksmith,
  author={Jattke, Patrick and Van Der Veen, Victor and Frigo, Pietro and Gunter, Stijn and Razavi, Kaveh},
  booktitle={2022 IEEE Symposium on Security and Privacy (SP)}, 
  title={BLACKSMITH: Scalable Rowhammering in the Frequency Domain}, 
  year={2022},
  volume={},
  number={},
  pages={716-734},
  doi={10.1109/SP46214.2022.9833772}}

@inproceedings {ZenHammer,
author = {Patrick Jattke and Max Wipfli and Flavien Solt and Michele Marazzi and Matej B{\"o}lcskei and Kaveh Razavi},
title = {{ZenHammer}: Rowhammer Attacks on {AMD} Zen-based Platforms},
booktitle = {33rd USENIX Security Symposium (USENIX Security 24)},
year = {2024},
isbn = {978-1-939133-44-1},
address = {Philadelphia, PA},
pages = {1615--1633},
url = {https://www.usenix.org/conference/usenixsecurity24/presentation/jattke},
publisher = {USENIX Association},
month = aug
}

@inproceedings{UncoverRowhammer,
author = {Hassan, Hasan and Tugrul, Yahya Can and Kim, Jeremie S. and van der Veen, Victor and Razavi, Kaveh and Mutlu, Onur},
title = {Uncovering In-DRAM RowHammer Protection Mechanisms:A New Methodology, Custom RowHammer Patterns, and Implications},
year = {2021},
isbn = {9781450385572},
publisher = {Association for Computing Machinery},
address = {New York, NY, USA},
url = {https://doi.org/10.1145/3466752.3480110},
doi = {10.1145/3466752.3480110},
booktitle = {MICRO-54: 54th Annual IEEE/ACM International Symposium on Microarchitecture},
pages = {1198–1213},
numpages = {16},
location = {Virtual Event, Greece},
series = {MICRO '21}
}

@article{ImageNet,
author = {Russakovsky, Olga and Deng, Jia and Su, Hao and Krause, Jonathan and Satheesh, Sanjeev and Ma, Sean and Huang, Zhiheng and Karpathy, Andrej and Khosla, Aditya and Bernstein, Michael and Berg, Alexander C. and Fei-Fei, Li},
title = {ImageNet Large Scale Visual Recognition Challenge},
year = {2015},
issue_date = {December 2015},
publisher = {Kluwer Academic Publishers},
address = {USA},
volume = {115},
number = {3},
issn = {0920-5691},
url = {https://doi.org/10.1007/s11263-015-0816-y},
doi = {10.1007/s11263-015-0816-y},
journal = {Int. J. Comput. Vision},
month = dec,
pages = {211–252},
numpages = {42}
}

@article{AlexNet,
author = {Krizhevsky, Alex and Sutskever, Ilya and Hinton, Geoffrey E.},
title = {ImageNet classification with deep convolutional neural networks},
year = {2017},
issue_date = {June 2017},
publisher = {Association for Computing Machinery},
address = {New York, NY, USA},
volume = {60},
number = {6},
issn = {0001-0782},
url = {https://doi.org/10.1145/3065386},
doi = {10.1145/3065386},
journal = {Commun. ACM},
month = may,
pages = {84–90},
numpages = {7}
}

@article{VGG16,
  title={Very deep convolutional networks for large-scale image recognition},
  author={Simonyan, Karen and Zisserman, Andrew},
  journal={arXiv preprint arXiv:1409.1556},
  year={2014}
}

@inproceedings{ResNet50,
  title={Deep residual learning for image recognition},
  author={He, Kaiming and Zhang, Xiangyu and Ren, Shaoqing and Sun, Jian},
  booktitle={Proceedings of the IEEE conference on computer vision and pattern recognition},
  pages={770--778},
  year={2016}
}

@article{DenseNet,
  title={Densenet: Implementing efficient convnet descriptor pyramids},
  author={Iandola, Forrest and Moskewicz, Matt and Karayev, Sergey and Girshick, Ross and Darrell, Trevor and Keutzer, Kurt},
  journal={arXiv preprint arXiv:1404.1869},
  year={2014}
}

@inproceedings{InceptionV3,
  title={Rethinking the inception architecture for computer vision},
  author={Szegedy, Christian and Vanhoucke, Vincent and Ioffe, Sergey and Shlens, Jon and Wojna, Zbigniew},
  booktitle={IEEE Conference on Computer Vision and Pattern Recognition (CVPR)},
  year={2016}
}

@inproceedings{Rowhammer2014,
author = {Kim, Yoongu and Daly, Ross and Kim, Jeremie and Fallin, Chris and Lee, Ji Hye and Lee, Donghyuk and Wilkerson, Chris and Lai, Konrad and Mutlu, Onur},
title = {Flipping bits in memory without accessing them: an experimental study of DRAM disturbance errors},
year = {2014},
isbn = {9781479943944},
publisher = {IEEE Press},
booktitle = {Proceeding of the 41st Annual International Symposium on Computer Architecuture},
pages = {361–372},
numpages = {12},
location = {Minneapolis, Minnesota, USA},
series = {ISCA '14}
}

@INPROCEEDINGS{Eccploit,
  author={Cojocar, Lucian and Razavi, Kaveh and Giuffrida, Cristiano and Bos, Herbert},
  booktitle={2019 IEEE Symposium on Security and Privacy (SP)}, 
  title={Exploiting Correcting Codes: On the Effectiveness of ECC Memory Against Rowhammer Attacks}, 
  year={2019},
  volume={},
  number={},
  pages={55-71},
  doi={10.1109/SP.2019.00089}}

@inproceedings {HalfDouble,
author = {Andreas Kogler and Jonas Juffinger and Salman Qazi and Yoongu Kim and Moritz Lipp and Nicolas Boichat and Eric Shiu and Mattias Nissler and Daniel Gruss},
title = {{Half-Double}: Hammering From the Next Row Over},
booktitle = {31st USENIX Security Symposium (USENIX Security 22)},
year = {2022},
isbn = {978-1-939133-31-1},
address = {Boston, MA},
pages = {3807--3824},
url = {https://www.usenix.org/conference/usenixsecurity22/presentation/kogler-half-double},
publisher = {USENIX Association},
month = aug
}

@inproceedings {Deephammer,
author = {Fan Yao and Adnan Siraj Rakin and Deliang Fan},
title = {{DeepHammer}: Depleting the Intelligence of Deep Neural Networks through Targeted Chain of Bit Flips},
booktitle = {USENIX Security},
year = {2020},
url = {https://www.usenix.org/conference/usenixsecurity20/presentation/yao},
}

@misc{jedec_ddr5_prac,
author={\mbox JEDEC },
title={{JESD79-5C}},
howpublished={\url{https://www.jedec.org/document_search?search_api_views_fulltext=jesd79-5c}},
note= {Accessed: 2025-01-22}
}

@article{JEDEC-DDR4,
  title={{DDR4 SDRAM standard (JESD79-4B)}},
  author={JEDEC},
  year={2017},
}

@article{JEDEC-DDR5,
  title={{DDR5 Specification}},
  author={JESD79-5},
  year={2020},
}

@article{JEDEC-HBM3,
  title={{HBM3 Specification}},
  author={JESD238A},
  year={2023},
}

@misc{marketShare2,
author={Jon Peddie Research},
title={{NVIDIA Market Share}},
howpublished={\url{https://www.jonpeddie.com/news/shipments-of-graphics-add-in-boards-decline-in-q1-of-24-as-the-market-experiences-a-return-to-seasonality/}},
note         = {Accessed: 2025-01-22}}

@inproceedings{tol2023dontknock,
  title={Don't Knock! Rowhammer at the Backdoor of DNN Models},
  author={Tol, M Caner and Islam, Saad and Adiletta, Andrew J and Sunar, Berk and Zhang, Ziming},
  booktitle={53rd Annual IEEE/IFIP International Conference on Dependable Systems and Networks (DSN)},
  year={2023}
}

@inproceedings{chen2021proflip,
  title={Proflip: Targeted trojan attack with progressive bit flips},
  author={Chen, Huili and Fu, Cheng and Zhao, Jishen and Koushanfar, Farinaz},
  booktitle={Proceedings of the IEEE/CVF International Conference on Computer Vision (ICCV)},
  pages={7718--7727},
  year={2021}
}

@inproceedings{DNNFaultInjection,
author = {Liu, Yannan and Wei, Lingxiao and Luo, Bo and Xu, Qiang},
title = {Fault injection attack on deep neural network},
year = {2017},
booktitle = {Proceedings of the 36th International Conference on Computer-Aided Design (ICCAD)},
}

@inproceedings{CrushingFlips,
  author={Rakin, Adnan Siraj and He, Zhezhi and Fan, Deliang},
  booktitle={2019 IEEE/CVF International Conference on Computer Vision (ICCV)}, 
  title={Bit-Flip Attack: Crushing Neural Network With Progressive Bit Search}, 
  year={2019},
  volume={},
  number={},
  pages={1211-1220},
  doi={10.1109/ICCV.2019.00130}}

@inproceedings{li2024_1bitflip,
  title={Yes, One-Bit-Flip Matters! Universal DNN Model Inference Depletion with Runtime Code Fault Injection},
  author={Li, Shaofeng and Wang, Xinyu and Xue, Minhui and Zhu, Haojin and Zhang, Zhi and Gao, Yansong and Wu, Wen and Shen, Xuemin Sherman},
  booktitle={Proceedings of the 33th USENIX Security Symposium},
  year={2024}
}

@misc{ RAPIDS,
              title = {RAPIDS Memory Manager},
              author = {RAPIDS AI},
              howpublished = {\url{https://github.com/rapidsai/rmm}}
            }

@misc{nvidia_ptx,
  title     = {{PTX: Parallel Thread Execution ISA, Version 8.5}},
  author    = {{NVIDIA}},
  year      = {2025},
  howpublished      = { \url{https://docs.nvidia.com/cuda/parallel-thread-execution/index.html}},
  note         = {Accessed: 2025-01-22}
}

@misc{chipsncheeseGPUs,
      title={Measuring GPU Memory Latency}, 
      author={Chester Lam},
      year={2021},
      howpublished={\url{https://chipsandcheese.com/p/measuring-gpu-memory-latency}}, 
      note= {Accessed: 2025-01-22}
}

@inproceedings{mlperf,
  title={Mlperf inference benchmark},
  author={Reddi, Vijay Janapa and Cheng, Christine and Kanter, David and Mattson, Peter and Schmuelling, Guenther and Wu, Carole-Jean and Anderson, Brian and Breughe, Maximilien and Charlebois, Mark and Chou, William and others},
  booktitle={2020 ACM/IEEE 47th Annual International Symposium on Computer Architecture (ISCA)},
  year={2020}
}

@inproceedings{HBMRowhammer,
  title={An experimental analysis of RowHammer in HBM2 DRAM chips},
  author={Olgun, Ataberk and Osseiran, Majd and Ya{\u{g}}l{\i}k{\c{c}}{\i}, A Giray and Tu{\u{g}}rul, Yahya Can and Luo, Haocong and Rhyner, Steve and Salami, Behzad and Luna, Juan Gomez and Mutlu, Onur},
  booktitle={2023 53rd Annual IEEE/IFIP International Conference on Dependable Systems and Networks-Supplemental Volume (DSN-S)},
  pages={151--156},
  year={2023},
  organization={IEEE}
}

@article{advexamples,
  title={Explaining and harnessing adversarial examples},
  author={Goodfellow, Ian J and Shlens, Jonathon and Szegedy, Christian},
  journal={arXiv preprint arXiv:1412.6572},
  year={2014}
}

@inproceedings{modelinv,
  title={Model inversion attacks that exploit confidence information and basic countermeasures},
  author={Fredrikson, Matt and Jha, Somesh and Ristenpart, Thomas},
  booktitle={Proceedings of the 22nd ACM SIGSAC conference on computer and communications security},
  pages={1322--1333},
  year={2015}
}

@article{jailbreakingLLMs,
  title={Jailbreaking black box large language models in twenty queries},
  author={Chao, Patrick and Robey, Alexander and Dobriban, Edgar and Hassani, Hamed and Pappas, George J and Wong, Eric},
  journal={arXiv preprint arXiv:2310.08419},
  year={2023}
}

@inproceedings{wang2020dramdig,
  title={Dramdig: A knowledge-assisted tool to uncover dram address mapping},
  author={Wang, Minghua and Zhang, Zhi and Cheng, Yueqiang and Nepal, Surya},
  booktitle={2020 57th ACM/IEEE Design Automation Conference (DAC)},
  year={2020}
}

@INPROCEEDINGS{PRIDE,
  author={Jaleel, Aamer and Saileshwar, Gururaj and Keckler, Stephen W. and Qureshi, Moinuddin},
  booktitle={2024 ACM/IEEE 51st Annual International Symposium on Computer Architecture (ISCA)}, 
  title={PrIDE: Achieving Secure Rowhammer Mitigation with Low-Cost In-DRAM Trackers}, 
  year={2024},
  volume={},
  number={},
  pages={1157-1172},
  doi={10.1109/ISCA59077.2024.00087}}

@inproceedings{RRS,
author = {Saileshwar, Gururaj and Wang, Bolin and Qureshi, Moinuddin and Nair, Prashant J.},
title = {Randomized row-swap: mitigating Row Hammer by breaking spatial correlation between aggressor and victim rows},
year = {2022},
booktitle = {27th ACM International Conference on Architectural Support for Programming Languages and Operating Systems (ASPLOS)}
}

@INPROCEEDINGS{Blockhammer,
  author={Yağlikçi, A. Giray and Patel, Minesh and Kim, Jeremie S. and Azizi, Roknoddin and Olgun, Ataberk and Orosa, Lois and Hassan, Hasan and Park, Jisung and Kanellopoulos, Konstantinos and Shahroodi, Taha and Ghose, Saugata and Mutlu, Onur},
  booktitle={HPCA}, 
  title={BlockHammer: Preventing RowHammer at Low Cost by Blacklisting Rapidly-Accessed DRAM Rows}, 
  year={2021},
  }

@inproceedings{Hydra,
author = {Qureshi, Moinuddin and Rohan, Aditya and Saileshwar, Gururaj and Nair, Prashant J.},
title = {Hydra: enabling low-overhead mitigation of row-hammer at ultra-low thresholds via hybrid tracking},
year = {2022},
isbn = {9781450386104},
publisher = {Association for Computing Machinery},
address = {New York, NY, USA},
url = {https://doi.org/10.1145/3470496.3527421},
doi = {10.1145/3470496.3527421},
booktitle = {Proceedings of the 49th Annual International Symposium on Computer Architecture},
pages = {699–710},
numpages = {12},
location = {New York, New York},
series = {ISCA '22}
}

@INPROCEEDINGS{TWiCe,
  author={Lee, Eojin and Kang, Ingab and Lee, Sukhan and Suh, G. Edward and Ahn, Jung Ho},
  booktitle={2019 ACM/IEEE 46th Annual International Symposium on Computer Architecture (ISCA)}, 
  title={TWiCe: Preventing Row-hammering by Exploiting Time Window Counters}, 
  year={2019},
  volume={},
  number={},
  pages={385-396},
  doi={}}

@inproceedings{Drammer,
author = {van der Veen, Victor and Fratantonio, Yanick and Lindorfer, Martina and Gruss, Daniel and Maurice, Clementine and Vigna, Giovanni and Bos, Herbert and Razavi, Kaveh and Giuffrida, Cristiano},
title = {{Drammer: Deterministic Rowhammer Attacks on Mobile Platforms}},
year = {2016},
url = {https://doi.org/10.1145/2976749.2978406},
booktitle = {2016 ACM SIGSAC Conference on Computer and Communications Security (CCS)},
}

@inproceedings {FFS,
author = {Kaveh Razavi and Ben Gras and Erik Bosman and Bart Preneel and Cristiano Giuffrida and Herbert Bos},
title = {Flip Feng Shui: Hammering a Needle in the Software Stack},
booktitle = {25th USENIX Security Symposium (USENIX Security)},
year = {2016},
url = {https://www.usenix.org/conference/usenixsecurity16/technical-sessions/presentation/razavi},
}

@inproceedings{RowhammerJS,
  title={Rowhammer. js: A remote software-induced fault attack in javascript},
  author={Gruss, Daniel and Maurice, Cl{\'e}mentine and Mangard, Stefan},
  booktitle={Detection of Intrusions and Malware, and Vulnerability Assessment: 13th International Conference, DIMVA 2016, San Sebasti{\'a}n, Spain, July 7-8, 2016, Proceedings 13},
  pages={300--321},
  year={2016},
  organization={Springer}
}

@inproceedings {ThrowHammer,
author = {Andrei Tatar and Radhesh Krishnan Konoth and Elias Athanasopoulos and Cristiano Giuffrida and Herbert Bos and Kaveh Razavi},
title = {Throwhammer: Rowhammer Attacks over the Network and Defenses},
booktitle = {2018 USENIX Annual Technical Conference (USENIX ATC)},
year = {2018},
url = {https://www.usenix.org/conference/atc18/presentation/tatar},
}

@inproceedings{MOAT,
  title={Moat: Securely mitigating rowhammer with per-row activation counters},
  author={Qureshi, Moinuddin and Qazi, Salman},
  booktitle={Proceedings of the 30th ACM International Conference on Architectural Support for Programming Languages and Operating Systems, Volume 1},
  pages={698--714},
  year={2025}
}

@INPROCEEDINGS{AQUA,
  author={Saxena, Anish and Saileshwar, Gururaj and Nair, Prashant J. and Qureshi, Moinuddin},
  booktitle={2022 55th IEEE/ACM International Symposium on Microarchitecture (MICRO)}, 
  title={AQUA: Scalable Rowhammer Mitigation by Quarantining Aggressor Rows at Runtime}, 
  year={2022},
  volume={},
  number={},
  pages={108-123},
  doi={10.1109/MICRO56248.2022.00022}}

@inproceedings {Aegis,
author = {Jialai Wang and Ziyuan Zhang and Meiqi Wang and Han Qiu and Tianwei Zhang and Qi Li and Zongpeng Li and Tao Wei and Chao Zhang},
title = {Aegis: Mitigating Targeted Bit-flip Attacks against Deep Neural Networks},
booktitle = {32nd USENIX Security Symposium (USENIX Security 23)},
year = {2023},
isbn = {978-1-939133-37-3},
address = {Anaheim, CA},
pages = {2329--2346},
url = {https://www.usenix.org/conference/usenixsecurity23/presentation/wang-jialai},
publisher = {USENIX Association},
month = aug
}

@INPROCEEDINGS{RADAR,
  author={Li, Jingtao and Rakin, Adnan Siraj and He, Zhezhi and Fan, Deliang and Chakrabarti, Chaitali},
  booktitle={2021 Design, Automation \& Test in Europe Conference \& Exhibition (DATE)}, 
  title={RADAR: Run-time Adversarial Weight Attack Detection and Accuracy Recovery}, 
  year={2021},
  volume={},
  number={},
  pages={790-795},
  doi={10.23919/DATE51398.2021.9474113}}

@inproceedings {NeuroPots,
author = {Qi Liu and Jieming Yin and Wujie Wen and Chengmo Yang and Shi Sha},
title = {{NeuroPots}: Realtime Proactive Defense against {Bit-Flip} Attacks in Neural Networks},
booktitle = {32nd USENIX Security Symposium (USENIX Security)},
year = {2023},
url = {https://www.usenix.org/conference/usenixsecurity23/presentation/liu-qi},
}

@inproceedings{DRAMScope,
  author={Nam, Hwayong and Baek, Seungmin and Wi, Minbok and Kim, Michael Jaemin and Park, Jaehyun and Song, Chihun and Kim, Nam Sung and Ahn, Jung Ho},
  booktitle={2024 ACM/IEEE 51st Annual International Symposium on Computer Architecture (ISCA)}, 
  title={DRAMScope: Uncovering DRAM Microarchitecture and Characteristics by Issuing Memory Commands}, 
  year={2024},
  volume={},
  number={},
  pages={1097-1111},
  doi={10.1109/ISCA59077.2024.00083}}

@misc{nvidia_cuda_samples,
  author       = {NVIDIA Corporation},
  title        = {CUDA Samples},
  year         = {2023},
  howpublished = {\url{https://github.com/NVIDIA/cuda-samples}},
  note         = {Accessed: 2025-01-22}
}

@inproceedings{GPUPT1,
author = {Jang, Sungbin and Park, Junhyeok and Kwon, Osang and Lee, Yongho and Hong, Seokin},
title = {Rethinking Page Table Structure for Fast Address Translation in GPUs: A Fixed-Size Hashed Page Table},
year = {2024},
isbn = {9798400706318},
publisher = {Association for Computing Machinery},
address = {New York, NY, USA},
url = {https://doi.org/10.1145/3656019.3676900},
doi = {10.1145/3656019.3676900},
booktitle = {Proceedings of the 2024 International Conference on Parallel Architectures and Compilation Techniques},
pages = {325–337},
numpages = {13},
location = {Long Beach, CA, USA},
series = {PACT '24}
}

@misc{webGPU,
  author       = {Google Chrome},
  title        = {WebGPU: Unlocking modern GPU access in the browser},
  year         = {2023},
  howpublished = {\url{https://developer.chrome.com/blog/webgpu-io2023}},
  note         = {Accessed: 2025-01-22}
}

@inproceedings{RamBleed,
  title={Rambleed: Reading bits in memory without accessing them},
  author={Kwong, Andrew and Genkin, Daniel and Gruss, Daniel and Yarom, Yuval},
  booktitle={2020 IEEE Symposium on Security and Privacy (SP)},
  pages={695--711},
  year={2020},
  organization={IEEE}
}

@inproceedings{RowPress,
  title={Rowpress: Amplifying read disturbance in modern dram chips},
  author={Luo, Haocong and Olgun, Ataberk and Ya{\u{g}}l{\i}k{\c{c}}{\i}, Abdullah Giray and Tu{\u{g}}rul, Yahya Can and Rhyner, Steve and Cavlak, Meryem Banu and Lindegger, Jo{\"e}l and Sadrosadati, Mohammad and Mutlu, Onur},
  booktitle={50th International Symposium on Computer Architecture (ISCA)},
  year={2023}
}

@misc{GDDR7ECC,
  author       = {AnandTech},
  title        = {{JEDEC Publishes GDDR7 Memory Spec: Next-Gen Graphics Memory Adds Faster PAM3 Signaling and On-Die ECC}},
  year         = {2024},
  howpublished = {\url{https://www.anandtech.com/show/21287/jedec-publishes-gddr7-specifications-pam3-ecc-higher-density
}},
  note         = {Accessed: 2025-01-22}
}

@misc{HBM3ECC,
  author       = {JEDEC},
  title        = {{JEDEC Publishes HBM3 Update to High Bandwidth Memory (HBM) Standard}},
  year         = {2022},
  howpublished = {\url{https://www.jedec.org/news/pressreleases/jedec-publishes-hbm3-update-high-bandwidth-memory-hbm-standard }},
  note         = {Accessed: 2025-01-22}
}

@inproceedings{IMTSullivan,
  title={Implicit memory tagging: No-overhead memory safety using alias-free tagged ecc},
  author={Sullivan, Michael B and Ziad, Mohamed Tarek Ibn and Jaleel, Aamer and Keckler, Stephen W},
  booktitle={50th Annual International Symposium on Computer Architecture (ISCA)},
  year={2023}
}

@inproceedings{ASAN,
  title={{AddressSanitizer: A fast address sanity checker}},
  author={Serebryany, Konstantin and Bruening, Derek and Potapenko, Alexander and Vyukov, Dmitriy},
  booktitle={{USENIX Annual Technical Conference (USENIX ATC)}},
  year={2012}
}

@inproceedings{GrandPwning,
  title={{Grand pwning unit: Accelerating microarchitectural attacks with the GPU}},
  author={Frigo, Pietro and Giuffrida, Cristiano and Bos, Herbert and Razavi, Kaveh},
  booktitle={2018 ieee symposium on security and privacy (sp)},
  pages={195--210},
  year={2018},
  organization={IEEE}
}

@inproceedings{Siloz,
  title={Siloz: Leveraging DRAM Isolation Domains to Prevent Inter-VM Rowhammer},
  author={Loughlin, Kevin and Rosenblum, Jonah and Saroiu, Stefan and Wolman, Alec and Skarlatos, Dimitrios and Kasikci, Baris},
  booktitle={29th Symposium on Operating Systems Principles (SOSP)},
  year={2023}
}

@inproceedings{van2018guardion,
  title={GuardION: Practical mitigation of DMA-based rowhammer attacks on ARM},
  author={Van der Veen, Victor and Lindorfer, Martina and Fratantonio, Yanick and Pillai, Harikrishnan Padmanabha and Vigna, Giovanni and Kruegel, Christopher and Bos, Herbert and Razavi, Kaveh},
  booktitle={International Conference on Detection of Intrusions and Malware, and Vulnerability Assessment},
  pages={92--113},
  year={2018},
  organization={Springer}
}

@inproceedings{brasser2017can,
  title={CAn’t touch this: Software-only mitigation against Rowhammer attacks targeting kernel memory},
  author={Brasser, Ferdinand and Davi, Lucas and Gens, David and Liebchen, Christopher and Sadeghi, Ahmad-Reza},
  booktitle={{26th USENIX Security Symposium (USENIX Security 17)}},
  pages={117--130},
  year={2017}
}

@inproceedings{konoth2018zebram,
  title={ZebRAM: comprehensive and compatible software protection against rowhammer attacks},
  author={Konoth, Radhesh Krishnan and Oliverio, Marco and Tatar, Andrei and Andriesse, Dennis and Bos, Herbert and Giuffrida, Cristiano and Razavi, Kaveh},
  booktitle={{13th USENIX - (OSDI 18)}},
  pages={697--710},
  year={2018}
}

@inproceedings{bock2019rip,
  title={RIP-RH: Preventing rowhammer-based inter-process attacks},
  author={Bock, Carsten and Brasser, Ferdinand and Gens, David and Liebchen, Christopher and Sadeghi, Ahamd-Reza},
  booktitle={Proceedings of the 2019 ACM Asia Conference on Computer and Communications Security},
  pages={561--572},
  year={2019}
}

@inproceedings{ProTRR,
  title={{Protrr: Principled yet optimal in-dram target row refresh}},
  author={Marazzi, Michele and Jattke, Patrick and Solt, Flavien and Razavi, Kaveh},
  booktitle={{IEEE Symposium on Security and Privacy (SP)}},
  pages={735--753},
  year={2022},
  organization={IEEE}
}

@inproceedings{Shadow,
  title={{SHADOW: Preventing Row Hammer in DRAM with Intra-Subarray Row Shuffling}},
  author={Wi, Minbok and Park, Jaehyun and Ko, Seoyoung and Kim, Michael Jaemin and Kim, Nam Sung and Lee, Eojin and Ahn, Jung Ho},
  booktitle={{HPCA}},
  year={2023},
}

@inproceedings{SRS,
  title={Scalable and Secure Row-Swap: Efficient and Safe Row Hammer Mitigation in Memory Systems},
  author={Woo, Jeonghyun and Saileshwar, Gururaj and Nair, Prashant J},
  booktitle={HPCA},
  year={2023},
}

@inproceedings{REGA_SP23,
  title={{REGA: Scalable Rowhammer Mitigation with Refresh-Generating Activations}},
  author={Marazzi, Michele and Solt, Flavien and Jattke, Patrick and Takashi, Kubo and Razavi, Kaveh},
  booktitle={{IEEE Symposium on Security and Privacy (SP)}},
  year={2023}
}

@inproceedings{park2020graphene,
  title={Graphene: Strong yet Lightweight Row Hammer Protection},
  author={Park, Yeonhong and Kwon, Woosuk and Lee, Eojin and Ham, Tae Jun and Ahn, Jung Ho and Lee, Jae W},
  booktitle={2020 53rd Annual IEEE/ACM International Symposium on Microarchitecture (MICRO)},
  pages={1--13},
  year={2020},
  organization={IEEE}
}

@INPROCEEDINGS {mithril,
author = {M. Kim and J. Park and Y. Park and W. Doh and N. Kim and T. Ham and J. W. Lee and J. Ahn},
booktitle = {2022 IEEE International Symposium on High-Performance Computer Architecture (HPCA)},
title = {Mithril: Cooperative Row Hammer Protection on Commodity DRAM Leveraging Managed Refresh},
year = {2022},
volume = {},
issn = {},
pages = {1156-1169},
doi = {10.1109/HPCA53966.2022.00088},
url = {https://doi.ieeecomputersociety.org/10.1109/HPCA53966.2022.00088},
publisher = {IEEE Computer Society},
address = {Los Alamitos, CA, USA},
month = {apr}
}

@misc{NVIDIA_MIG, title={NVIDIA Multi-Instance GPU and NVIDIA Virtual Compute Server}, 
howpublished={\url{https://www.nvidia.com/content/dam/en-zz/Solutions/design-visualization/solutions/resources/documents1/Technical-Brief-Multi-Instance-GPU-NVIDIA-Virtual-Compute-Server.pdf}},
 author={NVIDIA}, year={2020},
note         = {Accessed: 2025-01-22}}

@article{NVIDIA_CC,
author = {Dhanuskodi, Gobikrishna and Guha, Sudeshna and Krishnan, Vidhya and Manjunatha, Aruna and O'Connor, Michael and Nertney, Rob and Rogers, Phil},
title = {Creating the First Confidential GPUs: The team at NVIDIA brings confidentiality and integrity to user code and data for accelerated computing.},
year = {2023},
issue_date = {July/August 2023},
publisher = {Association for Computing Machinery},
address = {New York, NY, USA},
volume = {21},
number = {4},
issn = {1542-7730},
url = {https://doi.org/10.1145/3623393.3623391},
doi = {10.1145/3623393.3623391},
journal = {Queue},
month = sep,
pages = {68–93},
numpages = {26}
}

@article{MIGnificient, title={MIGnificient: Fast, Isolated, and GPU-Enabled Serverless Functions}, url={https://sc24.supercomputing.org/proceedings/poster/poster_files/post236s2-file3.pdf}, journal={SC ’24: Proceedings of the International Conference for High Performance Computing, Networking, Storage, and Analysis}, author={Copik, Marcin and Calotoiu, Alexandru and Zhou, Pengyu and University of Toronto and Tobler, Lukas and Hoefler, Torsten and ETH Z¨urich and AYES} }

@misc{TimeSlicing, title={{GPU Time Slicing}}, author={NVIDIA Run:ai Docs}, howpublished={\url{https://docs.run.ai/v2.17/Researcher/scheduling/GPU-time-slicing-scheduler/}}, language={en},  note         = {Accessed: 2025-01-22} }

@misc{GKE, 
title= {{Share GPUs across workloads with GPU time-sharing}},
author={Google},
howpublished={\url{https://cloud.google.com/kubernetes-engine/docs/how-to/timesharing-gpus}},
note= {Accessed: 2025-01-22}
}

@misc{RAPIDS_endors, 
title={Fast, Flexible Allocation for NVIDIA CUDA with RAPIDS Memory Manager}, 
howpublished={\url{https://developer.nvidia.com/blog/fast-flexible-allocation-for-cuda-with-rapids-memory-manager/}}, 
author={Harris, Mark and Harris, Mark},
note= {Accessed: 2025-01-22}
}

@article{HBM2eECC,
  title={{A 16-GB 640-GB/s HBM2E DRAM with a data-bus window extension technique and a synergetic on-die ECC scheme}},
  author={Chun, Ki Chul and Kim, Yong Ki and Ryu, Yesin and Park, Jaewon and Oh, Chi Sung and Byun, Young Yong and Kim, So Young and Shin, Dong Hak and Lee, Jun Gyu and Ho, Byung-Kyu and others},
  journal={IEEE Journal of Solid-State Circuits},
  volume={56},
  number={1},
  pages={199--211},
  year={2020},
  publisher={IEEE}
}

@misc{ECCDisable1, 
title= {Impact of enabling ECC on power and performance},
author={NVIDIA Developer Forums},
howpublished={\url{https://forums.developer.nvidia.com/t/impact-of-enabling-ecc-on-power-and-performance/174567}}, 
note= {Accessed: 2025-01-22}
}

@misc{ECCDisable2, 
title= {ECC on vs ECC off},
author={NVIDIA Developer Forums},
howpublished={\url{https://forums.developer.nvidia.com/t/ecc-on-vs-ecc-off/20315}},
note= {Accessed: 2025-01-22}
}

@inproceedings{QPRAC,
  title={Qprac: Towards secure and practical prac-based rowhammer mitigation using priority queues},
  author={Woo, Jeonghyun and Lin, Shaopeng Chris and Nair, Prashant J and Jaleel, Aamer and Saileshwar, Gururaj},
  booktitle={2025 IEEE International Symposium on High Performance Computer Architecture (HPCA)},
  year={2025}
}

@inproceedings{Chronus,
  title={Chronus: Understanding and Securing the Cutting-Edge Industry Solutions to DRAM Read Disturbance},
  author={Canpolat, O{\u{g}}uzhan and Ya{\u{g}}l{\i}k{\c{c}}{\i}, A Giray and Oliveira, Geraldo F and Olgun, Ataberk and Bostanc{\i}, Nisa and Yuksel, Ismail Emir and Luo, Haocong and Ergin, O{\u{g}}uz and Mutlu, Onur},
  booktitle={2025 IEEE International Symposium on High Performance Computer Architecture (HPCA)},
  year={2025}
}

\appendix
%\section{Appendices}
%\TODO{Bank Set Explanation here}

\section*{Appendix}

\begin{appendices}

\section{Synchronized Many-Sided Hammer Kernel}
%Multi-Warp/Multi-Thread
%K-warps x M-threads Hammer Kernel} 
\label{app:Nwarp_Kthread}
%We leverage similar nomenclature defined in \cref{sec:sync} in this sample CUDA kernel. 

%As described in \cref{sec:sync}, we perform
Our synchronized many-sided hammering uses a CUDA kernel with $k$ warps and $m$ threads per warp, \OTHERCHANGE{hammering} \verb|k| $\times$ \verb|m| \OTHERCHANGE{rows}.
%The aggressor addresses are in \verb|addr_arr|.
%with its size defined by \verb|k| $\times$ \verb|m|. 
The kernel is launched as \verb|hammer<<<1, 1024>>>| to ensure a sufficient number of warps\OTHERCHANGE{. Each warp contains 32 threads, of which the first $m$ threads are used for hammering.}
%Our indexing assumes every 32 threads belong to the same warp, and 

\begin{comment}
As described in \cref{sec:sync}, we perform synchronized many-sided hammering using a CUDA kernel with $k$ warps and $m$ threads per warp.
The total \OTHERCHANGE{hammered rows} are \verb|k| $\times$ \verb|m|.
The aggressor row addresses are in \verb|addr_arr|.
%with its size defined by \verb|k| $\times$ \verb|m|. 
The kernel is launched as \verb|hammer<<<1, 1024>>>| to ensure the sufficient number of warps allocated in the SM. Our indexing assumes every 32 threads belong to the same warp, and we activate only the first $m$ threads in each warp for hammering.
\end{comment}

\begin{lstlisting}[caption={$k$-warps, $m$-threads per warp Hammering Kernel},label={lstlisting:synchammering}]
void hammer(size_t **aggr_addr, size_t it, size_t k, size_t m, size_t round, size_t delay) {
  size_t dummy, dummy_sum;
  size_t warpId = threadIdx.x / 32;
  size_t threadId_in_warp = threadIdx.x % 32;

  if (warpId < k && threadId_in_warp < m) {
    size_t *addr = *(aggr_addr + threadId_in_warp + warpId * m);
    asm volatile("discard.global.L2 [%0], 128;"
                ::"l"(addr));
    __syncthreads();
    /* Hammers performed for 128ms */
    for (;count--;) {
      /* perform 'round' ACTs before sync delay */
      for (size_t i = round; i--;) {
        asm volatile("discard.global.L2 [%0], 
                128;" ::"l"(addr));
        /* Aggressor Row Activation*/
        asm volatile("ld.u64.global.volatile %0, 
                [%1];" : "=l"(dummy) : "l"(addr));
        /* Ordered memory access within threads */
        __threadfence_block();
      }
      /* Delay Added to Synchronize to REF */
      for (size_t i = delay; i--;)
        dummy_sum += dummy;
    }
  }
}
\end{lstlisting}
\vspace{-0.1in}

\section{Virtual Address to DRAM Row Mapping}\label{app:non-linear}
\begin{figure}[ht]
\centering
\includegraphics[width=3.3in,height=\paperheight,keepaspectratio]
{"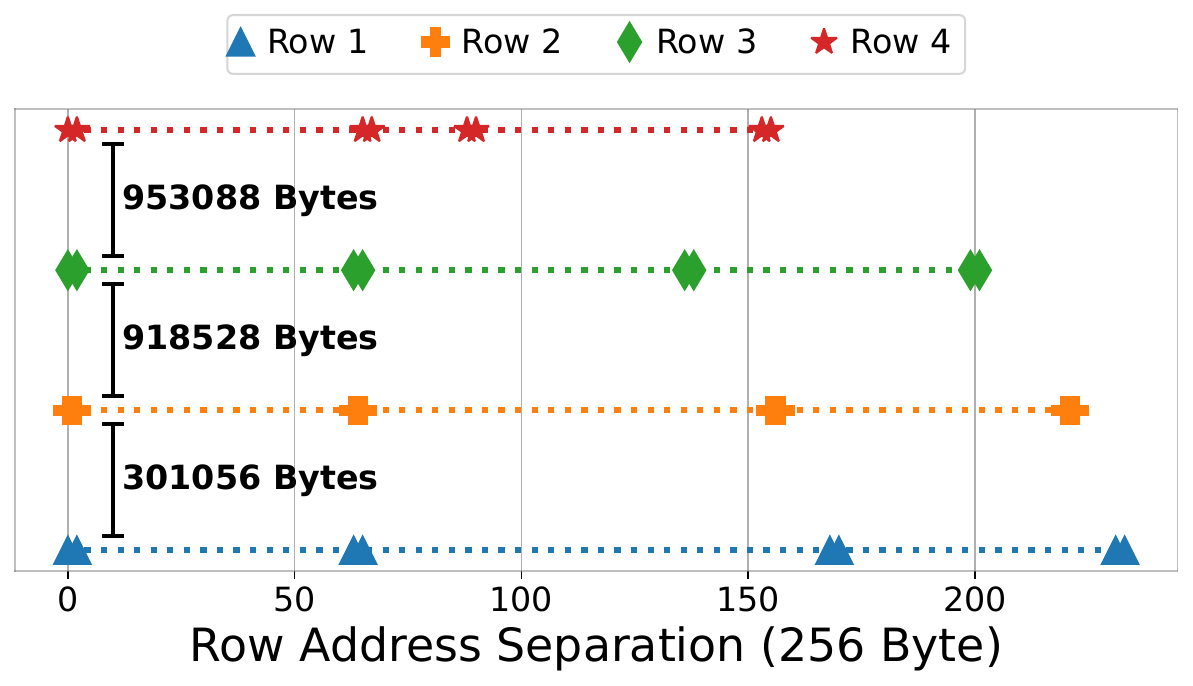"}
\caption{Mapping of virtual addresses 
%of a large 47GB array 
to DRAM rows in a given bank, for four different rows (Rows 1 to 4) at 256-byte granularity. Each 256 byte chunk is one point on the graph.} 
%We observe that 8 chunks map to a 2KB DRAM row.
%We hypothesize that the mapping uses a \OTHERCHANGE{complex} hashing function.}
%2KB DRAM rows are separated into 256B chunks for virtual memory mapping. Row address mappings are offset by a non-linear value with respect to the previous mapping position. }
\label{fig:non-linear}
\end{figure}

A 2KB DRAM row is divided into 256-byte chunks, with each chunk mapped to a different physical bank and row. 
As we cannot access GPU physical memory, we allocate a large array in virtual memory (47GB out of 48GB), and identify the 256B chunks in virtual addresses (VA) that map to unique rows, as shown in \cref{fig:non-linear}.
A 256-byte chunk maps to a new DRAM row within a bank only after all the chunks of the prior row have been mapped. However, the distance between successive rows in a bank does not appear to have any obvious trends. This suggests that \OTHERCHANGE{a complex function might be used to map memory addresses} to DRAM bank/rows.
%of a non-linear hashing function in the mapping of virtual to physical memory.

\section{Characterizing Bit-Flip Data Patterns}
\label{subsec:data_pat_appdx}
% \begin{figure}[htbp]
%     \centering
%     \begin{subfigure}[b]{\linewidth}
%         \centering
%         \includegraphics[width=0.95\linewidth]{Figures/data_pat_0_to_1.pdf}
%         \caption{A $0 \rightarrow 1$ bit-flip at the $6^\text{th}$ bit ($D_1$)}
%         \label{fig:pat_0_to_1}
%     \end{subfigure}
    
%     \vspace{0.3cm} % Adjust vertical space between figures
    
%     \begin{subfigure}[b]{\linewidth}
%         \centering
%         \includegraphics[width=0.56\linewidth]{Figures/data_pat_1_to_0.pdf}
%         \caption{A $1 \rightarrow 0$ bit-flip at the $4^\text{th}$ bit ($A_1$)}
%         \label{fig:pat_1_to_0}
%     \end{subfigure}

%     \caption{Heatmap displaying percentage of hammers that triggered bit-flips while varying victim and aggressor data patterns. We sort the victim and aggressor data patterns in descending order of bit-flip rate. Patterns that did not induce any bit-flips are excluded.}
%     \label{fig:data_pat}
% \end{figure}

\begin{figure}[htbp]
    \centering
    \begin{subfigure}[c]{0.61\linewidth}
        \centering
        \includegraphics[width=\linewidth]{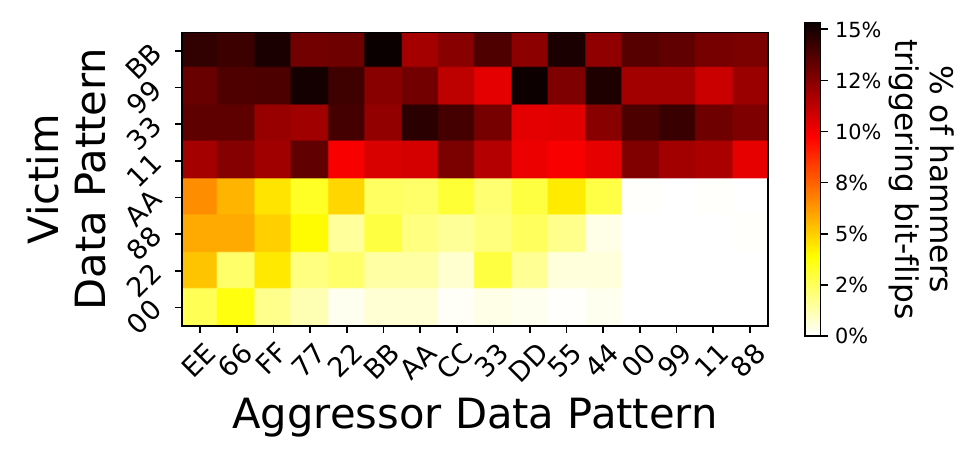}
        \caption{A $0 \rightarrow 1$ bit-flip \\at the $6^\text{th}$ bit ($D_1$)}
        \label{fig:pat_0_to_1}
    \end{subfigure}
    \hfill
    \begin{subfigure}[c]{0.38\linewidth}
        \centering
        \includegraphics[width=\linewidth]{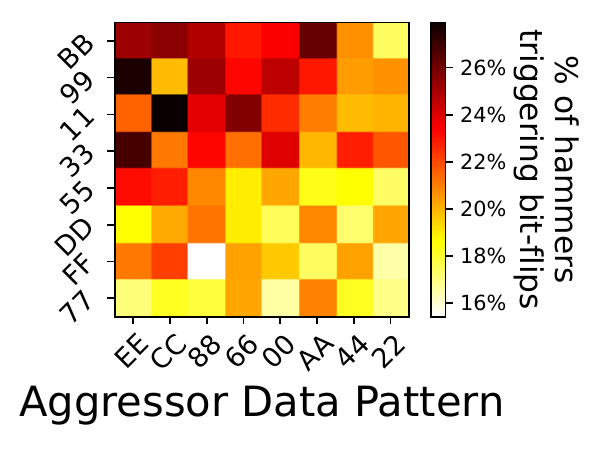}
        \caption{A $1 \rightarrow 0$ bit-flip \\at the $4^\text{th}$ bit ($A_1$)}
        \label{fig:pat_1_to_0}
    \end{subfigure}
    
    \caption{Heatmap showing fraction of hammers triggering bit-flips with varying victim and aggressor data patterns.}
    %We sort the victim and aggressor data patterns in descending order of bit-flip rate. Patterns that did not induce any bit-flips are excluded.}
    \label{fig:data_pat}
\end{figure}

\noindent
We analyze the impact of data patterns on bit-flip frequency by varying the victim and aggressor data from \texttt{0x00} to \texttt{0xFF}, with a step size of \texttt{0x11}, using a 24-sided pattern on our A6000 GPU.
We measure the bit flip frequency as the data values vary for a $0 \rightarrow 1$ flip ($D_1$) and a $1 \rightarrow 0$ flip ($A_1$), as shown in \cref{fig:pat_0_to_1} and \cref{fig:pat_1_to_0} respectively. 
%We measure the bit flip frequency while varying the victim and aggressor data from \texttt{0x00} to \texttt{0xFF}, with a step size of \texttt{0x11}, using a 24-sided pattern.
%\cref{fig:pat_0_to_1} and \cref{fig:pat_1_to_0} show the percentage of hammers causing a bit flip
\begin{comment}
For our A6000 GPU, we use two bit-flips we discovered, a $0 \rightarrow 1$ ($D_1$) and a $1 \rightarrow 0$ flip ($A_1$), and the respective aggressor patterns that resulted in the bit-flip, and analyzed the effect of data values in the victim and aggressor rows on bit-flip frequency. We varied victim and aggressor data patterns from \texttt{0x00} to \texttt{0xFF}, with a step size of \texttt{0x11}. For each combination, we launched 24-sided hammers.
%with sequential aggressor patterns, targeting known bit-flip locations.
The percentage of hammers causing bit-flip across the data patterns for the $0 \rightarrow 1$ and $1 \rightarrow 0$ flips are shown in \cref{fig:pat_0_to_1} and \cref{fig:pat_1_to_0}, respectively. 
In both figures, we see a strong correlation between aggressor and victim data and the frequency of bit-flips.
\end{comment}

\smallskip
\noindent\textbf{Aggressor Data Influence.}
The aggressor bits directly above and below the victim bit strongly influence the rate of bit-flips, with the highest rate achieved when these bits are the inverse of the victim bit. For example, in the $1 \rightarrow 0$ flip ($A_1$, 4th bit), when victim data is \texttt{0x55} (\texttt{0b010\textbf{1}0101}), the aggressor data with \texttt{0xEE} (\texttt{0b111\textbf{0}1110}) or \texttt{0xCC} (\texttt{0b110\textbf{0}1100}) achieves high bit-flip rates.  
For the $0 \rightarrow 1$ flip ($D_1$, 6th bit), additionally, the aggressor bit diagonally opposite the vulnerable victim bit has an influence.
For instance, when victim data is \texttt{0xAA} (\texttt{0b1\textbf{0}101010}), even the aggressor data \texttt{0x33} (\texttt{0b0\textbf{01}10011}) triggers bit flips in addition to \texttt{0xEE} (\texttt{0b1\textbf{1}101110}).
%This leads to more effective aggressor data patterns for  $0 \rightarrow 1$ flips. 

%Conversely, for $0 \rightarrow 1$ flips, aggressor bits diagonally adjacent to the victim bit also contribute to bit-flip occurrences, resulting in more effective patterns.

\smallskip
\noindent\textbf{Victim Data Influence.}
The victim data has a stronger influence on the rate of bit-flips than aggressor data. For both the $1 \rightarrow 0$ flip ($A_1$, 4th bit)  and the $0 \rightarrow 1$ flip ($D_1$, 6th bit), the victim data of \texttt{0xBB} (\texttt{0b\textbf{1}0\textbf{1}1\textbf{1}011}) and \texttt{0x99} (\texttt{0b\textbf{1}0\textbf{0}1\textbf{1}001}) are the most effective. Interestingly, for the $0 \rightarrow 1$ flip ($D_1$, 6th bit), ($D_1$), \texttt{0x11} (\texttt{0b\textbf{0}0\textbf{0}10001}), which surrounds the victim bit with two bits of the same charge, is more effective than the gridded pattern \texttt{0xAA} (\texttt{0b\textbf{1}0\textbf{1}01010}), where the victim bit is surrounded by two bits of the opposite charge.

\end{appendices}

%%%%%%%%%%%%%%%%%%%%%%%%%%%%%%%%%%%%%%%%%%%%%%%%%%%%%%%%%%%%%%%%%%%%%%%%%%%%%%%%
\end{document}